%% file: alchemi_pca.tex
\documentclass[twocolumn]{aastex631}
\usepackage{comment,amsmath,longtable, bm} % bm is for bold-math

\newcommand{\propyne}{CH$_3$CCH}
\newcommand{\twco}{$^{12}$CO}
\newcommand{\ttco}{$^{13}$CO}
\newcommand{\ceto}{C$^{18}$O}
\newcommand{\nthp}{N$_2$H$^+$}
\newcommand{\hcop}{HCO$^+$}
\newcommand{\hocp}{HOC$^+$}
\newcommand{\methanol}{CH$_3$OH}
\newcommand{\hocop}{HOCO$^+$}
\newcommand{\hctn}{HC$_3$N}
\newcommand{\hcfn}{HC$_5$N}
\newcommand{\cthp}{C$_3$H$^+$}
\newcommand{\hcnhp}{HCNH$^+$}

\newcommand{\waterp}{H$_3$O$^+$}

\newcommand{\kms}{km\,s$^{-1}$}
\newcommand{\cc}{cm$^{-3}$}

\newcommand{\naoj}{National Astronomical Observatory of Japan, 2-21-1 Osawa, Mitaka, Tokyo 181-8588, Japan}
\newcommand{\asiaa}{Institute of Astronomy and Astrophysics, Academia Sinica, 11F of AS/NTU
Astronomy-Mathematics Building, No.1, Sec. 4, Roosevelt Rd, Taipei 10617, Taiwan}
\newcommand{\sokendai}{Astronomical Science Program, Graduate Institute for Advanced Studies, SOKENDAI, 2-21-1 Osawa, Mitaka, Tokyo 181-1855, Japan}
\newcommand{\nmimt}{New Mexico Institute of Mining and Technology, 801 Leroy Place, Socorro, NM 87801, USA}
\newcommand{\nraos}{National Radio Astronomy Observatory, PO Box O, 1003 Lopezville Road, Socorro, NM 87801, USA}
\newcommand{\esoc}{European Southern Observatory, Alonso de C\'ordova, 3107, Vitacura, Santiago 763-0355, Chile}
\newcommand{\jao}{Joint ALMA Observatory, Alonso de C\'ordova, 3107, Vitacura, Santiago 763-0355, Chile}
\newcommand{\oso}{Department of Space, Earth and Environment, Chalmers University of Technology, Onsala Space Observatory, SE-439 92 Onsala, Sweden}
\newcommand{\mpifr}{Max-Planck-Institut f\"ur Radioastronomie, Auf dem H\"ugel 69, 53121 Bonn, Germany}
\newcommand{\kau}{Astronomy Department, Faculty of Science, King Abdulaziz
  University, P.~O.~Box 80203, Jeddah 21589, Saudi Arabia}
\newcommand{\keio}{Department of Physics, Faculty of Science and Technology, Keio University, 3-14-1 Hiyoshi, Yokohama, Kanagawa 223--8522 Japan}
\newcommand{\nraoc}{National Radio Astronomy Observatory, 520 Edgemont Road, Charlottesville, VA  22903-2475, USA}

\newcommand{\shibaura}{Materials Science and Engineering, College of Engineering, Shibaura Institute of Technology, 3-7-5 Toyosu, Koto-ku, Tokyo 135-8548, Japan}

\newcommand{\cisc}{Centro de Astrobiolog\'ia (CAB), CSIC-INTA, Ctra. de Ajalvir Km. 4, Torrej\'on de Ardoz, 28850 Madrid, Spain}

\newcommand{\leiden}{Leiden Observatory, Leiden University, P.O. Box 9513, 2300 RA Leiden, The Netherlands}

\newcommand{\ioa}{Institute of Astronomy, Graduate School of Science,
The University of Tokyo, 2-21-1 Osawa, Mitaka, Tokyo 181-0015, Japan}

\newcommand{\nihon}{Department of Physics, General Studies,
College of Engineering, Nihon University, Tamura-machi,
Koriyama, Fukushima 963-8642, Japan}

\newcommand{\utastro}{Department of Astronomy, The University of Tokyo, 7-3-1 Hongo, Bunkyo-ku, Tokyo 113-0033, Japan}
\newcommand{\ciera}{Center for Interdisciplinary Exploration and Research in Astrophysics (CIERA) and Department of Physics and Astronomy, Northwestern University, Evanston, IL 60208, USA}
\newcommand{\chalmers}{Department of Space, Earth and Environment, Chalmers University of Technology, SE-412 96 G\"oteborg, Sweden}
\shorttitle{ALCHEMI PCA}
\shortauthors{Harada et al.}
\graphicspath{{./}{figures/}}

\defcitealias{2021A&A...656A..46M}{MMH21}
\begin{document}

%\title{ALCHEMI: ATLAS of unblended transitions in NGC~253 and principal component analysis}
\title{The ALCHEMI atlas: principal component analysis reveals starburst evolution in NGC~253}
\input{ALCHEMICollabAuthList}



\begin{abstract}
%background
Molecular lines are powerful diagnostics of the physical and chemical properties of the interstellar medium (ISM). These ISM properties, which affect future star formation, are expected to differ in starburst galaxies from those of more quiescent galaxies.
We investigate the ISM properties in the central molecular zone of the nearby starburst galaxy NGC~253 using the ultra-wide millimeter spectral scan survey from the ALMA Large Program ALCHEMI. 
We present an atlas of velocity-integrated images at a 1\farcs6 resolution of 148 unblended transitions from 44 species, including the first extragalactic detection of HCNH$^+$ and the first interferometric images of C$_3$H$^+$, NO, HCS$^+$. We conduct a principal component analysis (PCA) on these images to extract correlated chemical species and to identify key groups of diagnostic transitions. To the best of our knowledge, our dataset is currently the largest astronomical set of molecular lines to which PCA has been applied. 
The PCA can categorize transitions coming from different physical components in NGC~253 such as i) young starburst tracers characterized by high-excitation transitions of \hctn\ and complex organic molecules (COMs) versus tracers of on-going star formation (radio recombination lines) and high-excitation transitions of CCH and CN tracing PDRs, ii) tracers of cloud-collision-induced shocks (low-excitation transitions of \methanol, HNCO, \hocop, and OCS) versus shocks from star-formation-induced outflows (high-excitation transitions of SiO),  as well as iii) outflows showing emission from \hocp, CCH, \waterp, CO isotopologues, HCN, \hcop, CS, and CN.
Our findings show these intensities vary with galactic dynamics, star formation activities, and stellar feedback. 
\end{abstract}



\section{Introduction} \label{sec:intro}
The physical properties of molecular gas determine where and how future stars will be forming. A powerful way to investigate the gas properties is via the study of molecular line emission; the physical properties and processes taking place during star formation affect rates of various chemical reactions and thus leave an imprint on the chemical properties of molecular clouds, star-forming regions, and protoplanetary systems. 

\sloppypar{While the Milky Way provides a good example of a galaxy with a relatively quiescent star formation activity, it is of great interest to extend molecular inventories to extragalactic sources and to probe the effects of extreme starbursts and active galactic nuclei (AGNs), which cannot be studied in our Galaxy due to the absence of such extreme activities \citep{2005ApJ...618..259M,2006ApJS..164..450M,2011A&A...527A..36M,2015AA...579A.101A,2018PASJ...70....7N,2019PASJ...71S..20T}. The Atacama Large Millimeter/submillimeter Array (ALMA) has enabled high angular resolution astrochemical observations targeting regions with extreme activity in the AGN-host NGC~1068 \citep{2014PASJ...66...75T} and NGC~1097 \citep{2015A&A...573A.116M}, the starburst galaxies NGC~253 \citep{2015ApJ...801...63M} and M83 \citep{2019ApJ...884..100H}, AGN-starburst composite NGC 4945 \citep{2018A&A...615A.155H,2023MNRAS.519L..68B}, the ultra-/luminous infrared galaxies (U/LIRGs) NGC~4418 \citep{2015A&A...582A..91C}, NGC~3256 \citep{2018ApJ...855...49H}, and Arp~220 \citep{2021ApJ...923..240S}.\footnote{See Table 1 of \citet{2021A&A...656A..46M} for the references of extragalactic spectral line scan studies.}}

Previous observations found that the center of NGC~253 is particularly chemically rich \citep[e.g., ][]{2006ApJS..164..450M,2015AA...579A.101A,2015ApJ...801...63M,2017ApJ...849...81A,2021A&A...656A..46M}. It is a barred spiral galaxy at a distance of 3.5 Mpc \citep{2005MNRAS.361..330R}. It hosts $\sim 2\,M_{\odot}$\, yr$^{-1}$ of star formation within the central molecular zone (CMZ) extending to several hundred parsecs in size \citep{leroy_alma_2015,2015MNRAS.450L..80B}. The CMZ of NGC~253 is rich in molecular gas \citep{sakamoto_star-forming_2011,leroy_alma_2015,leroy_forming_2018}. It also hosts outflows launched from the energetic starburst \citep{1985ApJ...299..312T,bolatto_suppression_2013,2017ApJ...835..265W,2019ApJ...881...43K,2021ApJ...912....4L}.

To investigate further the physical and chemical properties of the CMZ in NGC~253, and build upon its molecular richness, we conducted wide-frequency coverage observations as an ALMA large program, ALCHEMI \citep[AL{\scriptsize MA} Comprehensive High-resolution Extragalactic Molecular Inventory;][hereafter \citetalias{2021A&A...656A..46M}]{2021A&A...656A..46M}. Thanks to ALCHEMI, a strong influence of cosmic rays has been found by \citet{2021A&A...654A..55H} from the CCH fractional abundance, by \citet{2021ApJ...923...24H} from the HOC$^+$ fractional abundance and HCO$^+$/HOC$^+$ abundance ratio, by \citet{2022ApJ...931...89H} using the H$_3$O$^+$/SO ratio, and by \citet{2022ApJ...939..119B} to explain the low HCN/HNC ratio even in high-temperature regions. Further findings from ALCHEMI show evidence for strong PDRs in the star-forming regions \citep{2021ApJ...923...24H}, as well as the presence of shocks at the orbital intersections by \citet{2022AA...663A..33H} from Class I methanol masers, by \citet{2022ApJ...938...80H} from \hocop, and by \citet{2023arXiv230312685H} with HNCO and SiO. This survey also made the first extragalactic detection of a phosphorus-bearing molecule, PN \citep{2022A&A...659A.158H}. \citet{2023Tanaka_HB} make use of the multi-transition and multi-molecule data to accurately map the physical conditions (gas densities, temperatures, and molecular column densities) using hierarchical Bayesian analysis and found high densities compared to those in the center of our own Galaxy.

In addition to the chemical richness, chemical differentiation (i.e., variation of chemistry) within galactic centers provides useful information. Such chemical differentiation illuminates the changing physical conditions, variations in dynamics, energetics, and evolution of the ISM. Correlations between the chemistries and these conditions establish their relationships. Such data of spatially-resolved chemistry tend to have very large datasets. For their interpretation, statistical methods such as principal component analysis (PCA) are useful in reducing the data size into a handful of components that still retain most of the information. PCA has been used widely in astronomical studies including the interpretation of molecular line intensities from multiple species, although they are relatively limited in terms of the number of transitions observed and included in their analysis \citep[e.g., ][]{1997ApJ...482..245U,2005ApJ...618..259M,2011A&A...528A..30C, 2015A&A...582A..91C, 2017A&A...599A.100G, 2022ApJ...935..155S, Gorski2023}.  

In this paper, we conduct a PCA on the ALCHEMI data to derive dominant components within the plethora of lines detectable in the interstellar medium (ISM) of NGC~253. We aim to find a set of molecular transitions that trace physical conditions such as column densities, excitation conditions, star formation activities, and shocks through this analysis. To the best of our knowledge, this is currently the largest collection of molecular lines in a given astrophysical object to which PCA has been applied. Not least, the dataset is also characterized by a high degree of homogeneity, in terms of observations and calibration procedures, angular and velocity resolutions, and sensitivity, all of which make the ALCHEMI dataset a perfect statistical sample to run PCA. This type of analysis allows for a comprehensive understanding of the ALCHEMI dataset while previous ALCHEMI papers mostly focused on a few particular species.
%Our data have homogeneous beam size and sensitivity in the brightness temperature scale from the same instrument.

This paper is organized as follows. In Section \ref{sec:obs}, we describe the observations, line identification, and data products. In Section \ref{sec:mom0}, we report features in the velocity-integrated images. The results of PCA are discussed in Section \ref{sec:pca}. The implications of results are discussed in Section \ref{sec:discuss} and Section \ref{sec:summary} summarizes the main outcome of the work.

\section{Observations and data products}\label{sec:obs}

The ALCHEMI survey (Project codes: 2017.1.00161.L, 2018.1.00162.S) includes all the ALMA Bands 3--7 frequency coverage (84--375\,GHz; $\lambda = 3.6$--$0.8$ mm) except for the small spectral regions affected by telluric absorption lines. The observed area extends over $50\arcsec \times 20\arcsec$ ($830 \times 330 \mbox{ pc}^2$ on the sky), which covers most of the CMZ. The data products are convolved into a common 1\farcs6 beam (=27\,pc), and the maximum recoverable scale is equal to or greater than $15\arcsec$. The rms noise levels for a 10\,km\,s$^{-1}$ channel are $\sim 20$\,mK, but exact values for each spectral window can be found in \citetalias{2021A&A...656A..46M} where a complete description of this survey, as well as the details on calibration and imaging, are provided. 

\subsection{Line Identification}\label{sec:lineID}
In order to identify spectral transitions in the dataset, we extracted the spectra from a single pixel at position $\alpha_{J2000}=0^{\rm h} 47^{\rm m} 33.33^{\rm s}$, $\delta_{J2000}=-25^\circ 17' 15\farcs73$. This corresponds to GMC 6 in \citet{leroy_alma_2015} (see Figure \ref{fig:rgb} for the location) and is the position with the brightest molecular emission of most detected species \citep{2022A&A...659A.158H}. 
This position has the most complex emission spectra and therefore is used as the reference position for the purposes of line blending evaluation.
Line identification was done by fitting simulated spectra under the assumption of local thermodynamic equilibrium (LTE) from all the commonly observed species, based on previous extragalactic and Galactic surveys, using MADCUBA\footnote{Madrid Data Cube Analysis on ImageJ is a software program developed
at the Center of Astrobiology (CAB) in Madrid; https://cab.inta-csic.es/madcuba/ } \citep{2019A&A...631A.159M}. 
The model on which this line identification is based includes a total of 330 entries, including the most common rare isotopologues ($^{13}$C, $^{18}$O, $^{17}$O, $^{15}$N, $^{34}$S, $^{33}$S, $^{36}$S, $^{29}$Si, $^{30}$Si, $^{37}$Cl), vibrationally excited modes for relevant species, as well as hydrogen and helium recombination lines. In total, 134 species are detected. A line identification using the ACA resolution was already presented by \citetalias{2021A&A...656A..46M}, but a full presentation of the line identification including the 12-meter array and new detections in the extragalactic ISM will be provided in dedicated papers.

\subsection{Velocity-integrated images}\label{sec:mom0maps}
From the list of all transitions included in the modeled spectra used for the line identification, we produced velocity-integrated images of isolated lines. A transition is considered isolated if it meets one of the following two criteria. 1) A line has less than 10\,\% contamination compared with its intensity in fitted models assuming LTE. 2) A line has its line center separated by more than 210\,\kms, three times the maximum line width, from other detected lines close in frequency. This level of contamination is negligible considering the $\sim15\%$ calibration uncertainty \citepalias{2021A&A...656A..46M}. We also considered a transition as isolated if blended by the same molecule (e.g., CCH, CH$_3$CCH, CN), but used masks to include all the transitions as we describe in the next paragraph (see Appendix \ref{sec:app:blend} for the list of transitions). Our criterion for detection was having $>3\sigma_{\rm ch}$ intensity at their peak, where $\sigma_{\rm ch}$ is the rms value of a single channel. Our line identification was done at the most molecule-rich position, and the contamination is likely less than 10\% in other positions. We also eliminated lines with heavy contamination from unidentified lines through the visual inspection of spectra. Spectroscopic information that is used in this paper is listed in Appendix \ref{sec:tran}. We note that these criteria for isolated lines may be too strict in some cases, and some transitions not included in this study may contain only very little contamination. It is the reason this study omits some transitions used in prior ALCHEMI papers, in which the individual transitions were more thoroughly checked for contamination.

To create velocity-integrated (moment 0) images, we applied a 3D mask created from a reference transition with threshold cutoffs to a cube of each isolated transition. It is to exclude contamination from other transitions close in frequency in a similar way as in \citet{2021ApJ...923...24H}. We choose a reference transition with strong emission and determine the threshold cutoff based on the reference transition, then apply this cutoff to cubes of other transitions. We use the $^{12}$CO($J=$1--0) image cube as a reference image to create masks for transitions that are not \twco\ but have relatively strong emission (HCN, \hcop, HNC, CS, CCH, and CN). For other species, we used HCN($J=$1--0) as a reference transition. 
The \twco\ transitions have the highest S/N ratio in our dataset, and emission is detected in larger volumes of their cubes than for any other line.
%\twco\ transitions have the highest S/N ratio, and their cubes contain detected emissions in more regions than other transitions. 
Therefore, a \twco\ mask is suitable for collecting weak extended emission in transitions such as HCN. 
On the other hand, the HCN(1--0) mask is more suitable for weak lines to eliminate contamination from neighbors, because there is one location (GMC6) where the CO line is very broad.
%However, the HCN(1--0) mask is more suitable for eliminating contamination from neighboring transitions for weak transitions because there is a location (GMC6) where the CO line width is very large. 

To create the binary 0/1 masks, we first convolved the reference cubes to a lower angular resolution of 3\farcs2 to capture faint extended emission in the masks.
%because spatially neighboring regions may also contain detectable emission. 
We then set the mask value to 1 only at locations in the cubes with a $> 15 \sigma_{\rm ch}$ detection for the CO(1--0) mask or $> 5 \sigma_{\rm ch}$ detection for the HCN(1--0) mask, where $\sigma_{\rm ch}$ is the single-channel rms value in the 3\farcs2 resolution reference cube. 
%These masks were applied to transitions with weaker intensities. 
A 1\farcs6 resolution integrated intensity map is created as the input line cube multiplied by the mask cube on the same grid and integrated over velocity.  
%Note that masked cubes of these weaker transitions still have the angular resolution of 1\farcs6. 
The above cutoffs for the masks may seem unnecessarily high, but these masks should still include all regions with significant emission, because all the transitions except those of \twco\ are more than 10 times weaker than CO(1--0) in extended regions, and HCN(1--0) has more than 3 times higher S/N compared to the transitions to which the HCN mask was applied. We confirmed that our masking did not exclude target emission by visually inspecting the original and masked cubes. This high cutoff for CO(1--0) also helps to exclude nitrogen sulfide (NS) transitions at the rest frequency $f_{\rm rest}=115.154$ GHz near CO(1--0), whose rest frequency is at 115.271 GHz. 
We applied our masked integration to images within 400 \kms\ from the systemic velocity of 258 \kms.
The effective range of integration is narrower than 800 \kms\ because of the masking. 

For $^{12}$CO transitions, we created the masks from themselves (e.g., using CO 3--2 as the reference cube for CO 3--2).
We again convolved the reference cubes to 3\farcs2, twice the original beam size, and kept the positions with emission brighter than $2\sigma_{\rm ch}$. These masks were applied to the original 1\farcs6 resolution cubes. 

For transitions blended with the same species (e.g., CCH), we first made cubes with their velocity coordinates defined from one of the transitions (a velocity reference line), with enough width to cover all the detectable transitions of interest. Then, we created masks to include emissions from multiple transitions from the same species, using the known velocity shifts from the velocity reference line. The list of such transitions can be found in Appendix \ref{sec:app:blend}.

In addition to the 3D masking applied above, we also applied a threshold cutoff of $3\sigma$ to the integrated intensity images. The rms $\sigma$ of the integrated intensities is estimated as 
\begin{equation}
  \sigma = \sigma_{\rm ch}\Delta v \sqrt{N},  
\end{equation}
where $\sigma_{\rm ch}$ is the rms for one channel in the cube, $\Delta v$ is the channel width (10\,\kms), and $N$ is the number of channels used for integration. 

After removing transitions with peak intensities less than 10\,K\,\kms\ and Band 5 transitions with channels affected by severe artifacts (SiO 4--3, \nthp\ 2--1, see also Section \ref{sec:mom0}), we use 148 transitions of molecules or radio recombination lines (RRLs) from 44 species and two continuum images (one representative of synchrotron/free-free emission at the lower frequency range and the other representative of dust thermal emission at the higher frequency range) for PCA. The velocity-integrated images, original cubes, and masked cubes will be made public as a part of the high level data products to be made available through the ALMA science archive. In addition, a matrix of standardized intensities is also publicly available\footnote{\url{http://github.com/nanaseharada/alchemi_pca}}.

\section{Distribution of transition intensities}\label{sec:mom0}

\begin{figure*}[ht]
\centering{
\includegraphics[width=0.8\textwidth]{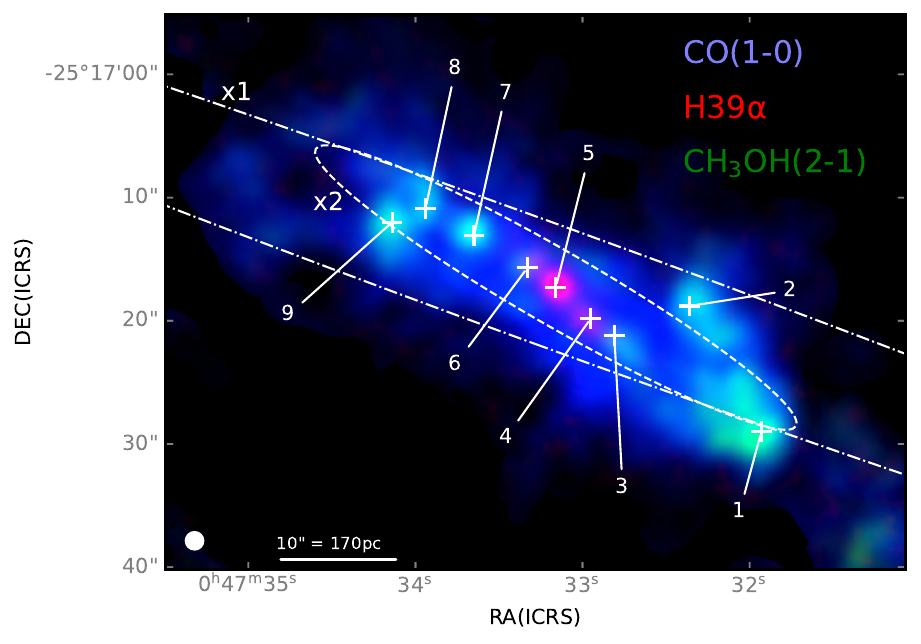}
\includegraphics[width=0.8\textwidth]{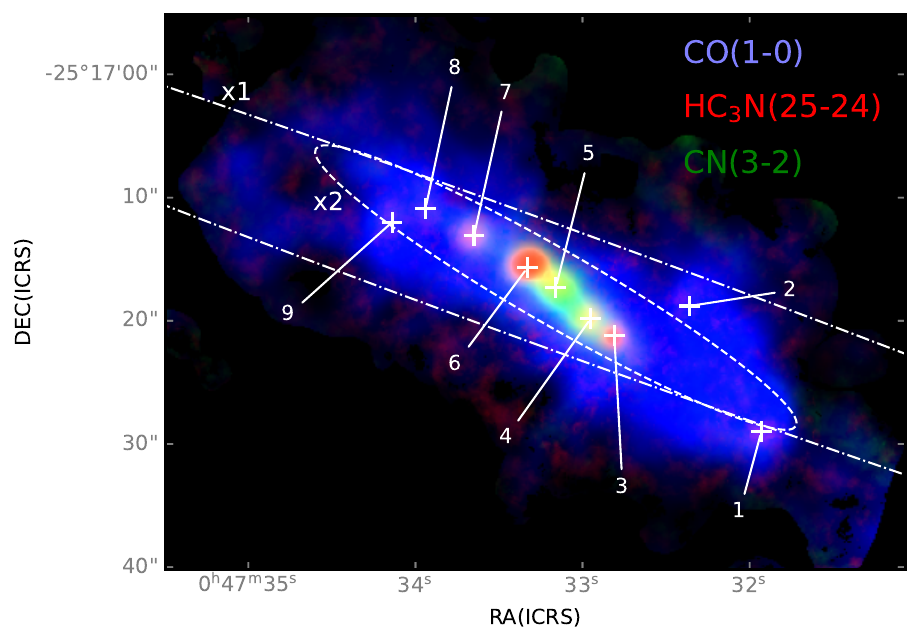}}
\caption{(Top) Velocity-integrated images of CO(1--0) in blue, H39$\alpha$ in red, and \methanol($2_K-1_K$) around the rest frequency of 96.7 GHz in green. Rough positions of parts of the $x_1$ orbits and the full $x_2$ orbits further inside are shown with dash-dotted and dashed lines, respectively. Note that these are just some examples of the presumably large families of possible $x_1$ and $x_2$ orbits. There likely exists another $x_2$ orbit connecting GMCs 3-6 or 3-7 \citep[see][]{2022ApJ...935...19L} almost fully edge-on. The synthetic beam size is shown at the left bottom corner as a white circle. (Bottom) The same as the top figure, but with \hctn(25--24) in red, and CN(3--2) in green. Intensities are scaled so that faint lines have similar dynamic ranges as other lines. We employed the algorithm by \citet{2004PASP..116..133L} to produce RGB images. \label{fig:rgb}}
\end{figure*}

We first describe locations in the CMZ of NGC~253 using the two color-composite (red-green-blue; RGB) images of selected transitions in Figure \ref{fig:rgb}. They show the contrast between images of CO(1--0), H39$\alpha$, \methanol($2_K-1_K$), \hctn(25--24), and CN(3--2). This choice of lines for the RGB channels is justified by the results of the PCA, from which we select these five representative lines as highlighting the extreme variance in the dataset, hence giving the RGB images high contrast. As we discuss later in Section  \ref{sec:projection}, these transitions represent the first three principal components (PC1, positive and negative PC2, positive and negative PC3). In this figure, we show the positions of giant molecular clouds (GMCs) that we refer to throughout the paper. They are close to the GMCs discussed in \citet{leroy_alma_2015}, but slightly modified to better match the molecular emission peaks in most species. Because the GMC peaks are hard to pinpoint from bright, ubiquitous, and extended transitions such as CO and HCN, we identified peaks from species that trace more specific locations such as SiO (GMCs 1, 7, 8, and 9), \hctn (GMCs 3, 4, 6), and H39$\alpha$ (GMC5). These positions are listed in Table \ref{tab:gmc_pos}.

Some of these clouds reside at orbital intersections of bar orbits (the $x_1$ orbits) and central orbits (the $x_2$ orbits) resulting from the stellar bar potential in NGC~253 \citep{2000PASJ...52..785S,2001ApJ...549..896D}. The gas on the bar orbits flows in from the northeast and southwest directions. GMCs 1, 2, 7, 8, and 9 have signs of shocks detected from Class I methanol masers \citep{2022AA...663A..33H}. Other clouds are locations of young massive star-cluster formation. An RRL, H40$\alpha$, has been detected in GMCs 3--6 \citep{2015MNRAS.450L..80B,2021ApJ...919..105M}. We note that the RRL in GMC3 is not obvious in our 1\farcs6 beam. Only high-angular resolution observations ($\sim$0\farcs2) by \citet{2021ApJ...919..105M} found weaker and narrower H40$\alpha$ in GMC 3 compared with GMCs 4--6.

%%%%%%%%%%%%%%%%%%%%%
\begin{deluxetable}{cccc} 
\tablecolumns{4} 
\tablewidth{0pc} 
%\tabletypesize{\scriptsize} 
\tablecaption{GMC positions and their properties\label{tab:gmc_pos}}
\tablehead{\colhead{GMC} &\colhead{RA(ICRS)} &\colhead{DEC(ICRS)}  &\colhead{Remarks} \\
\colhead{} & \colhead{0$^h$47$^m$} &\colhead{$-25^{\circ}$17$'$} &\colhead{}}
\startdata 
1 &31\fs93 &29\farcs0 & Class I methanol maser$\dagger$\\
2 &32\fs36 &18\farcs8 & Class I methanol maser$\dagger$\\
3 &32\fs81 &21\farcs2 & Clumps 1-3$\ddagger$\\
4 &32\fs95 &19\farcs8 & Clumps 4-7$\ddagger$\\
5 &33\fs16 &17\farcs3 & Clumps 8-13$\ddagger$\\
6 &33\fs33 &15\farcs7 & Clump 14$\ddagger$\\
&&&Line ID position$^\ast$ \\
7 &33\fs65 &13\farcs1 & Class I methanol maser$\dagger$\\
8 &33\fs94 &10\farcs9 & Class I methanol maser$\dagger$\\
9 &34\fs14 &12\farcs0 & Class I methanol maser$\dagger$\\
\enddata 
\tablecomments{$\dagger$\citet{2022AA...663A..33H} $\ddagger$\citet{leroy_forming_2018} $^\ast$ Position of the brightest molecular emission} 
\end{deluxetable} 
%%%%%%%%%%%%%%%%%%%%%%%%%%%%%%

%%%%%%%%%%%%%%%%%%%%%%
%%%%%%%%%%%%%%%%%%%%%

\begin{figure*}[ht]
\includegraphics[width=0.99\textwidth]{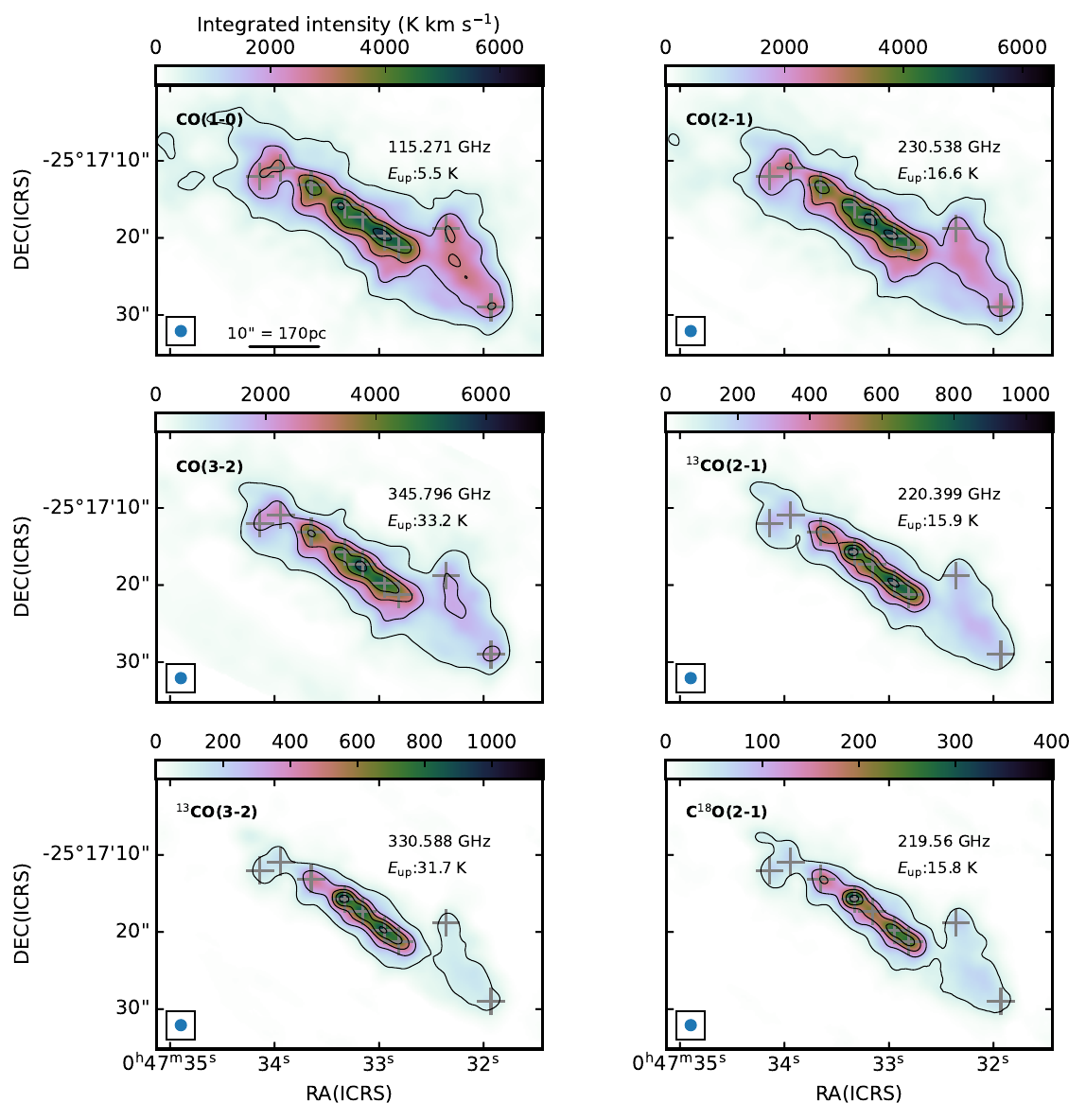}
\caption{
Velocity-integrated images integrated from within 400 \mbox{km s$^{-1}$} of the systemic velocity with masking as described in Sect.~\ref{sec:lineID}.
%Velocity-integrated images over the $[-400,400]~\rm km~s^{-1}$ range of transitions selected and imaged as described in Sect.~\ref{sec:lineID}. 
Contour levels are set at 10, 30, 50, 70, and 90\,\% of their peak values. Maximum values of color scales are set for 120\,\% of their peak values. We show more commonly observed species first, and transitions are in ascending order of frequency within the same species.
The transition names are in the upper left corner, and the line rest frequency and the energy level in the upper state are also in each panel. The 1\farcs6 beam is in the bottom left corner. The plus signs indicate GMC locations. The images are corrected for the primary beam response patterns of the ALMA antennas.
\label{fig:mom0_0}}
\end{figure*}

\begin{figure}[ht]
\includegraphics[width=0.99\textwidth]{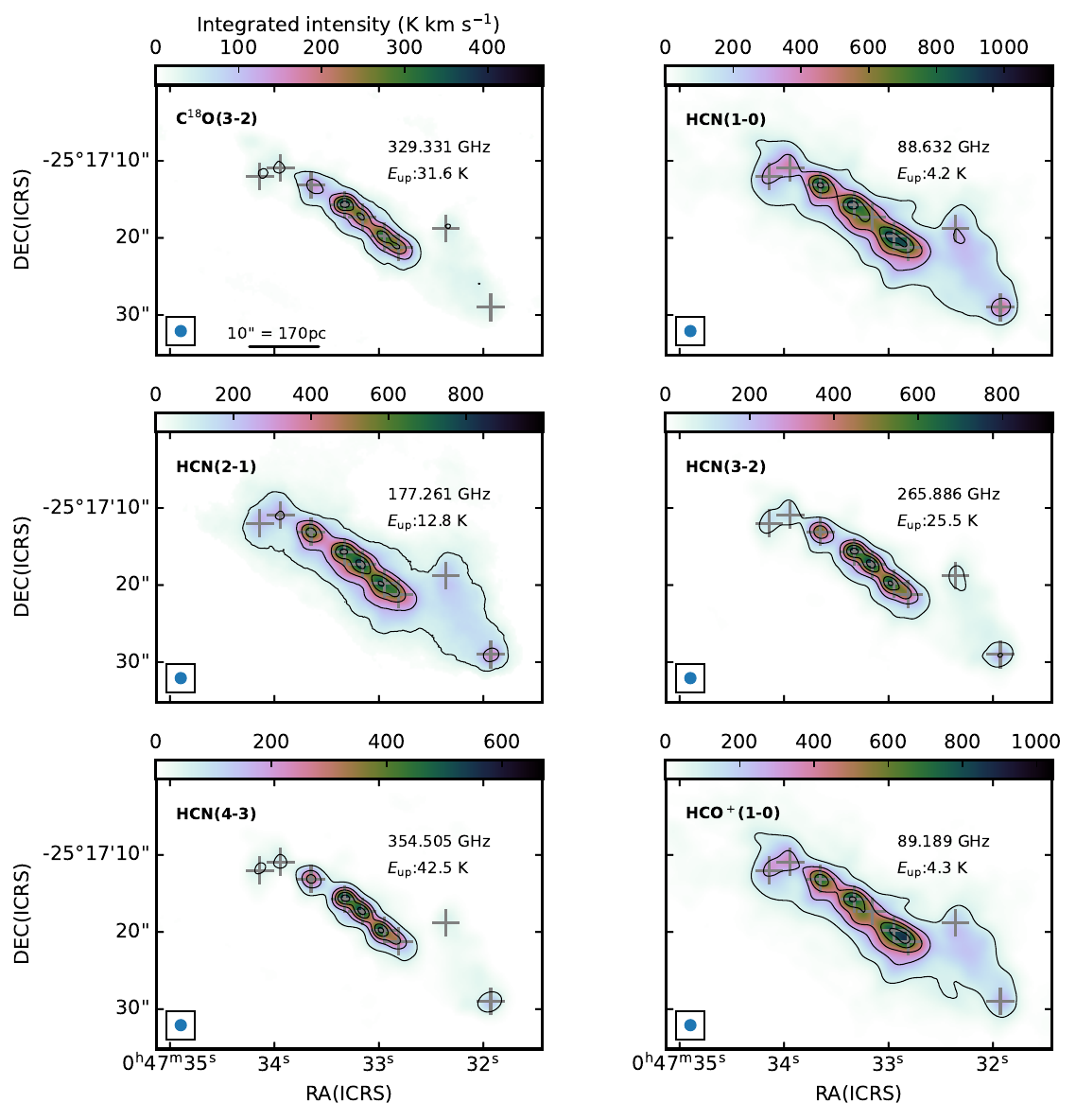}
\caption{Same as Figure \ref{fig:mom0_0}.
\label{fig:mom0_1}}
\end{figure}

\begin{figure}[ht]
\includegraphics[width=0.99\textwidth]{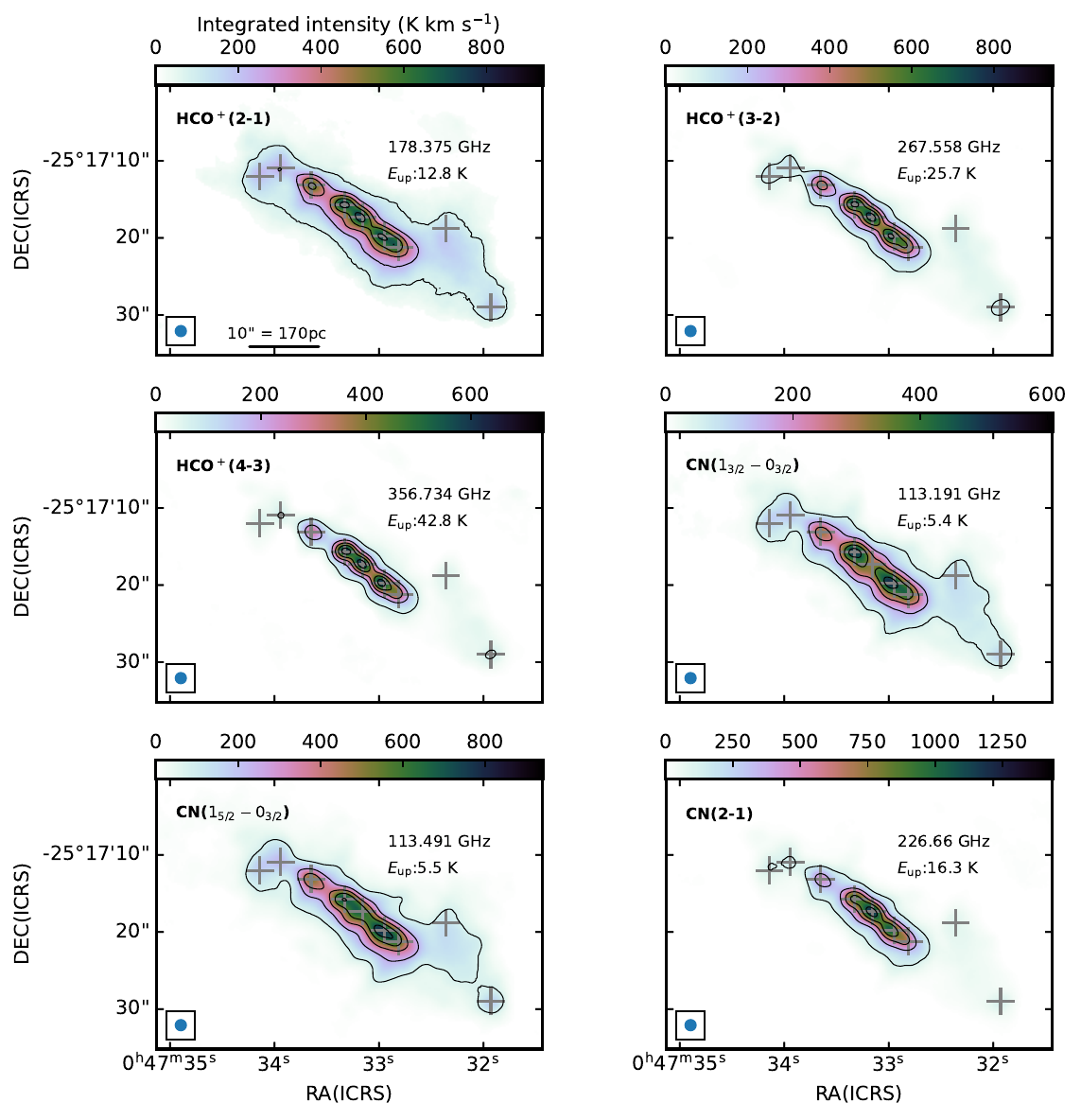}
\caption{Same as Figure \ref{fig:mom0_0}.\label{fig:mom0_2}}
\end{figure}

\begin{figure}[ht]
\includegraphics[width=0.99\textwidth]{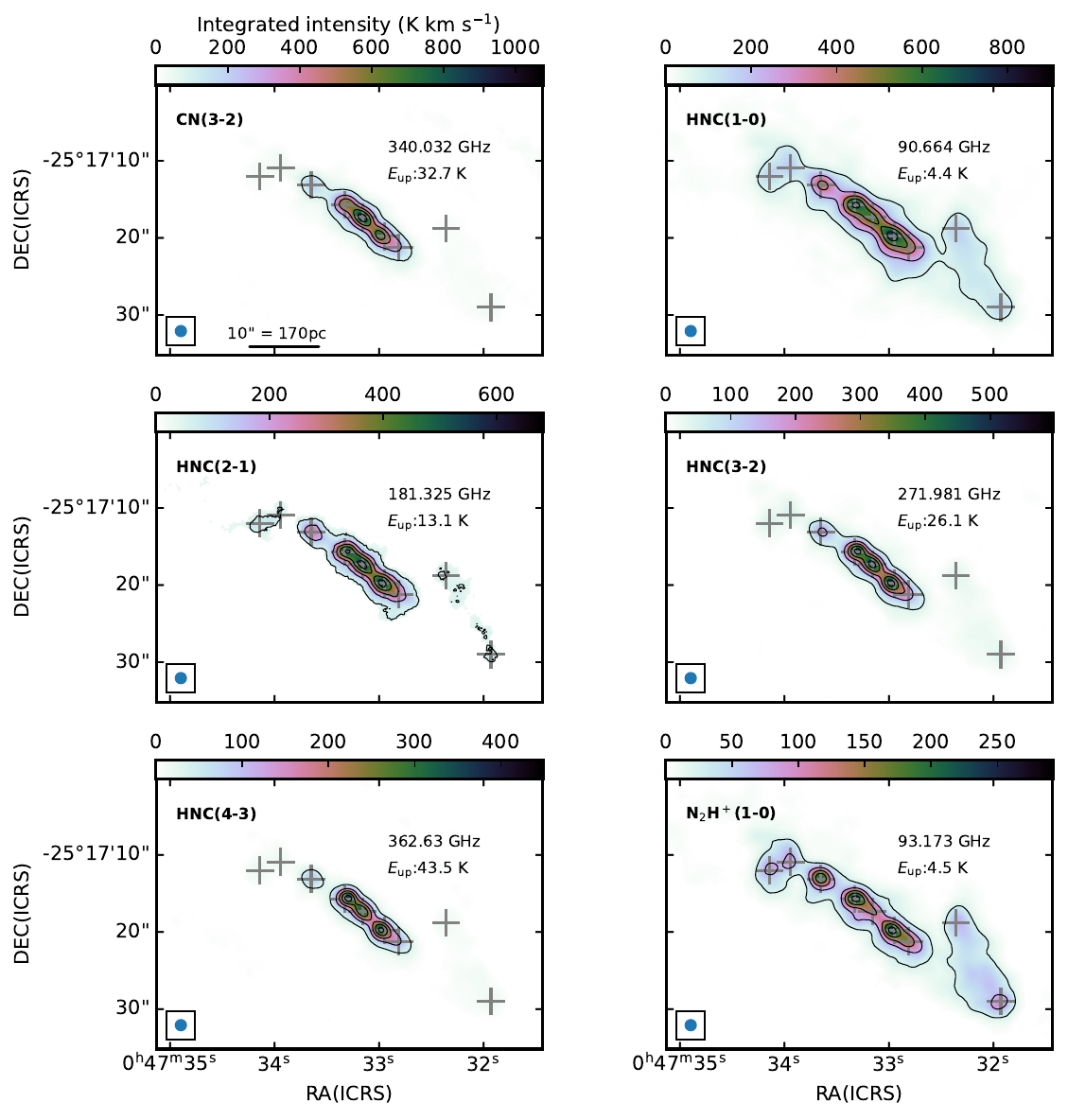}
\caption{Same as Figure \ref{fig:mom0_0}. \label{fig:mom0_3}}
\end{figure}

\begin{figure}[ht]
\includegraphics[width=0.99\textwidth]{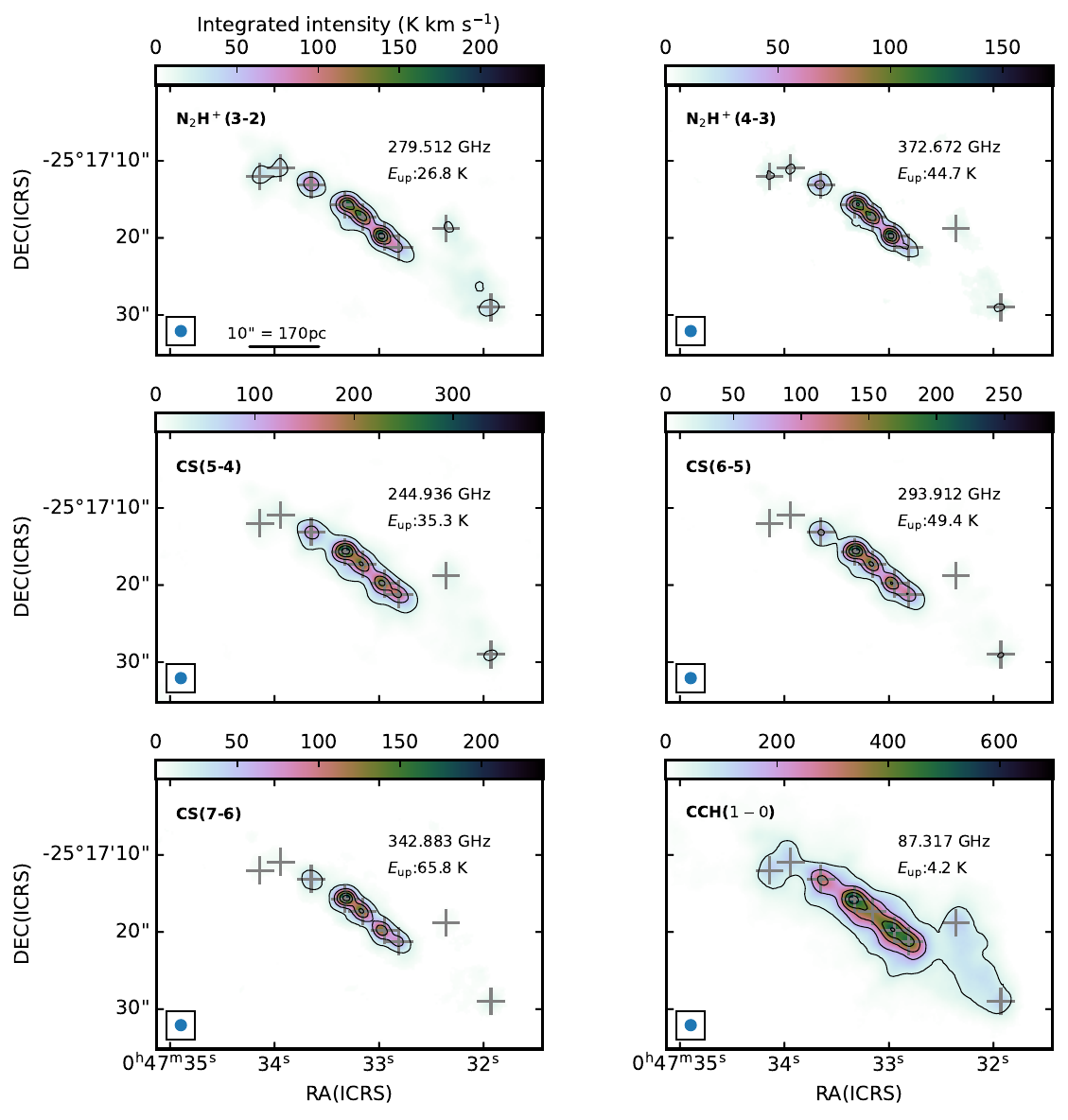}
\caption{Same as Figure \ref{fig:mom0_0}. \label{fig:mom0_4}}
\end{figure}

\begin{figure}[ht]
\includegraphics[width=0.99\textwidth]{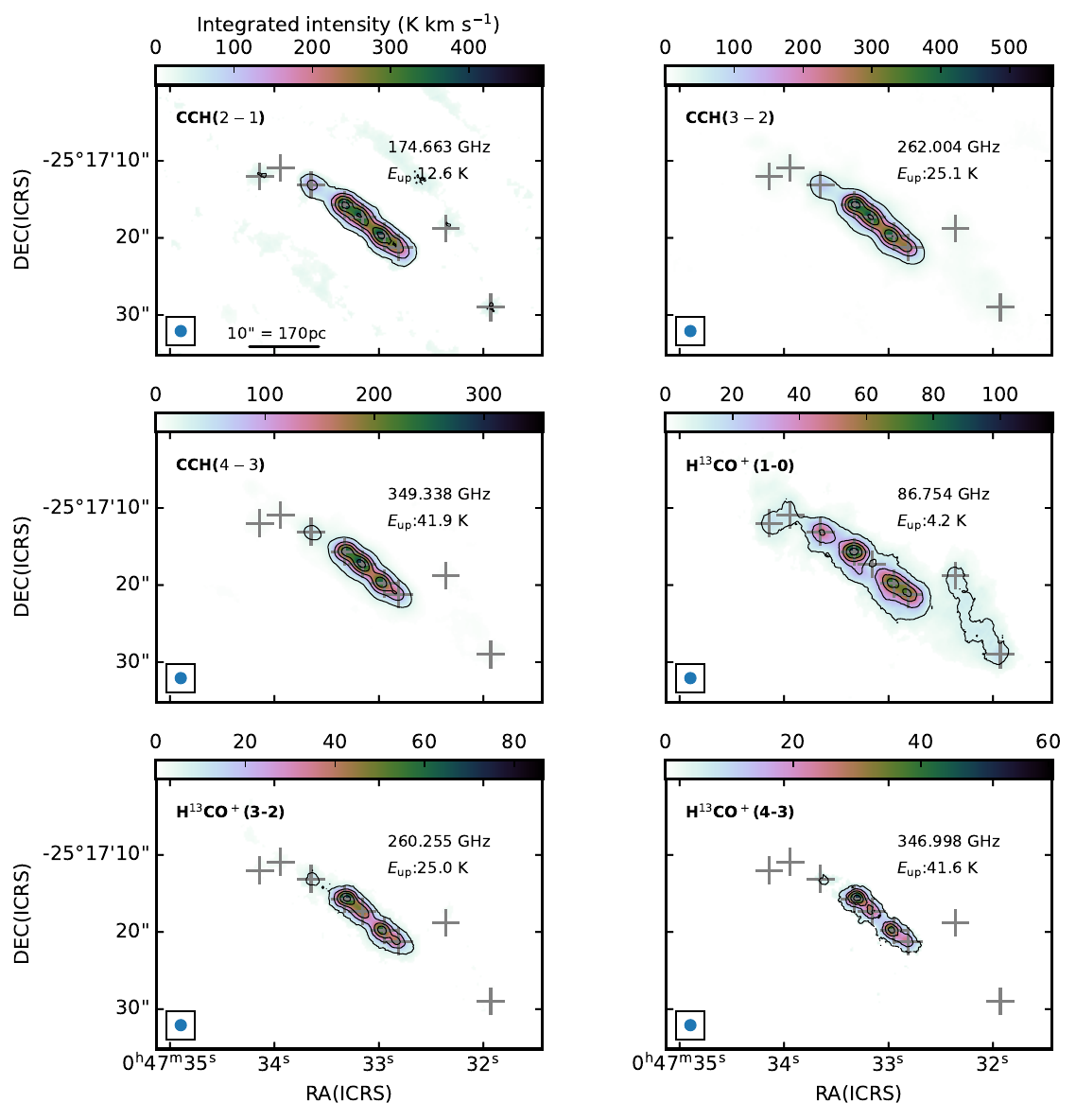}
\caption{Same as Figure \ref{fig:mom0_0}. \label{fig:mom0_5}}
\end{figure}

\begin{figure}[ht]
\includegraphics[width=0.99\textwidth]{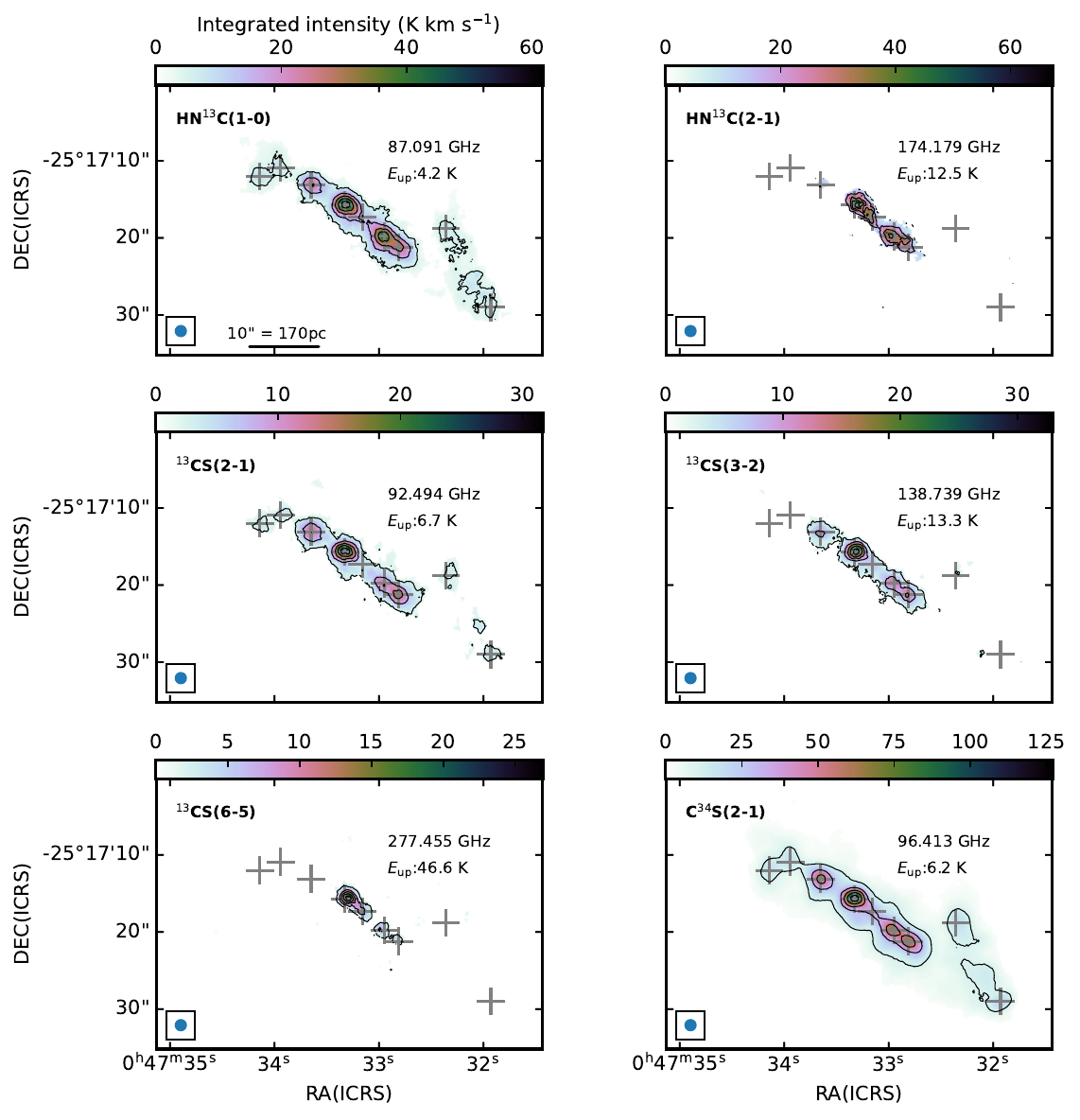}
\caption{Same as Figure \ref{fig:mom0_0}. \label{fig:mom0_6}}
\end{figure}

\begin{figure}[ht]
\includegraphics[width=0.99\textwidth]{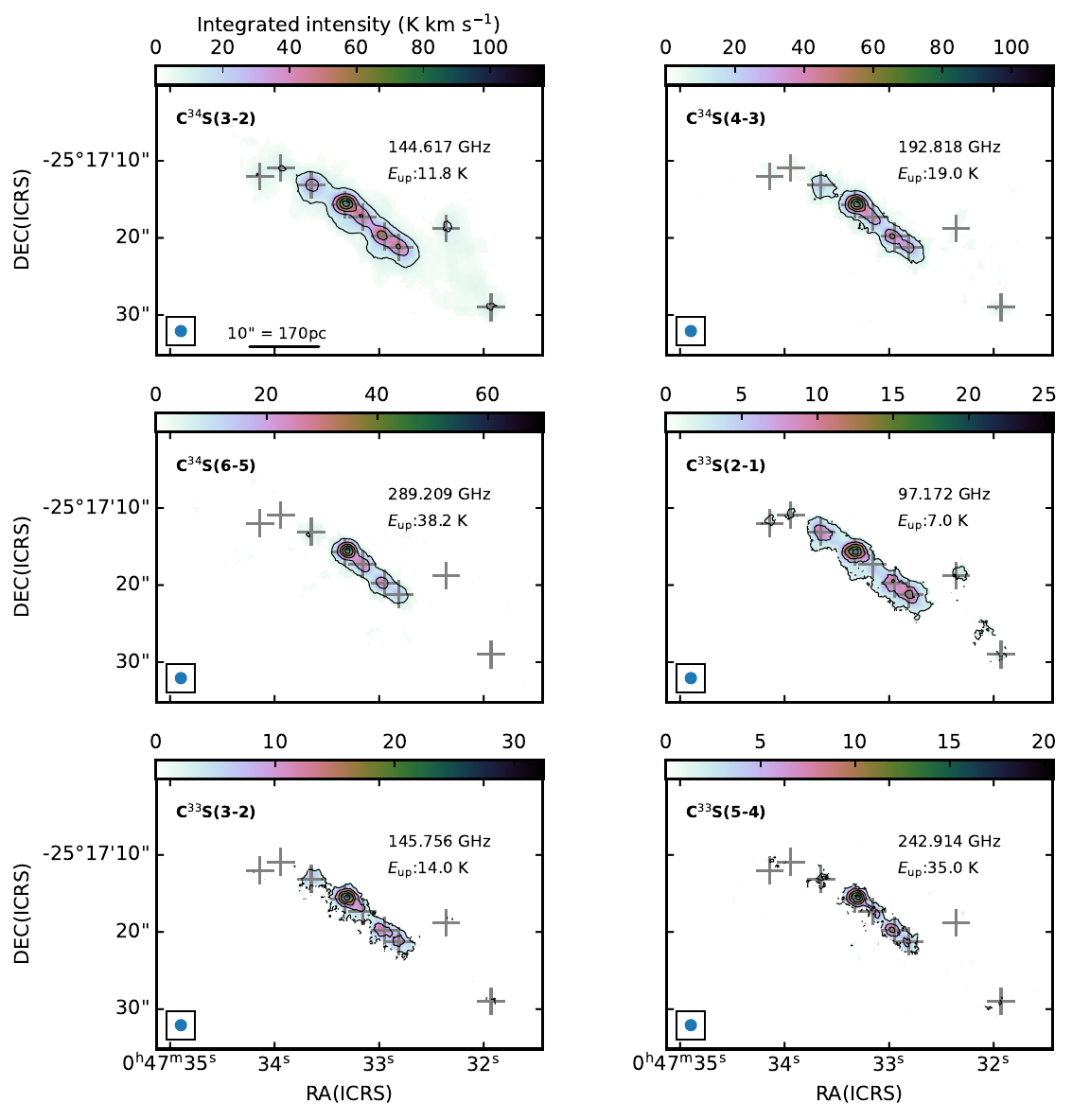}
\caption{Same as Figure \ref{fig:mom0_0}. \label{fig:mom0_7}}
\end{figure}

\begin{figure}[ht]
\includegraphics[width=0.99\textwidth]{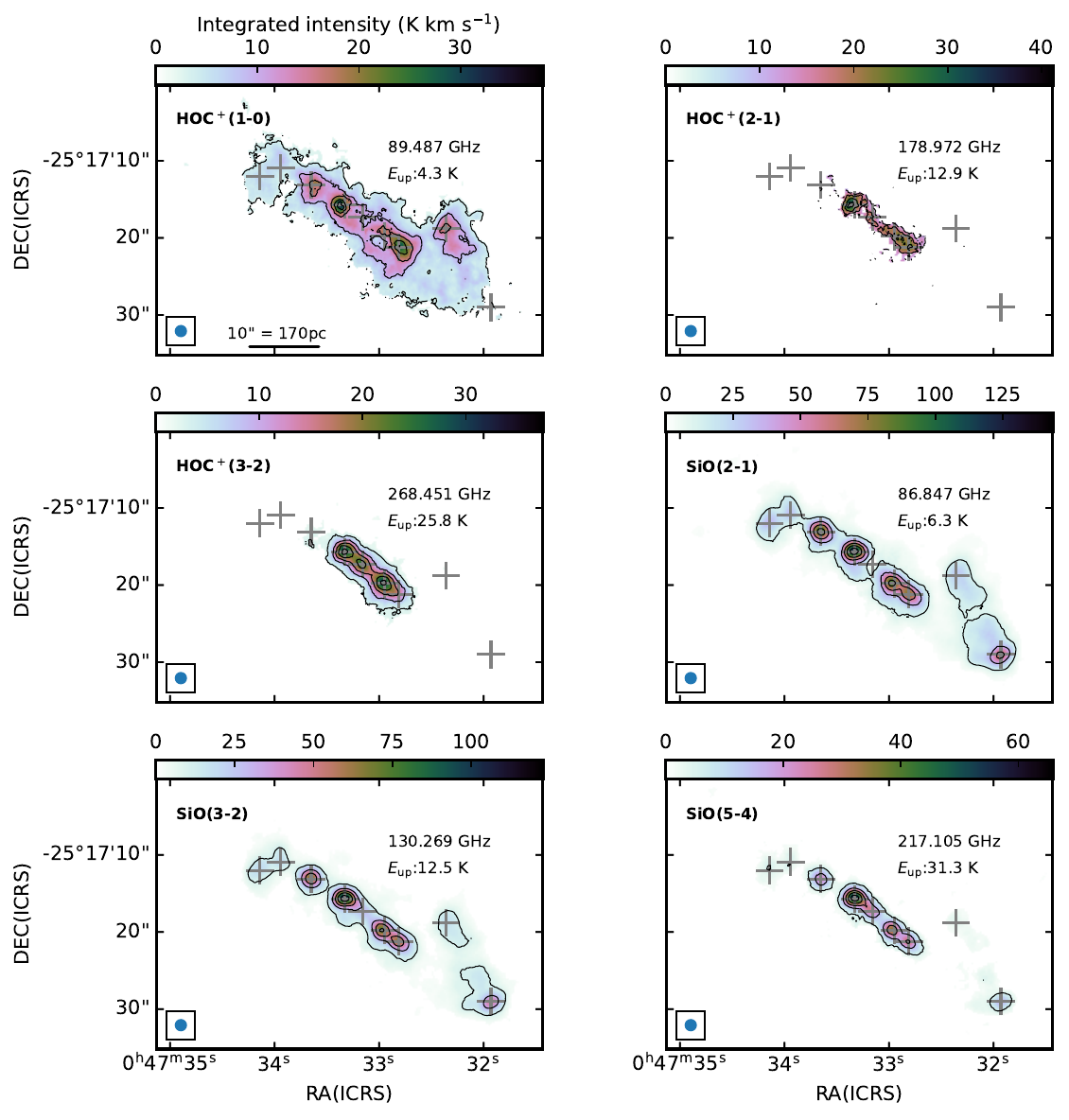}
\caption{Same as Figure \ref{fig:mom0_0}.\label{fig:mom0_8}}
\end{figure}

\begin{figure}[ht]
\includegraphics[width=0.99\textwidth]{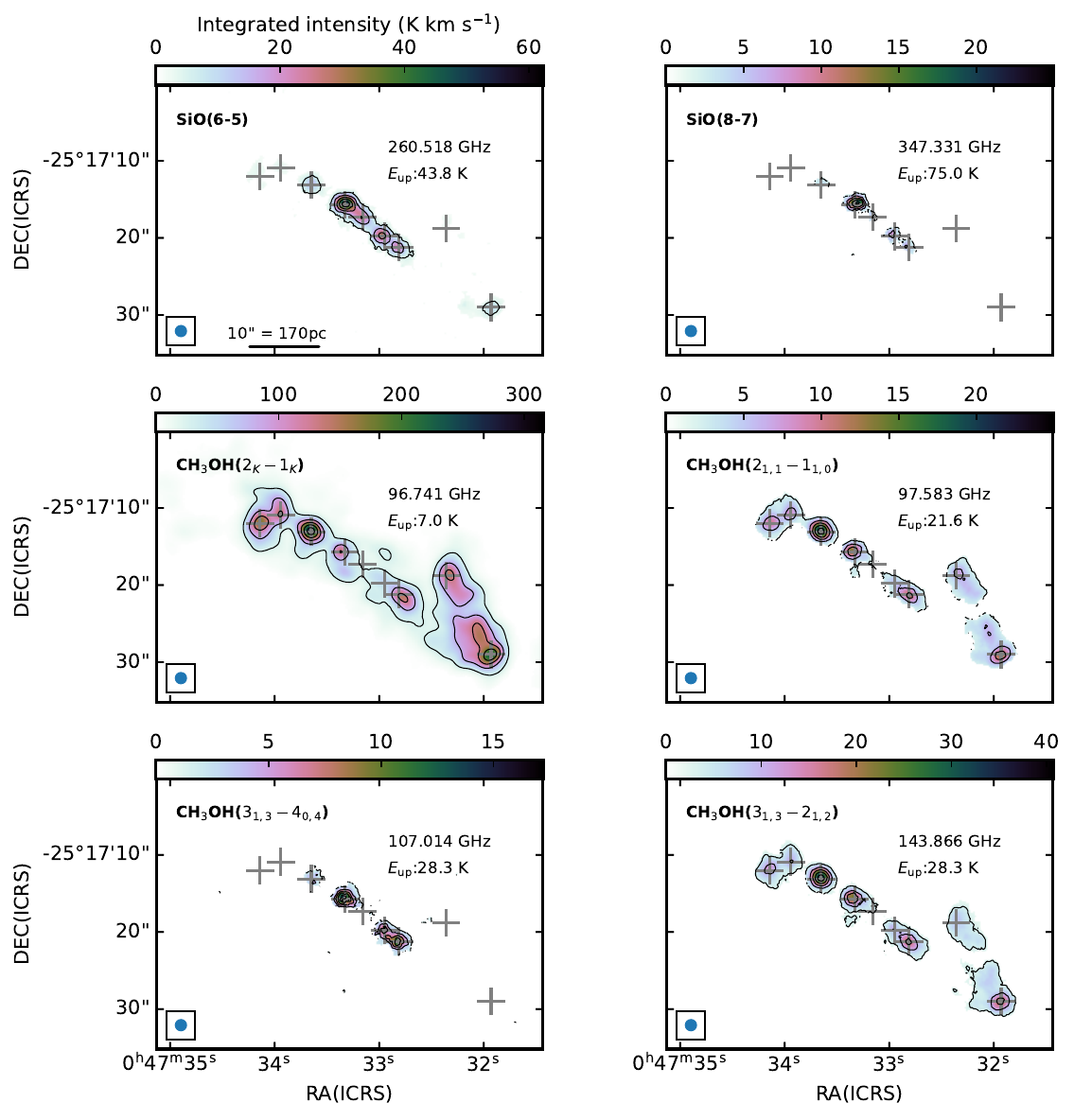}
\caption{Same as Figure \ref{fig:mom0_0}. \label{fig:mom0_9}}
\end{figure}

\begin{figure}[ht]
\includegraphics[width=0.99\textwidth]{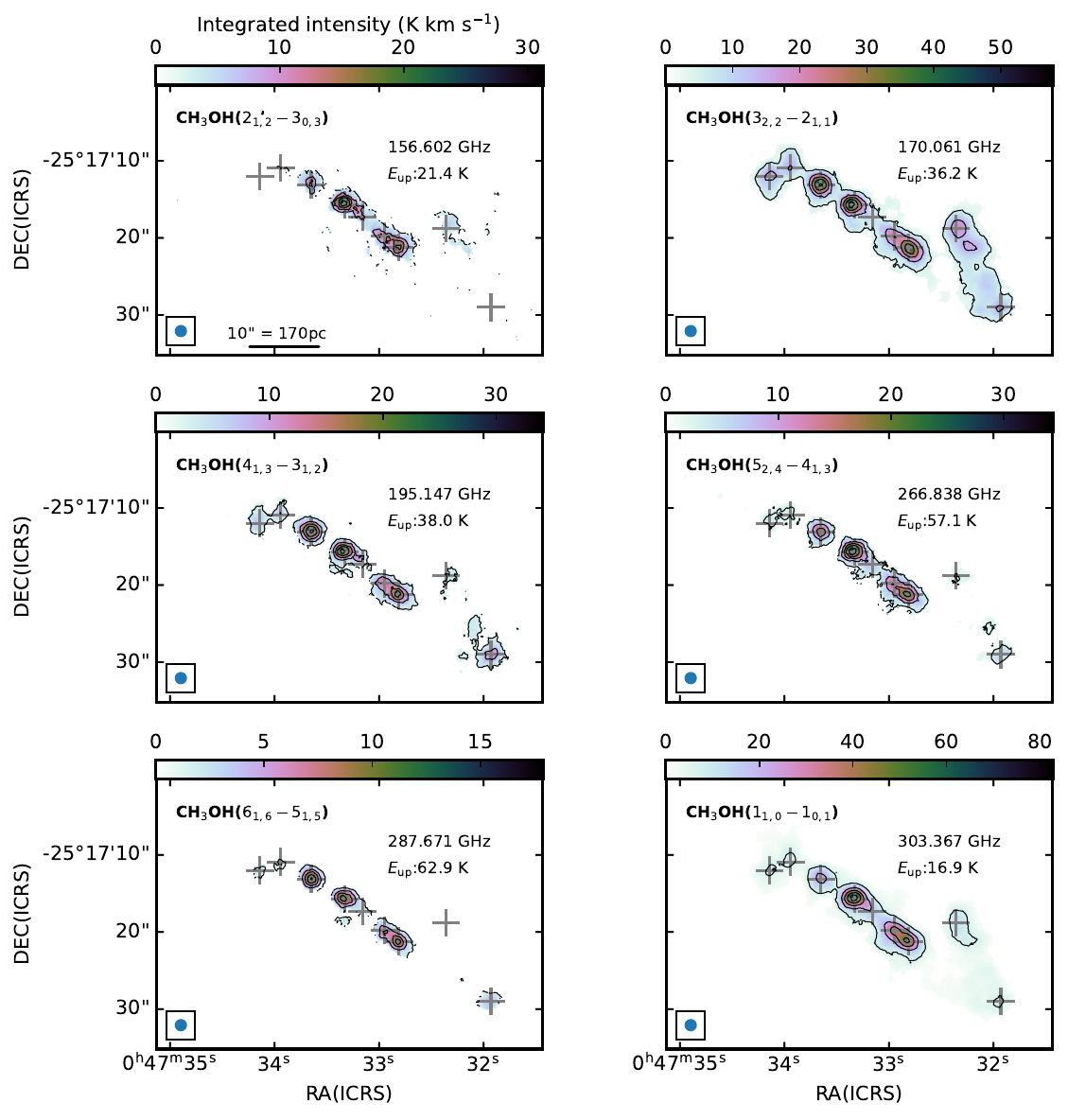}
\caption{Same as Figure \ref{fig:mom0_0}. \label{fig:mom0_10}}
\end{figure}

\begin{figure}[ht]
\includegraphics[width=0.99\textwidth]{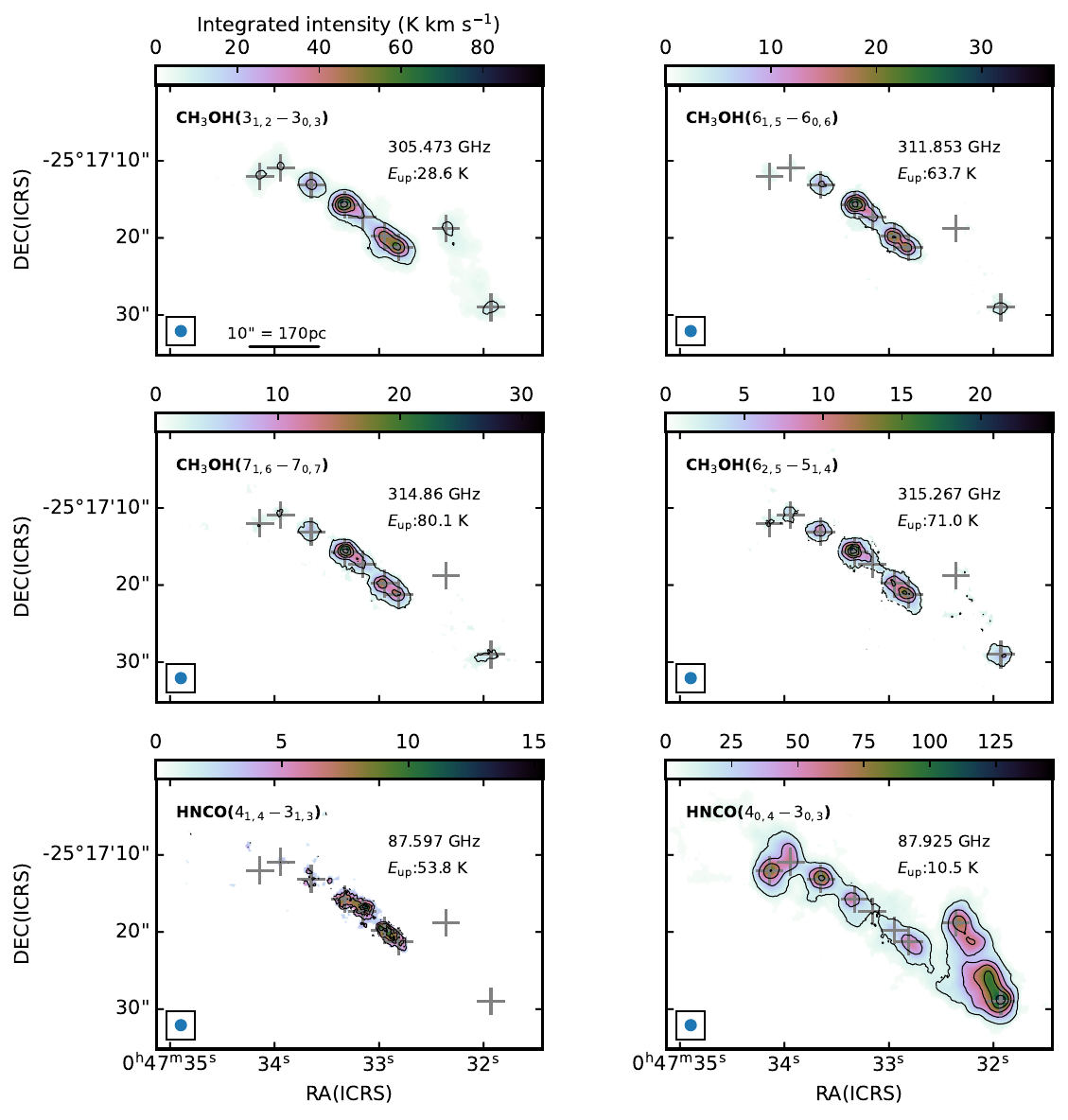}
\caption{Same as Figure \ref{fig:mom0_0}. \label{fig:mom0_11}}
\end{figure}

\begin{figure}[ht]
\includegraphics[width=0.99\textwidth]{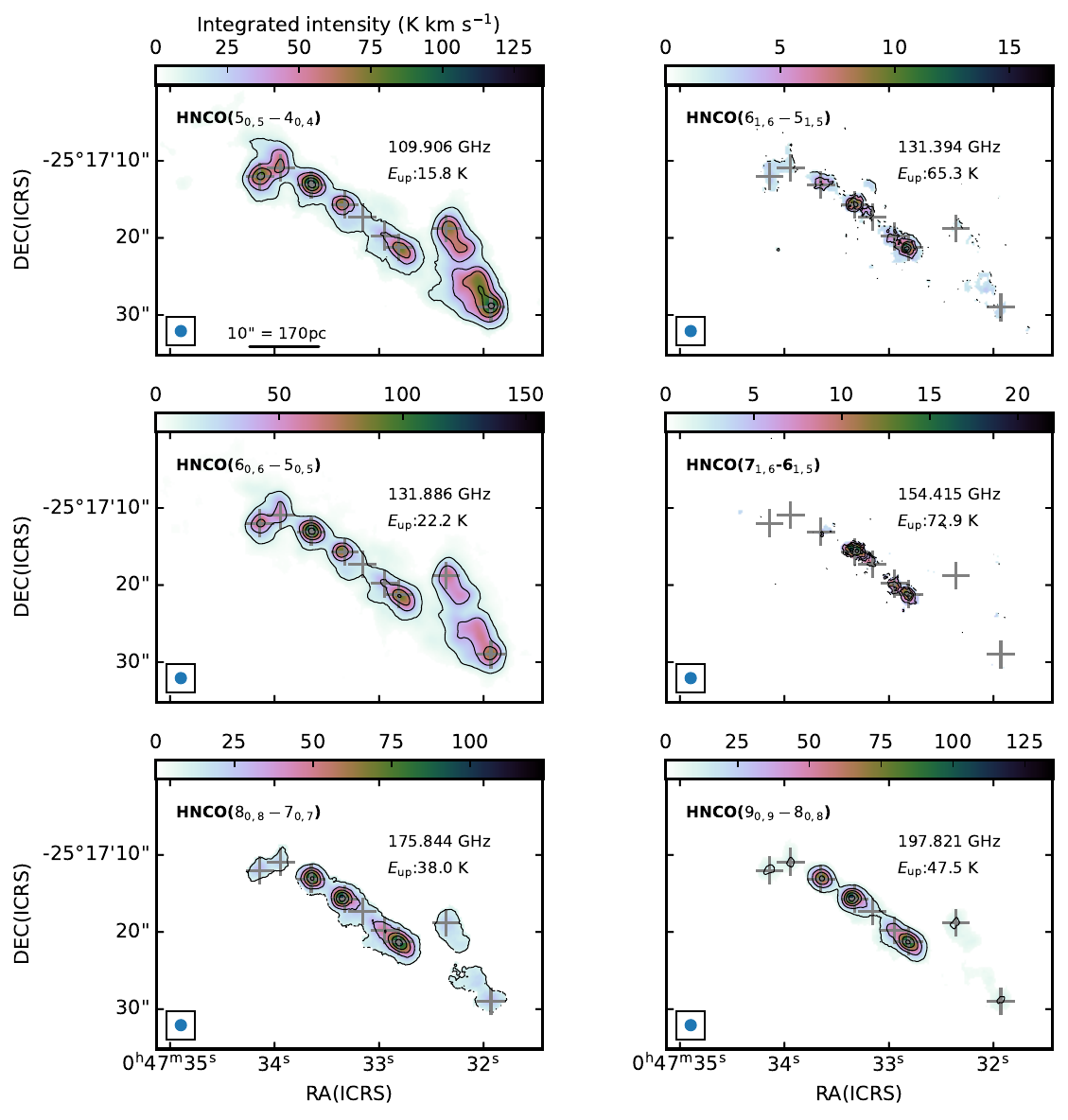}
\caption{Same as Figure \ref{fig:mom0_0}. \label{fig:mom0_12}}
\end{figure}

\begin{figure}[ht]
\includegraphics[width=0.99\textwidth]{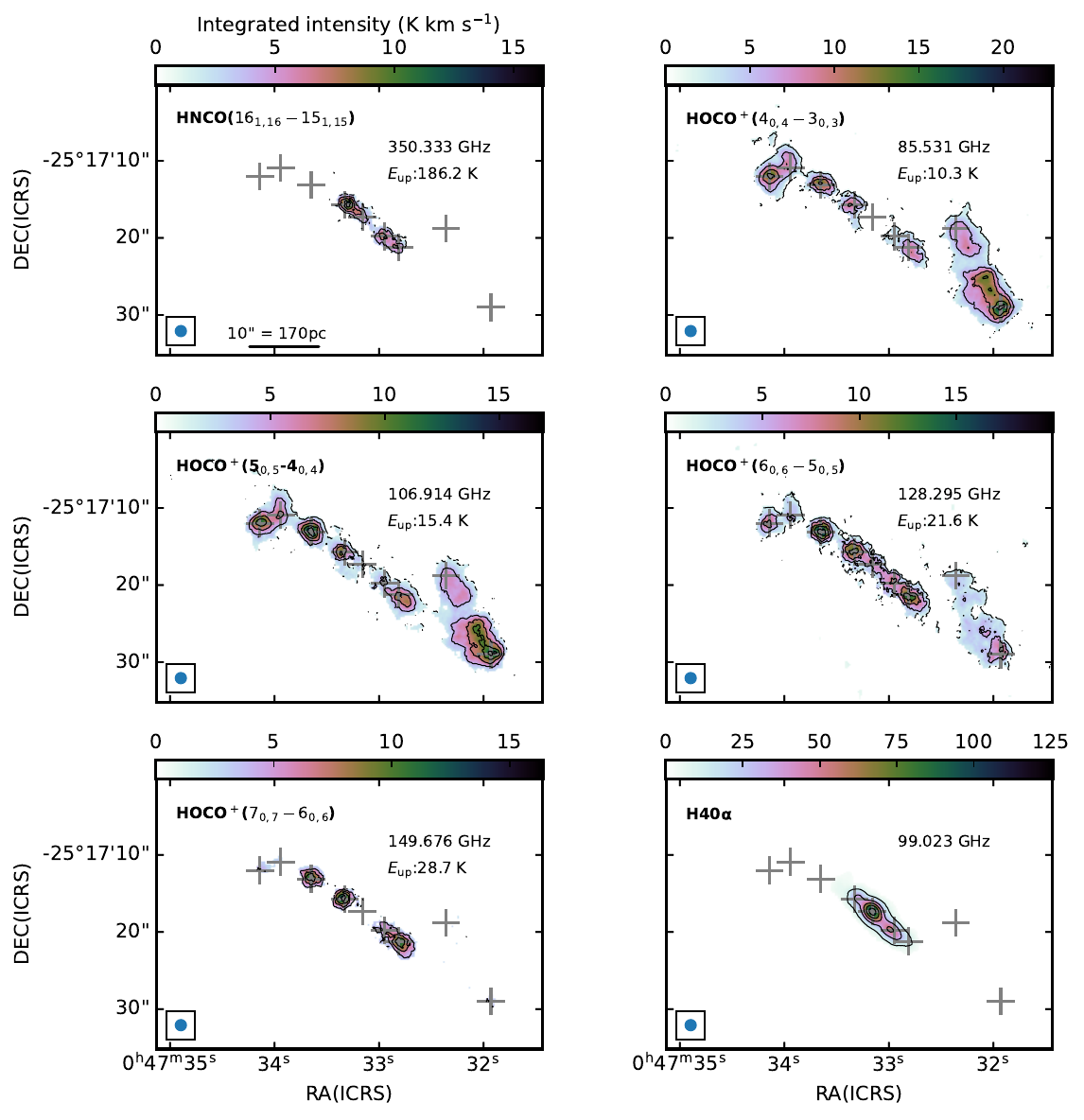}
\caption{Same as Figure \ref{fig:mom0_0}. \label{fig:mom0_13}}
\end{figure}

\begin{figure}[ht]
\includegraphics[width=0.99\textwidth]{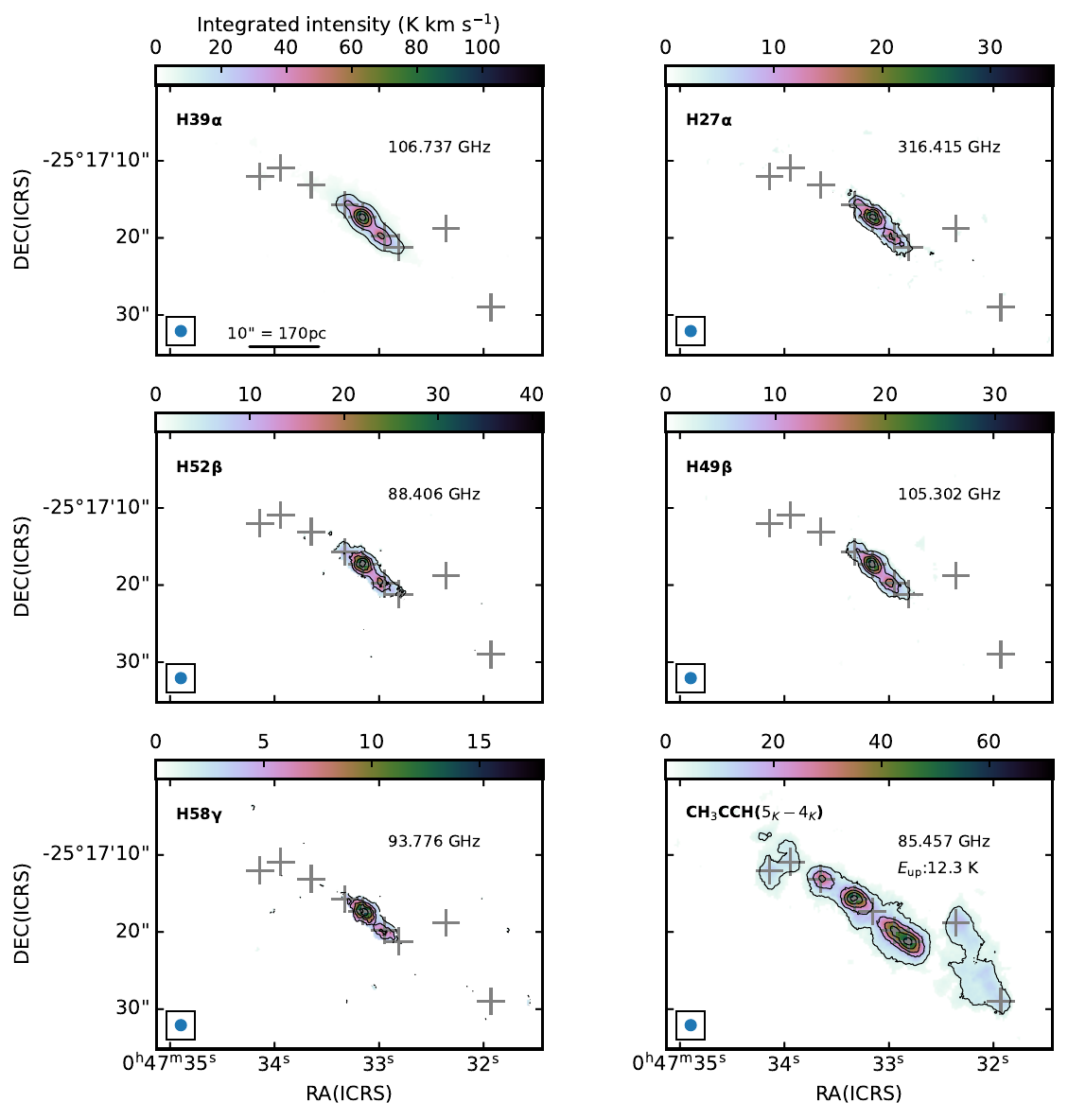}
\caption{Same as Figure \ref{fig:mom0_0}. \label{fig:mom0_14}}
\end{figure}

\begin{figure}[ht]
\includegraphics[width=0.99\textwidth]{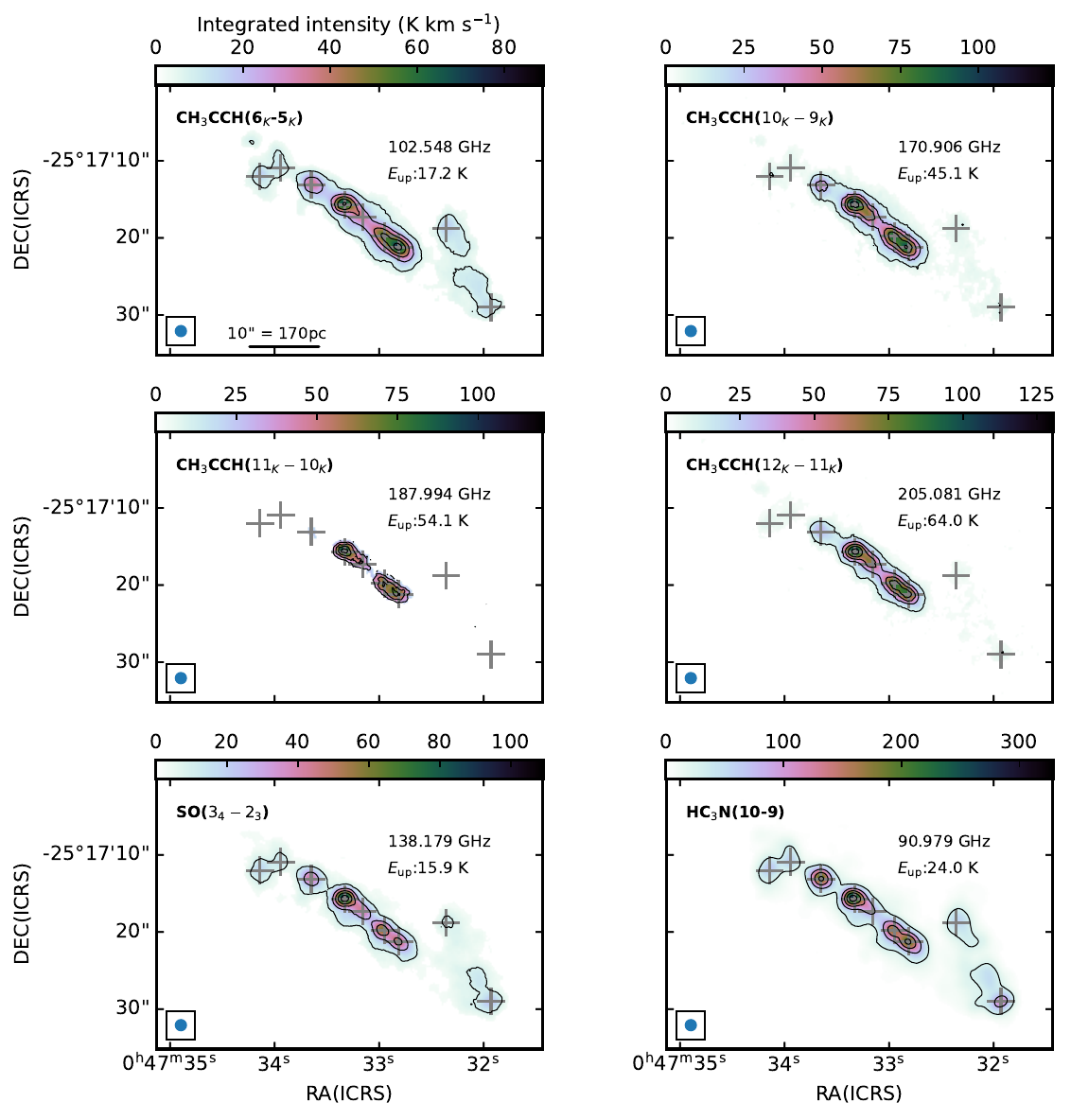}
\caption{Same as Figure \ref{fig:mom0_0}. \label{fig:mom0_15}}
\end{figure}

\begin{figure}[ht]
\includegraphics[width=0.99\textwidth]{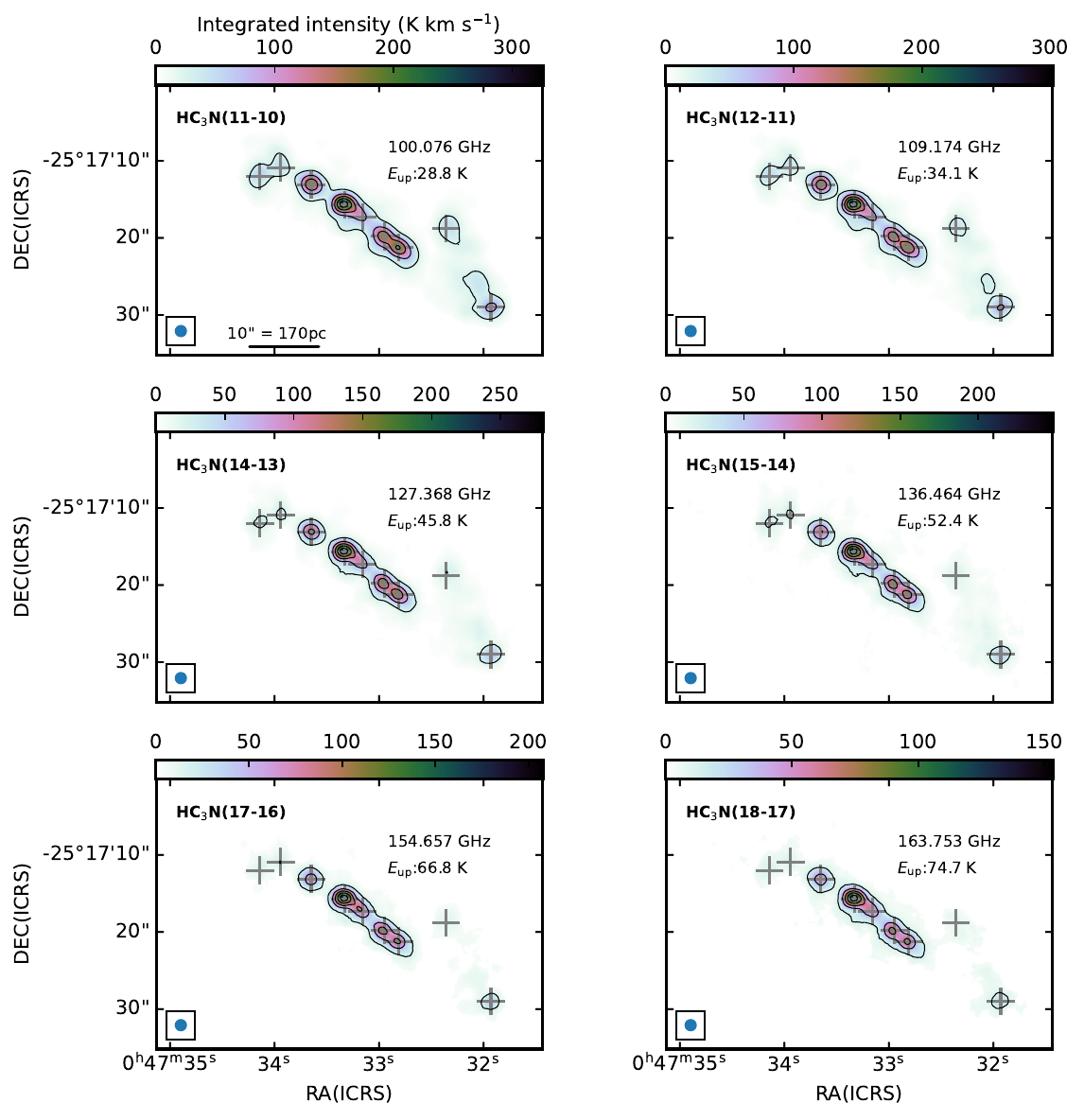}
\caption{Same as Figure \ref{fig:mom0_0}. \label{fig:mom0_16}}
\end{figure}

\begin{figure}[ht]
\includegraphics[width=0.99\textwidth]{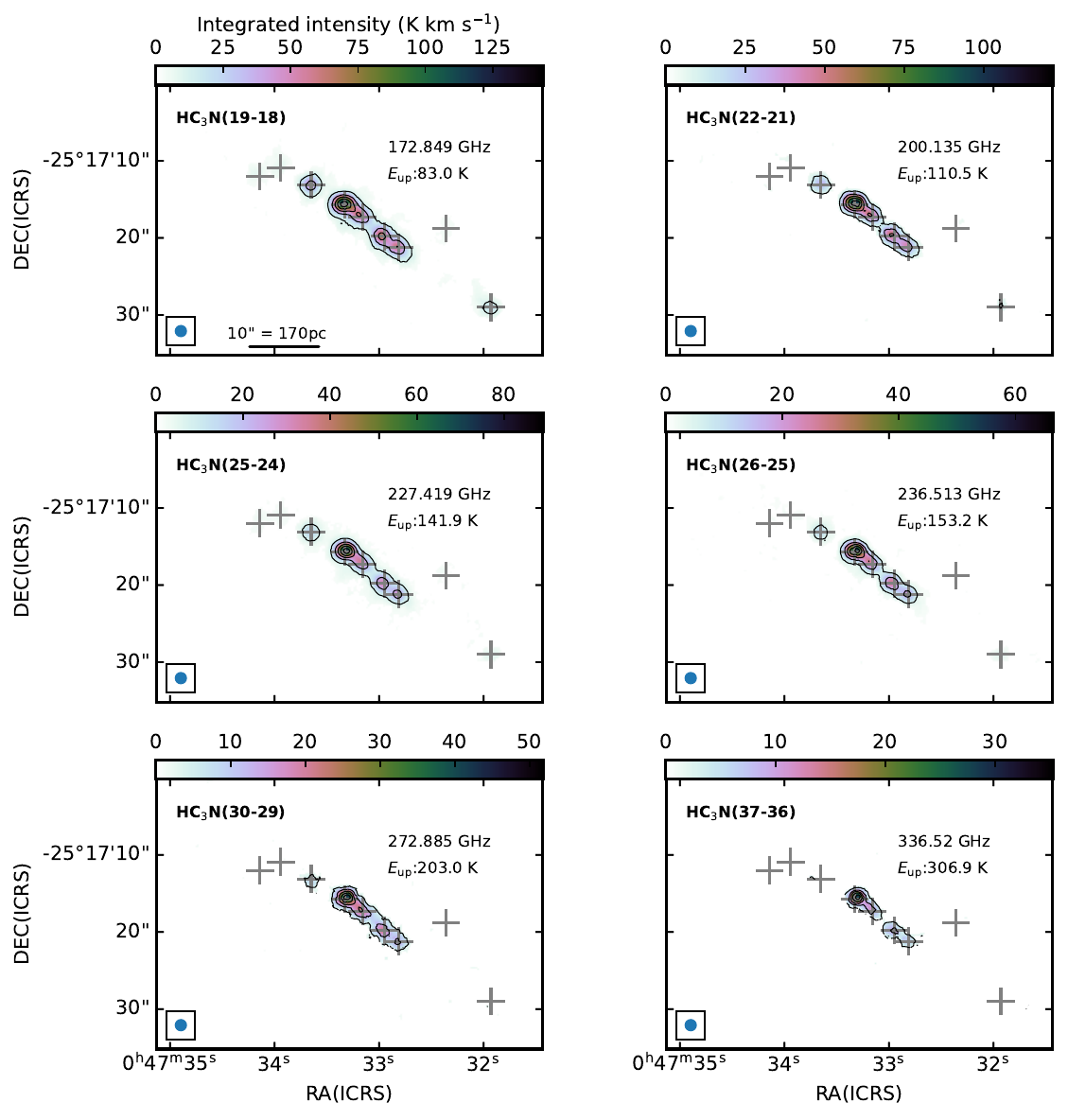}
\caption{Same as Figure \ref{fig:mom0_0}. \label{fig:mom0_17}}
\end{figure}

\begin{figure}[ht]
\includegraphics[width=0.99\textwidth]{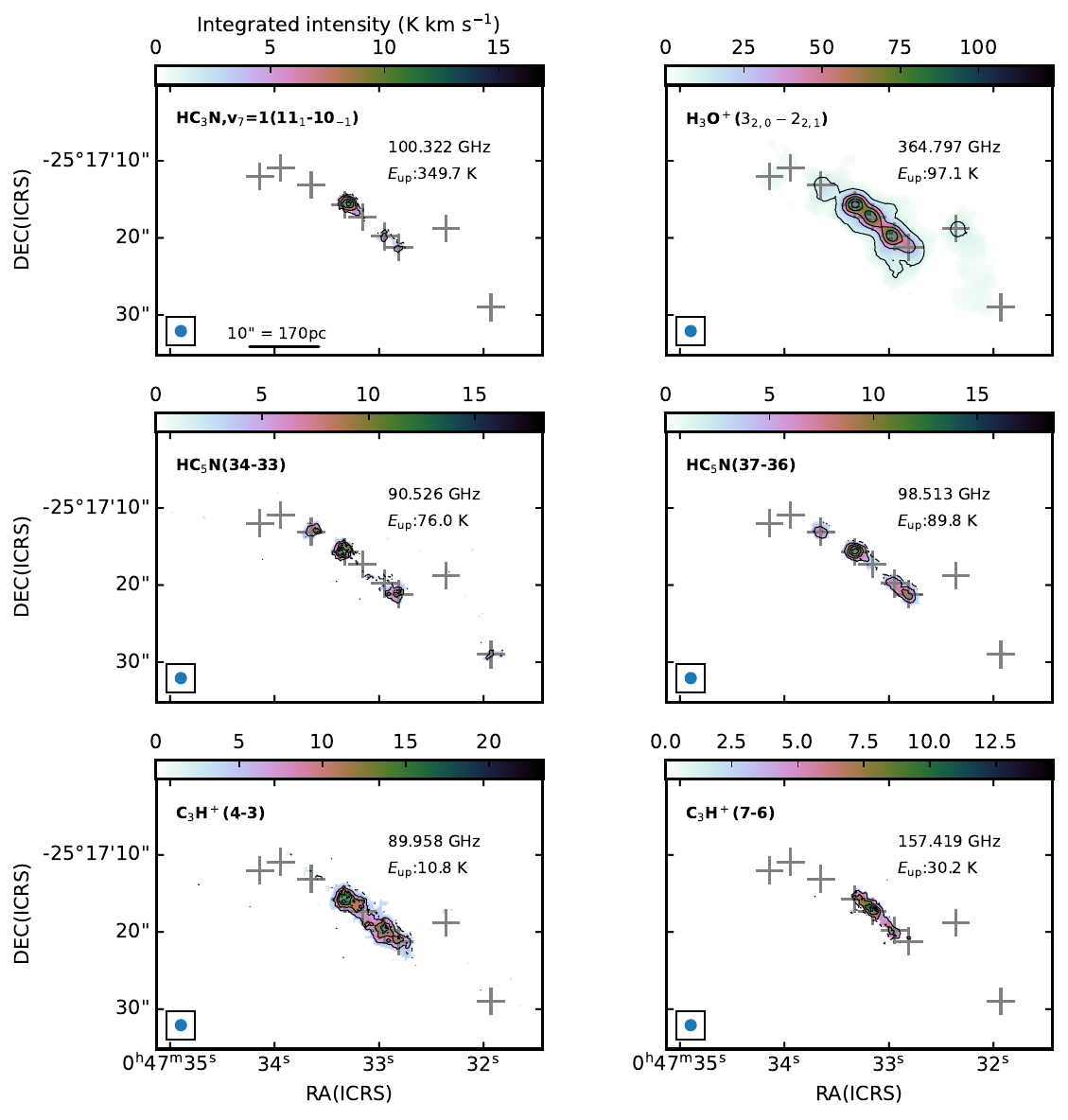}
\caption{Same as Figure \ref{fig:mom0_0}. \label{fig:mom0_18}}
\end{figure}

\begin{figure}[ht]
\includegraphics[width=0.99\textwidth]{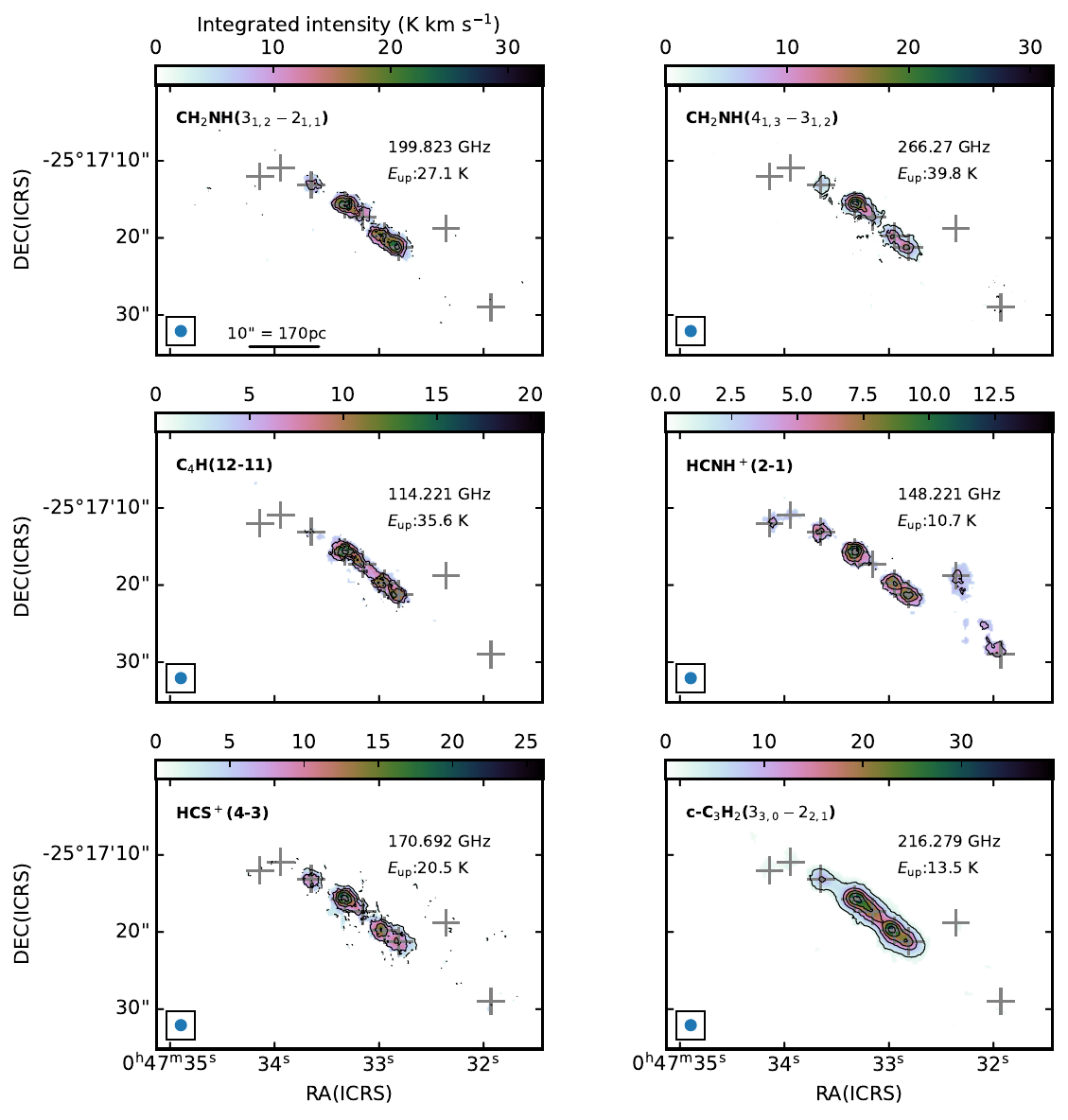}
\caption{Same as Figure \ref{fig:mom0_0}. \label{fig:mom0_19}}
\end{figure}

\begin{figure}[ht]
\includegraphics[width=0.99\textwidth]{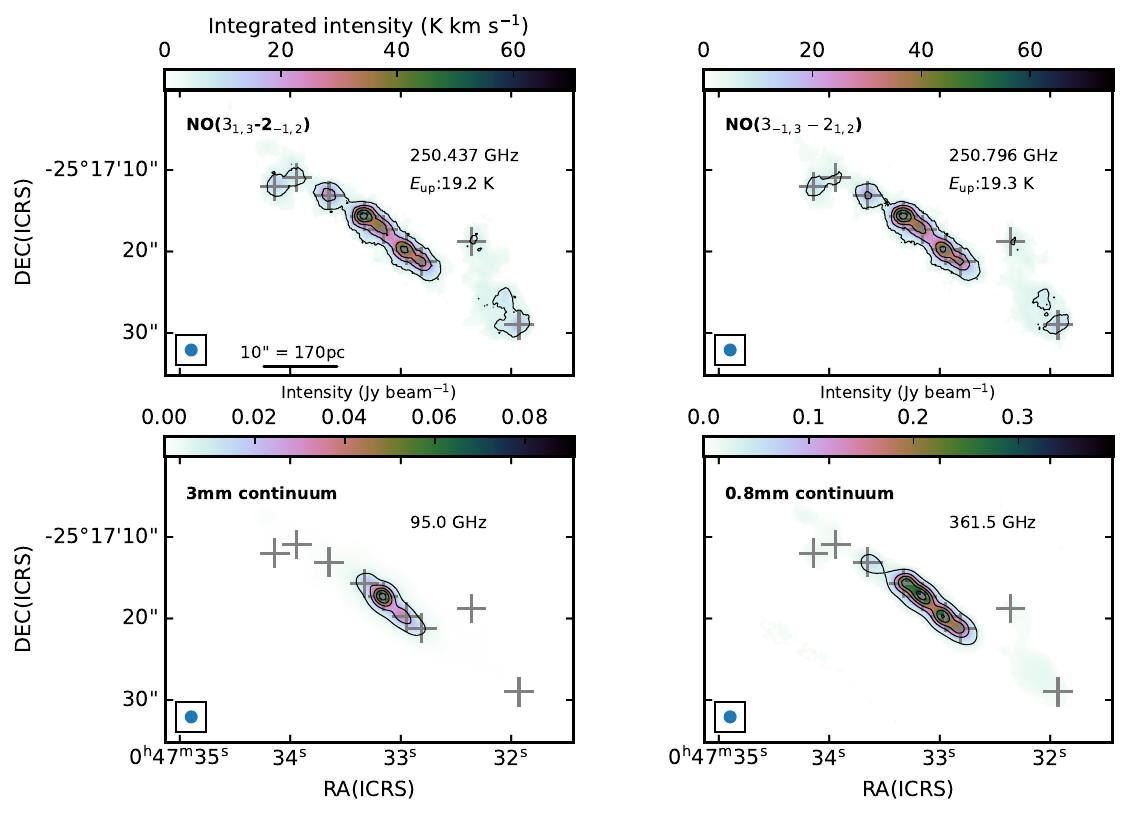}
\caption{Same as Figure \ref{fig:mom0_0}. \label{fig:mom0_20}}
\end{figure}

Figures \ref{fig:mom0_0}--\ref{fig:mom0_20} show the integrated intensity images created using the procedure described in the previous section. Continuum images at 95 GHz (3.2 mm) and 361.5 GHz (0.83 mm) are also included in the last two panels of Figure \ref{fig:mom0_20}. As already illustrated by \citet{2015ApJ...801...63M}, morphological differences are obvious from simple visual inspection. While distributions of some transitions are more extended, some are more concentrated around the central parts, and others are enhanced in the outskirts of the CMZ (GMCs 1, 2, 8, and 9) in NGC~253. 

In general, CO isotopologues ($^{12}$C$^{16}$O, $^{13}$C$^{16}$O, $^{12}$C$^{18}$O, and $^{12}$C$^{17}$O, hereafter \twco, \ttco, \ceto, and C$^{17}$O) exhibit more extended emission than other species. Higher-$J$ transitions are slightly more compact than $J=1-0$ transitions, likely because the gas is more excited near active star-forming regions. Rarer isotopologues of CO show more compact emission than the main isotopologue ($^{12}$C$^{16}$O) both due to the lower optical depths and lower S/N.

Transitions of species with high critical densities ($n_{\rm crit} > 10^4\,$\cc\ in optically thin cases) such as HCN, HCO$^+$, CS, \nthp, and CCH also show significantly extended emission in their $J=1-0$ transition, but higher-$J$ transitions are more compact, some of which have already been shown in our previous literature \citep{2021ApJ...923...24H,2021A&A...654A..55H,2022ApJ...939..119B}. We see this trend in CO isotopologues as well, but it is more prominent in these transitions with high critical densities. Similar to CO isotopologues, emission from rarer isotopologues of other molecules is more compact than from their main counterparts.\footnote{
\citet{2021ApJ...923..240S} reported similar size variations among various sub/millimeter emissions from their ALMA imaging spectral scans toward the nuclei of luminous infrared galaxies NGC 4418 and Arp 220. Taken together, these observations caution against the assumption of a common beam-filling factor in the excitation analysis of molecular lines for unresolved galactic nuclei.}

We find somewhat compact distributions in the 3-mm continuum emission, most of which should originate from free-free emission, and RRLs compared with other transitions mentioned above. They are expected to arise from the ionized gas, tracing the ongoing star formation \citep[e.g., ][]{2015MNRAS.450L..80B}. These types of emission and transitions such as HCN, \hcop, CCH, and CN with higher upper-state energies emit strongly in similar locations \citep[see also ][for CCH]{2021A&A...654A..55H}.

On the other hand, \methanol, HNCO, and \hocop\ transitions with low upper-state energies are enhanced in the outskirts of the NGC~253 CMZ (Figures \ref{fig:mom0_9}-\ref{fig:mom0_13}). Out of these species, \methanol\ and HNCO are known as tracers of slow shocks\footnote{Fast shocks could enhance methanol as well according to some modeling results, although there is also an observational result that suggests methanol destruction in fast shocks \citep{2014MNRAS.440.1844S}.}, and their enhancements are likely attributed to widespread shocks in those regions \citep{2015ApJ...801...63M,2023arXiv230312685H}. \citet{2022ApJ...938...80H} argued that \hocop\ should be tracing shock-evaporated CO$_2$, and its relation with shocks is suggested by the similarity with these species. A tracer of fast shocks, SiO, also shows some emission at the outskirts of the CMZ. Unlike weaker shock tracers, SiO transitions emit strongly near the starburst regions as well (Figures \ref{fig:mom0_8}-\ref{fig:mom0_9}).

 To the best of our knowledge, we also present the first extragalactic detection and/or first extragalactic interferometric images for certain species. For example, HCNH$^+$ (Figure \ref{fig:mom0_19}), protonated HCN or HNC (see Figures \ref{fig:mom0_1} and \ref{fig:mom0_3}), is detected for the first time outside the Milky Way. It shows a similar distribution as that of HCN (Figure \ref{fig:mom0_1}). One production pathway of \hcnhp\ is via protonation of HCN or HNC by proton exchange reactions with H$_3^+$ or \hcop. Another formation route is through HNC$^+$ or HCN$^+$ reacting with molecular hydrogen. A survey of high-mass star-forming regions suggested that \hcnhp\ is more abundant in cold starless cores compared with more evolved sources \citep{2021A&A...651A..94F}. \hcnhp(2--1) in the CMZ of NGC~253 has a similar distribution as that of HCN(1--0), but with a lower S/N (compare Figures \ref{fig:mom0_1} and \ref{fig:mom0_19}). It is distributed widely in the entire CMZ rather than being concentrated around the starburst region. This is consistent with the picture that  \hcnhp\ is more abundant in cold clouds. We also obtained the first extragalactic images of C$_3$H$^+$, NO, and HCS$^+$.  \cthp\ (Figure \ref{fig:mom0_18}) is a tracer of PDRs, and was first detected by \citet{2012A&A...548A..68P} in a PDR of the Horsehead nebula. With a lower fractional abundance, this species was also detected in TMC-1 \citep{2022A&A...657L..16C}. The first extragalactic detection of this species was made towards a molecular absorber at $z=0.89$ \citep{2020A&A...636L...7T}. Interstellar NO has been detected already several decades ago \citep{1978ApJ...224L..73L}. Its first extragalactic detection was made in NGC~253 by \citet{2003A&A...411L.465M}. Protonated CS, HCS$^+$ was detected for the first time by \citet{1981ApJ...246L..41T} rather ubiquitously in hot cores and cold clouds. Its first extragalactic detection was also in a molecular absorber at $z=0.89$ \citep{2013A&A...551A.109M}.

There are some moment 0 images that are used in PCA, but not shown in this paper because they will be included in dedicated publications, as indicated in Table \ref{tab:spec}. This is the case of the maps of sulfur-bearing species since M. Bouvier et al. (in preparation) will provide a comprehensive presentation to investigate the origin of the emission of various sulfur-bearing species and discuss physical processes dominate their release/presence onto the gas phase. Other transitions of CO, HCN, and \hcop\ contain rarer isotopologues such as $^{13}$C, $^{18}$O, $^{17}$O, and $^{15}$N. J. Butterworth et al. (in preparation) will discuss isotopic ratios through the lens of these isotopologues in order to investigate the possible relation between these ratios and the ages of SSCs observed in NGC~253. These images that are currently not included in this version of paper will be included later in the arXiv preprints after above papers are accepted.

We note that some of the transitions in Band 5 have poor image qualities influenced by artifacts (e.g., H$^{13}$CO$^+$ 2--1, \hocp\ 2--1). These transitions come from the datasets that were not fully observed and did not meet the homogeneous sensitivity requested in this survey \citepalias[see Section 3.2 of ][]{2021A&A...656A..46M}. This band has a water absorption line near the middle of the band, and transitions nearby are prone to increased noise levels. PCA results for these transitions are less certain.

%%%%%%%%%%%%%%%%%%%%%%
%%%%%%%%%%%%%%%%%%%%%

%%%%%%%%%%%%%%%%%%%%%
\clearpage
%%%%%%%%%%%%%%%%%%%%%
\clearpage
%%%%%%%%%%%%%%%%%%%%%
\clearpage
%%%%%%%%%%%%%%%%%%%%%
\clearpage
%%%%%%%%%%%%%%%%%%%%%
\clearpage
%%%%%%%%%%%%%%%%%%%%%
\clearpage
%%%%%%%%%%%%%%%%%%%%%
\clearpage
%%%%%%%%%%%%%%%%%%%%%
\clearpage
%%%%%%%%%%%%%%%%%%%%%
\clearpage
%%%%%%%%%%%%%%%%%%%%%
\clearpage
%%%%%%%%%%%%%%%%%%%%%
\clearpage
%\clearpage
%%%%%%%%%%%%%%%%%%%%%
\clearpage
%\clearpage
%%%%%%%%%%%%%%%%%%%%%
\clearpage
%\clearpage
%%%%%%%%%%%%%%%%%%%%%
\clearpage
%\clearpage
%%%%%%%%%%%%%%%%%%%%%
\clearpage
%\clearpage
%%%%%%%%%%%%%%%%%%%%%
\clearpage
%\clearpage
%%%%%%%%%%%%%%%%%%%%%
\clearpage
%\clearpage
%%%%%%%%%%%%%%%%%%%%%
\clearpage
%\clearpage
%%%%%%%%%%%%%%%%%%%%%
\clearpage
\clearpage
%%%%%%%%%%%%%%%%%%%%%

\section{Principal component analysis}\label{sec:pca}
Astronomical datasets often have a large number of variables that are correlated to some degree. It is a challenge to extract information from such data and there is a need for somewhat automatized statistical investigation. Dimensionality reduction is especially a useful technique in this regard, deriving a small number of components to reproduce  most of the variations within the data. Multiple methods of dimensionality reduction have been developed so far, including PCA, non-negative matrix factorization (NMF), and independent component analysis (ICA). PCA is one of the oldest and most commonly used with more than a century-old history \citep{Pearson1901}. As mentioned in Section \ref{sec:intro}, PCA has been most widely used for molecular line studies (see references in Section \ref{sec:intro}).
Moreover, this work aims to push the PCA analysis to the next step by applying it to a much larger set of molecular lines. %molecular inventory ever attempted with this technique. 

\subsection{The basics of PCA}\label{sec:pca_expl}
We begin by briefly describing the basics of PCA relevant to this work, leaving details and proofs to  
textbooks \citep[e.g.,][]{Jolliffe2002}. 
% PCA fundamental: The PC coordinate system and the reduction of data dimensionality
A data set of $p$ variables measured at $n$ targets  (e.g., intensities of $p$ emissions at each of the $n$ sky positions) can be viewed as $n$ data points in a $p$-dimensional data space.
PCA sets up an orthogonal coordinate system in the data space with the origin at the centroid of the data points and the first coordinate axis (PC1 axis) in the direction of 
the largest variance of the data point positions along the axis.
The PC2, PC3, ..., axes are successively defined as orthogonal to all previous axes and having the largest possible variance of the data positions along the axis.
%(If the variance is the same in multiple directions, choose any one. Arbitrarily choose the positive orientation of each PC axis.)
Coordinate values on a PC axis are called PC scores. 
The PC scores of the data points have a mean of zero for each PC and smaller variances for later PCs.
Accordingly, the relative positions of the data points can be described reasonably well by using only their first $q \; (\leq p)$ PC scores, since all the data points have approximately the same scores ($\approx 0$) for the rest of the PCs.
This description reduces the dimensionality of the data set from $p$ to $q$.

% Finding PC-k axis, \bm{a]_k, from the data matrix
Individual PC axes are found in the following way.
Let $x_{ij}$ be the value of the $j$-th variable measured at the $i$-th target.
The data matrix $\bm{X} = (x_{ij})_{1 \le i \le n, 1\le j \le p}$ has $x_{ij}$ in the $i$-th row, $j$-th column. 
We use the data matrix that is standardized for each variable; i.e., each column has its mean subtracted and then divided by its standard deviation.
This subtraction aligns the coordinate origin with the centroid of the data points.
The correlation matrix for the $p$ variables is then 
$ \bm{\Sigma} = \frac{1}{n} \bm{X}^T \bm{X} $, where $^T$ denotes the transpose.
The $p \times p$ real matrix is symmetric and non-negative definite, hence having $p$ eigenvalues that are real and non-negative. We denote them as $\lambda_1 \ge \cdots \ge \lambda_p \ge 0$.
%The $p$ eigenvalues of $\bm{\Sigma}$ are real and non-negative and are denoted as $\lambda_1 \ge \cdots \ge \lambda_p \ge 0$.
The direction vector of the PC$j$ axis is the unit eigenvector of $\bm{\Sigma}$ associated with the $j$-th largest eigenvalue $\lambda_j$, and 
we denote it as $\bm{a}_j = (a_{1j}, \cdots, a_{pj})^T$. 
They are orthonormal, i.e., $\bm{a}_i \cdot \bm{a}_j = \delta_{ij}$.
The variance of the PC$j$ scores is $\lambda_j$.

% Transformaiton between the data and PC coordinates, PC coefficients
The transformation between the data coordinates and the PC coordinates is as follows.
Let $z_{ij}$ be the PC$j$ score of the $i$-th target. 
The target is located at $\bm{z}_i = (z_{i1}, \cdots, z_{iq})$
in the $q$-dimensional PC coordinates 
and $\bm{x}_i = (x_{i1}, \cdots, x_{ip})$ 
in the standardized data coordinates.
Since $z_{ij}$ is the projection of $\bm{x}_i$ onto the PC$j$ axis along the unit vector $\bm{a}_j$, 
we have $z_{ij} = \bm{x}_i \bm{a}_j$ and $\bm{z}_i = \bm{x}_i \bm{A}$, 
where $\bm{A}$ is the matrix 
$
    \bm{A}
    = 
    (\bm{a}_1, \cdots, \bm{a}_q)
    =
    (a_{ij})_{1 \le i \le p, 1 \le j \le q}
$.
The element $a_{ij}$ is the direction cosine between the $i$-th data axis and the PC$j$ axis, and is usually referred to as a `coefficient' or `loading' for PC$j$.
If it is positive, a positive value of the $i$-th variable makes a positive contribution to the PC$j$ score, while a negative $a_{ij}$ indicates a negative contribution. 
%Note that $\bm{l}_i = (a_{i1}, \cdots, a_{iq})$, the $i$-th row of $\bm{A}$, 
%is the coefficient vector of the $i$-th variable. It 
%is in the direction of the original $i$-th data axis in the PC coordinates.

For the simplest case of two variables with a strong correlation, the PC1 axis would lie in the direction along the linear fit of the data points, while the PC2 axis would be in the direction perpendicular to it. Therefore, PC1 scores indicate the strength of the common properties in the two variables, while PC2 scores reflect the strength of the differential properties of the two variables.
%Therefore, PC1 explains the similarity between the two measurements while differences between them can be explained by PC2. 

PC scores are uncorrelated between different PCs because $\bm{\Sigma}$ is diagonal in the PC coordinates. Therefore, if variables are determined by underlying parameters with some correlations, then there would not be a one-to-one relationship between those parameters and the principal components.
We discuss this caveat when we apply PCA to the ALCHEMI data.

\subsection{Application of PCA to the ALCHEMI data}
% Hex Binning
We binned the velocity-integrated images into hexagonal pixels with a horizontal length of 0\farcs8 using Python's {\tt hexbin} function. The number of hex-pixels is $n=2723$. The choice of half the beam size is made for the Nyquist-like sampling. 
We filled the masked out channels/pixels with zero. 

% Normalization
We standardized our data for each variable (i.e., transition or emission) as described in Section \ref{sec:pca_expl}. The division by the standard deviation is a common procedure for a data set whose variables differ greatly in their variances. It also makes our PCA independent of the choice of data units between Jy and K. 
%Although we also tried another normalization method implemented by \citet{2017A&A...599A.100G}, we do not employ it here because our interest is in examining the differences between transitions that are more highlighted by our standard method. 

% Choice of variables and targets
The variables in our analysis are transitions and the targets are sky positions, as in \citet{1997ApJ...482..245U} or \citet{2005ApJ...618..259M}. With 148 lines and 2 continuum images, we have $p=150$ variables.
Instead, we could have used different transitions as targets and their intensities at the pixels as variables, as in facial recognition, where eigenvectors are eigenfaces. However, we did not use the approach to avoid the problem of high dimension and low sample size (i.e., $p > n$)\footnote{See \citet{2022arXiv220304535T} for how to properly perform PCA on such a dataset.}. 
%The matrix $\bm{A}$ has $p$ rows and $q$ columns, where $q$ is the number of principal components. 
It is also possible to conduct a PCA on image cubes, instead of integrated intensities. Nonetheless, we use integrated intensity images first, because integrated intensities tend to have higher signal-to-noise ratios. A PCA on image cubes, possibly with a lower number of transitions, is planned as a future work.
Consequently, there are PC$k$ scores for all hexagonal pixels, $\{z_{ik} \}_{i=1, \cdots, n}$.
We used the PCA package from the {\tt scikit-learn} project \citep{scikit-learn}.
%The number of dimension in our case is the number of transitions as in \citet{1997ApJ...482..245U} or \citet{2005ApJ...618..259M}. We note that we could instead use the number of pixels as our dimension like in cases of facial recognition where eigenvectors are eigenfaces. However, we do not use the latter approach to avoid high-dimension low-sample size problem\footnote{See \citet{2022arXiv220304535T} on how to conduct PCA properly when dealing with such a dataset.}. The matrix $A$ is a matrix with the size of the number of components times the number of transitions, while the $k$-th PC scores $z_k$ has a size of the number of hexagonal pixels $n$. We used the PCA package from the {\tt scikit-learn} project \citep{scikit-learn}.

\begin{figure}[]
\includegraphics[width=0.5\textwidth]{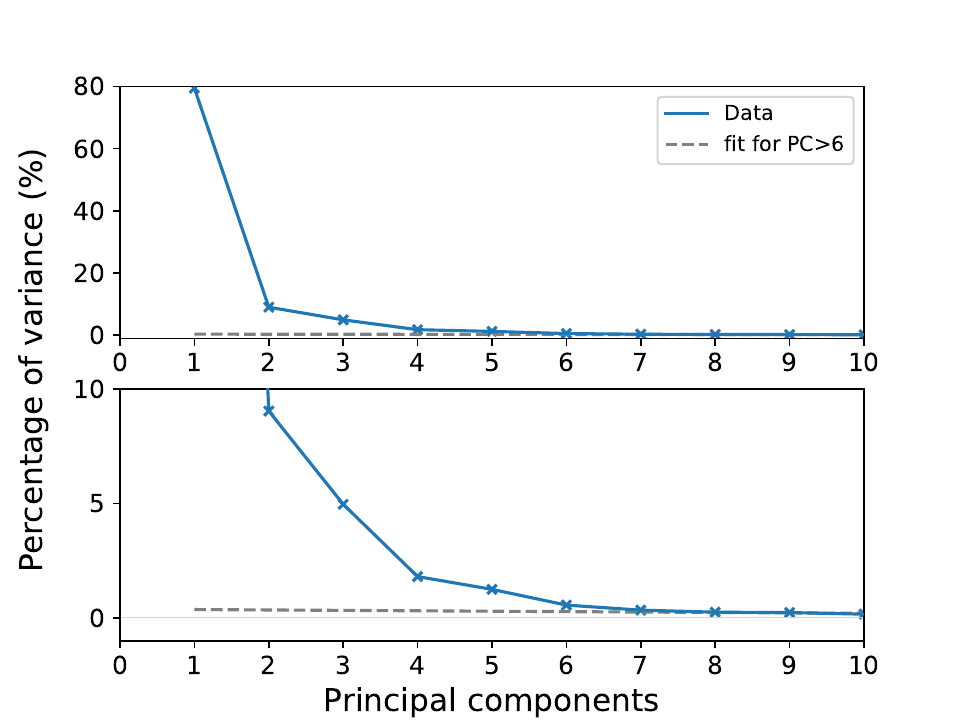}
\caption{(Top) A scree plot for our PCA. (Bottom) The same as the top figure, but for a narrower range of the y-axis.  \label{fig:scree}}
\end{figure}

% Deciding q from scree plot
The number of principal components $q$ must be chosen to retain most of the information in the data set. The information in the context of PCA is the configuration of the data points and is evaluated with the variances. For example, PC$k$ has the variance of $\lambda_k$ in its scores and therefore has less information for larger $k$.
Figure \ref{fig:scree}, called a scree plot, shows the fraction of the variance of each PC in the total variance, $f(k) \equiv \lambda_k/\sum_{j=1}^{p} \lambda_j$.
In this plot, the PCs that exceed the least-squares linear fit to the high-order PCs (PCs 6--20) are considered more significant.
The first three PCs have relatively large fractions of information (i.e., variance) 
$f(k) =$ 79.4, 9.0, and 5.0 \%, respectively, for $k=$1, 2, 3,
with a cumulative contribution of 93\% of the total.
PCs 4 and 5 also show some significance ($f(k) =$ 1.8 and 1.2 \%) exceeding the fit at PCs 6--20, each of which has $f(k) \le $ 0.6\%. 
As shown in  Section \ref{sec:projection}, the PC score maps for PC6 and higher are noisier than for PCs 1--5. 
%It is important to determine how many components of the PCA contain most of the variance in data. Figure \ref{fig:scree} shows the percentage of variance for each component, called a scree plot. In this plot, the PCs with higher variance than the least-square linear fit to high-order PCs (PCs6-20) tend to contain more significance. The first three components contain information that explains a relatively large amount of variance (79.4, 9.0, and 5.0 \%, respectively), which totals 93\%. PCs 4 and 5 also show some significance (1.8 and 1.2 \%) that is higher than we expected from the fit among PCs 6-20 (0.6\% or less). As shown in  Section \ref{sec:projection} the PC score maps of PC6 or higher are noisier than PCs1-5. 
Therefore, we only discuss PCs 1--5 in this paper; they have
$\sum_{j=1}^{5} \lambda_j/\sum_{j=1}^{p} \lambda_j = 96.4$\% of the total variance.

PCA is one of the unsupervised machine learning methods, which means that any known properties of data, or labels, are not used. This means that PCA results are derived independent of what kind of species the molecules are, or energy levels of the transitions. We will see below (in Section \ref{sec:pca_phys}) that our PCA could recover such information to some extent from the data alone.

%%%%%%%%%%%%%%%%%%
%%%%%%%%%%%%%%%%%%%%%

\subsection{PC score maps} \label{sec:projection}

%%%%%%%%%%%%%%%%%%%%%
\begin{figure*}[ht]
\centering{
\includegraphics[width=0.49\textwidth]{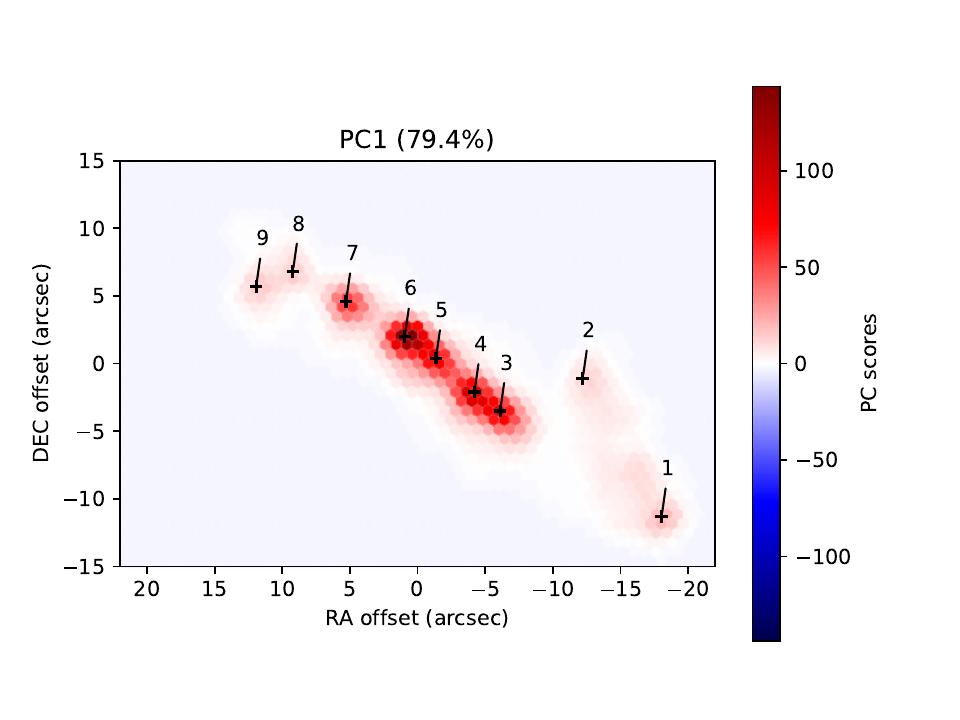}
\includegraphics[width=0.49\textwidth]{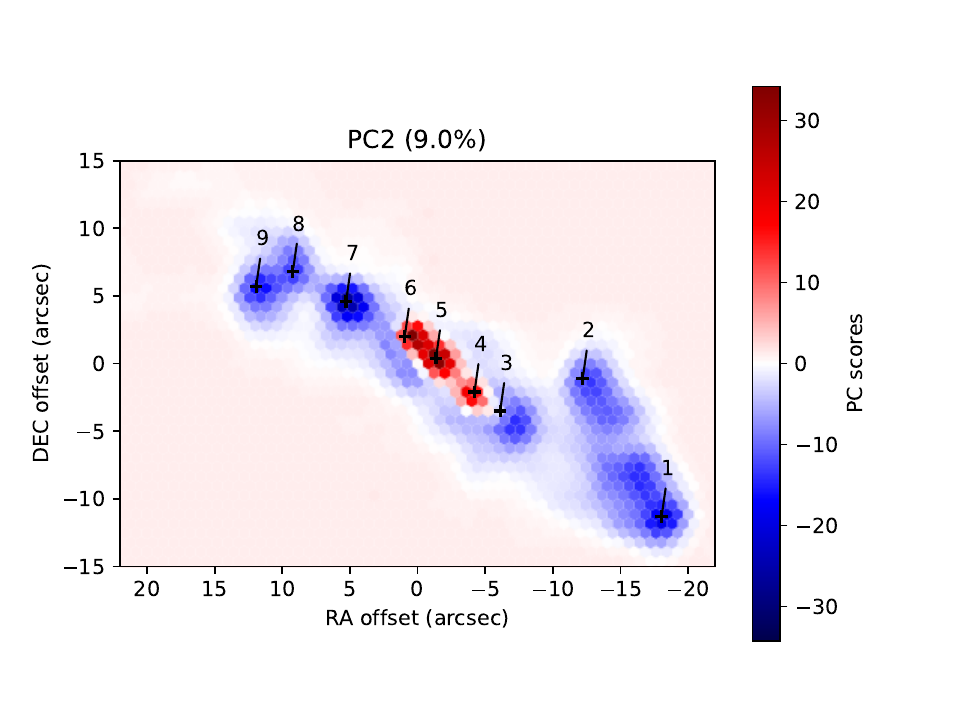}}
\centering{
\includegraphics[width=0.49\textwidth]{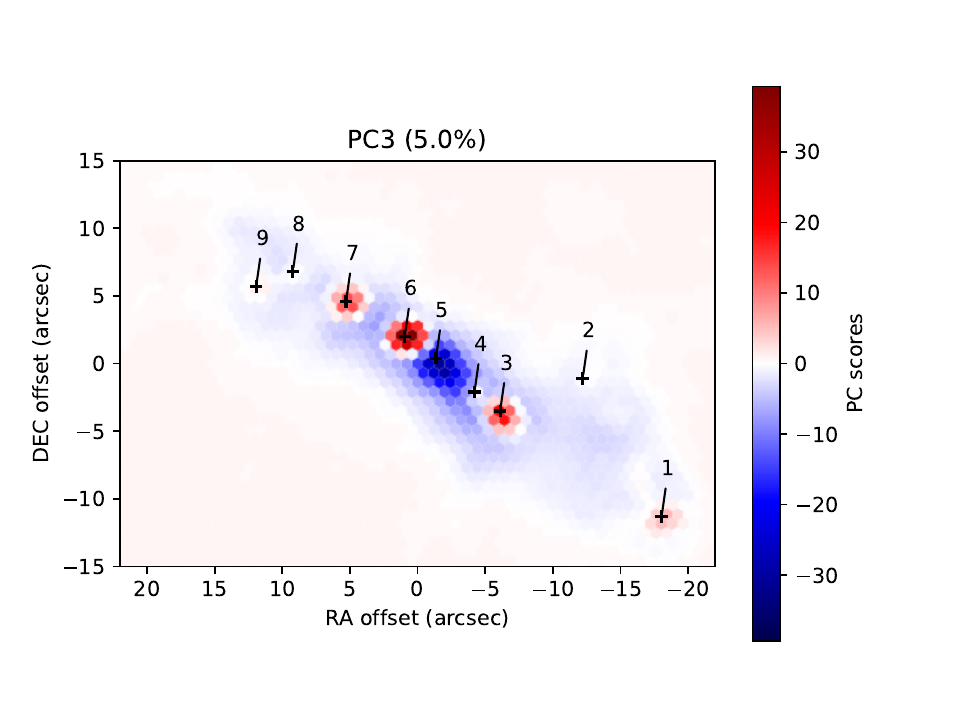}
\includegraphics[width=0.49\textwidth]{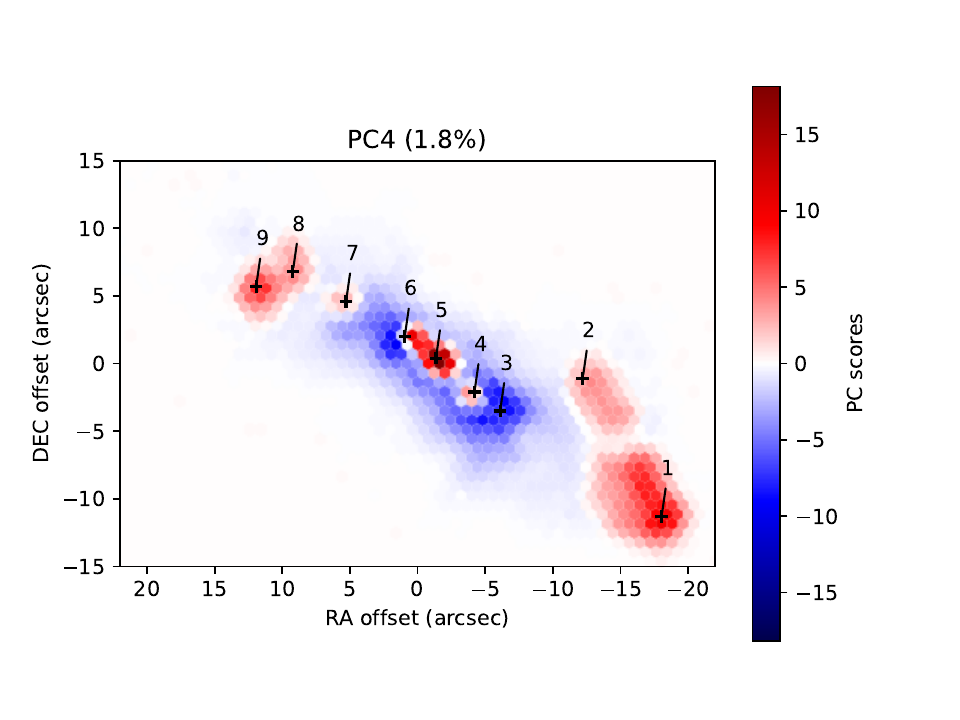}}
\centering{
\includegraphics[width=0.49\textwidth]{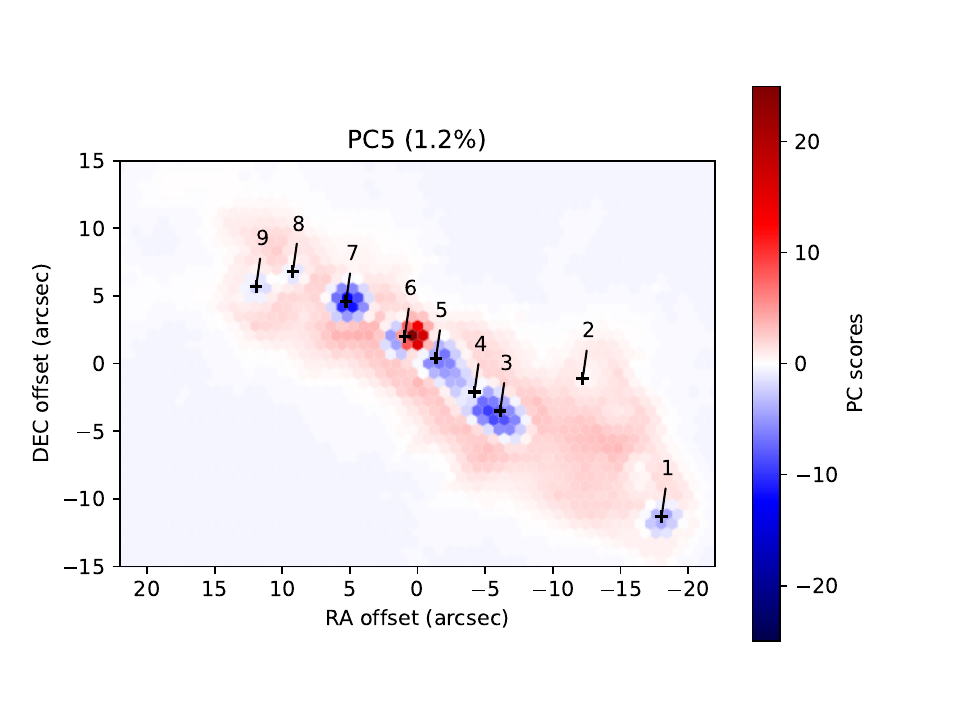}
\includegraphics[width=0.49\textwidth]{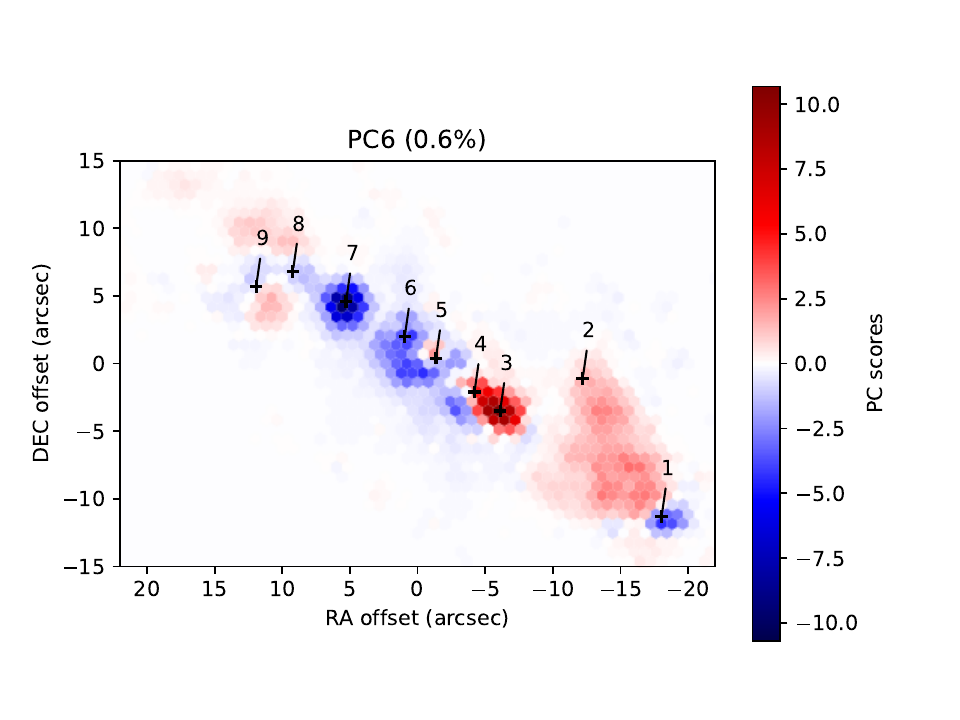}}
\caption{PC scores of the 1st to 6th principal components (i.e., $\{z_{ij}\}_{1\le i \le n}$ for $j=1-6$). GMC positions listed in Table~\ref{tab:gmc_pos} are shown with black crosses. The origin is at the phase center of the observation, $\alpha=00^h47^m33.26^s$, $\delta=-25^\circ 17^\prime 17.7^{\prime \prime}$ (ICRS). \label{fig:pcmap1}}
\end{figure*}
%%%%%%%%%%%%%%%%%%%%%
\begin{figure*}[ht]
\centering{
\includegraphics[width=0.49\textwidth]{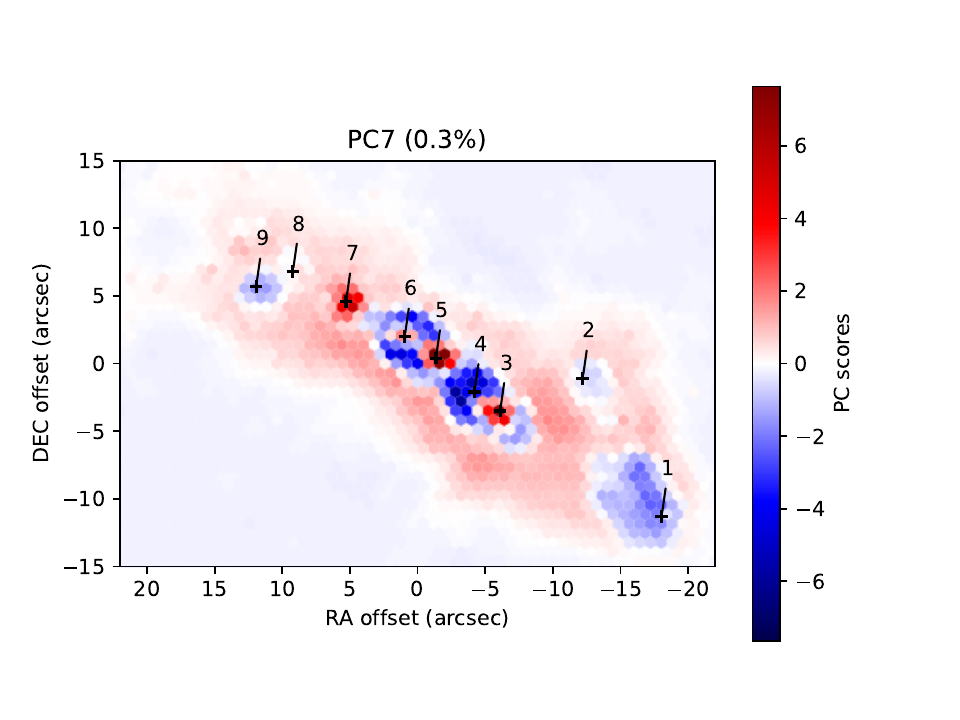}
\includegraphics[width=0.49\textwidth]{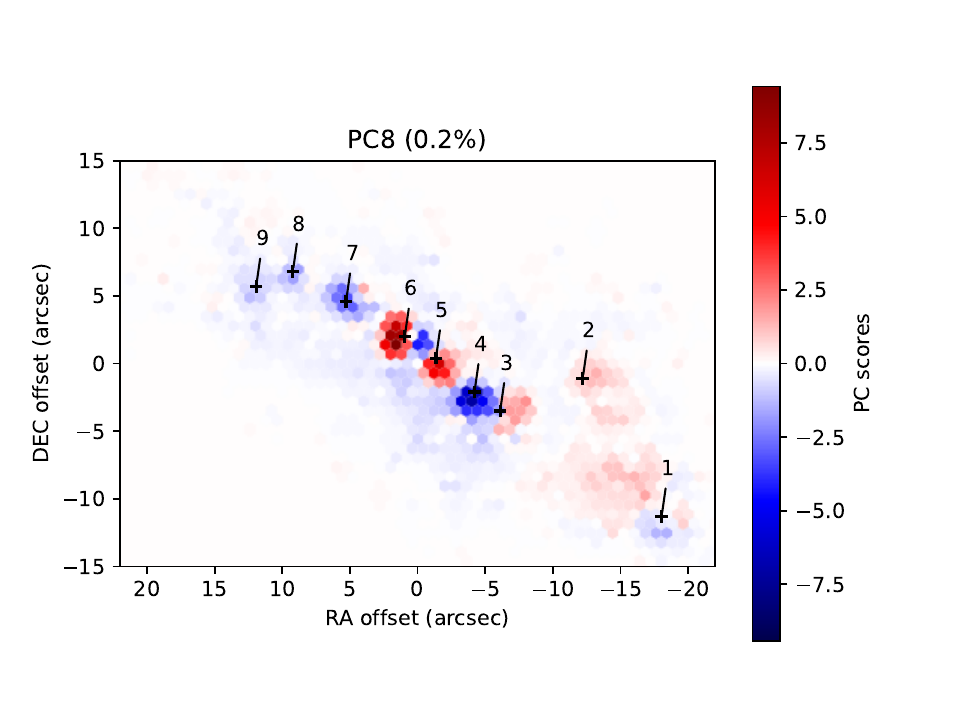}}
\centering{
\includegraphics[width=0.49\textwidth]{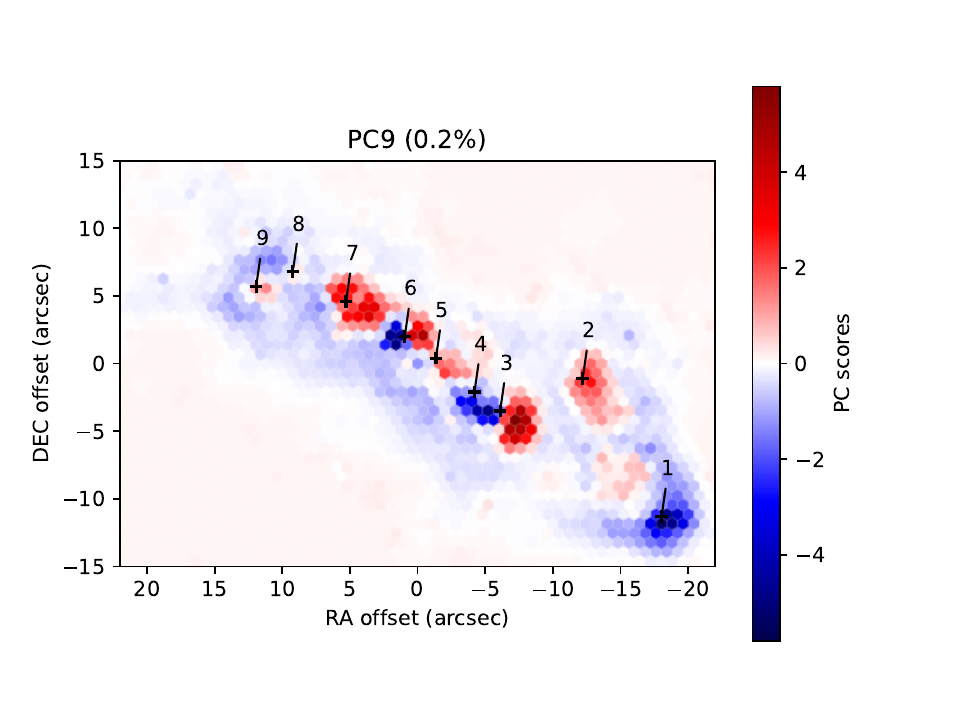}
\includegraphics[width=0.49\textwidth]{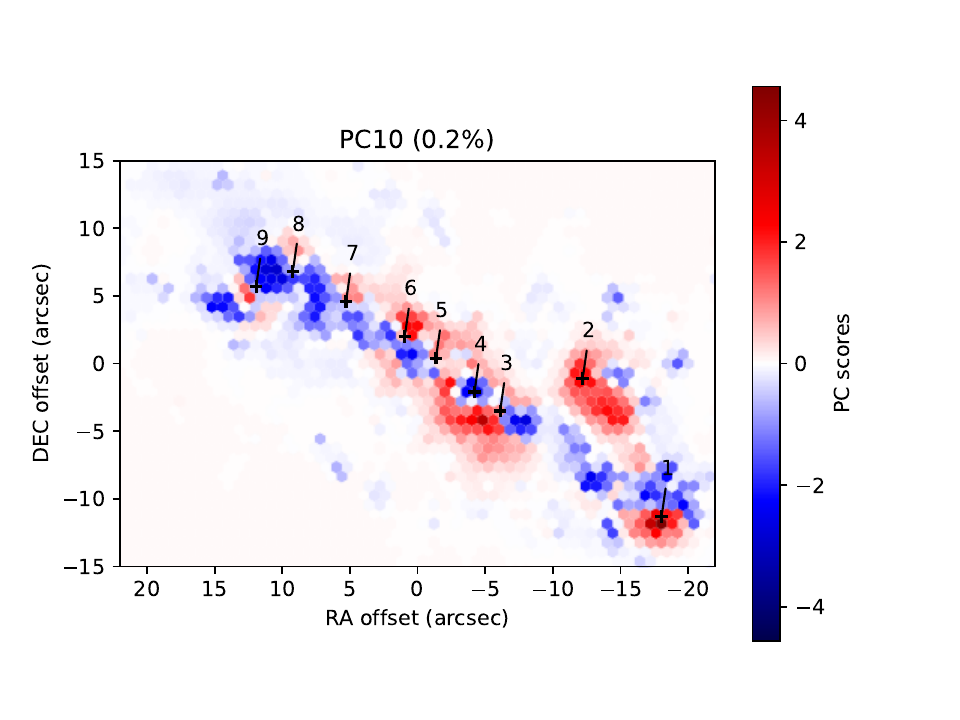}}
\caption{The same as \ref{fig:pcmap1}, but for PCs7-10. \label{fig:pcmap2}}
\end{figure*}
%%%%%%%%%%%%%%%%%%%%%

Figures \ref{fig:pcmap1} and \ref{fig:pcmap2} show the projection of data points onto the PC axes ($\{z_{ik} \}_{i=1, \cdots, n}$ for PC$k$ in the notation of Section \ref{sec:pca_expl}), or the maps of PC scores for the first ten PCs. Here we provide a brief overview of the relationship between the individual PCs and the emission of certain molecular species. Note that the PC scores are dimensionless and have little meaning in their absolute values. Only their relative magnitudes between different PCs and positions are relevant.
The PC1 score map (Figure \ref{fig:pcmap1} top left) shows a rough shape of the CMZ, similar to the distribution of CO isotopologues. The positive PC2 scores (Figure \ref{fig:pcmap1} top right) appear near the central starburst regions (GMC4-6), where the high-$J$ transitions of HCN emission and RRLs have high intensities. On the other hand, the negative PC2 scores are at the outskirts of the CMZ, where low-$J$ transitions of \methanol\ and HNCO are enhanced (GMC1-3,7-9). Note that the green (\methanol\ $2_K$--$1_K$) and red (H39$\alpha$) colors in Figure \ref{fig:rgb} (top) show remarkable similarity with blue (negative) and red (positive) colors in the PC2 score map (Figure \ref{fig:pcmap1}). High absolute values of the PC3 scores (Figure \ref{fig:pcmap1} middle left) appear around positions with positive values of the PC2 score map. Positive PC3 scores appear at the peak location of HC$_3$N vibrationally excited lines (GMC6) while the negative PC3 scores are at the peak of the RRL (GMC5). We again note that the green (CN 3--2) and red (\hctn\ 25--24) colors in Figure \ref{fig:rgb} (bottom) resemble the blue and red colors in the PC3 map.

As mentioned earlier, contributions of PCs 4 and 5 (Figure \ref{fig:pcmap1} middle right and bottom left) are less than those of PCs 1-3. They may contain multiple physical components because PCs (eigenvectors) are by definition orthogonal to each other while each physical component does not necessarily influence intensities orthogonally in dimensions of transition intensities. For example, PC4 has positive PC scores in most of the GMCs except for GMCs 3 and 7. The PC4 score map appears similar to that of PC2, except for GMCs in the outskirts (GMCs 1, 2, 7, 8, and 9). The PC4 score map also has an extended structure out of the galactic plane with the negative component. PC5 shows negative PC scores in most of the GMCs, but it has positive PC scores in GMC6 and diffuse extended structure. 

In principle, PC6 and higher-order maps still provide some information about this galaxy, not merely contributions due to noise. However,  PC score maps from PC6 to PC10 cannot be uniquely linked to the emission of individual species and therefore cannot easily be linked to specific physical processes (Figures \ref{fig:pcmap1} and \ref{fig:pcmap2}). 

\subsection{Relationship with physical parameters}\label{sec:pca_phys}
\begin{figure*}[ht]
\centering{
\includegraphics[width=0.8\textwidth]{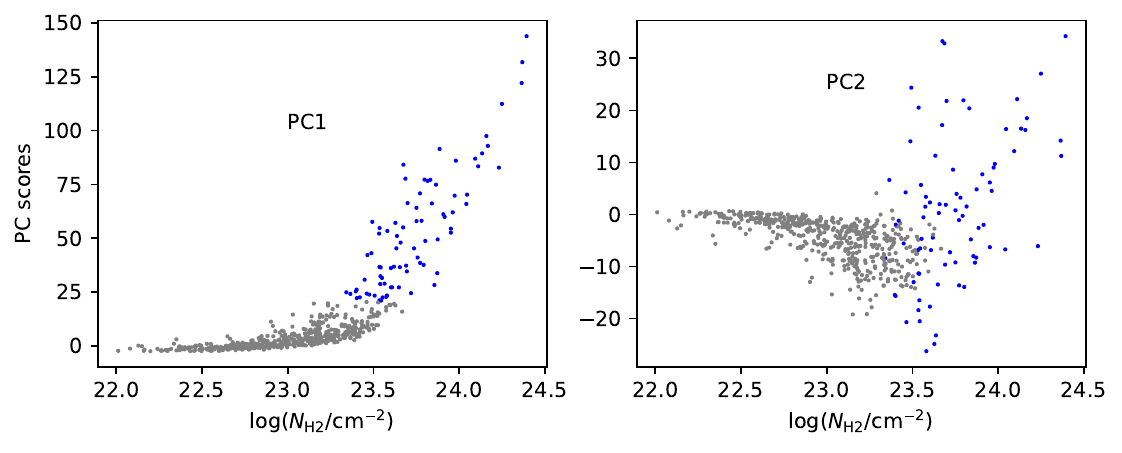}
\includegraphics[width=0.8\textwidth]{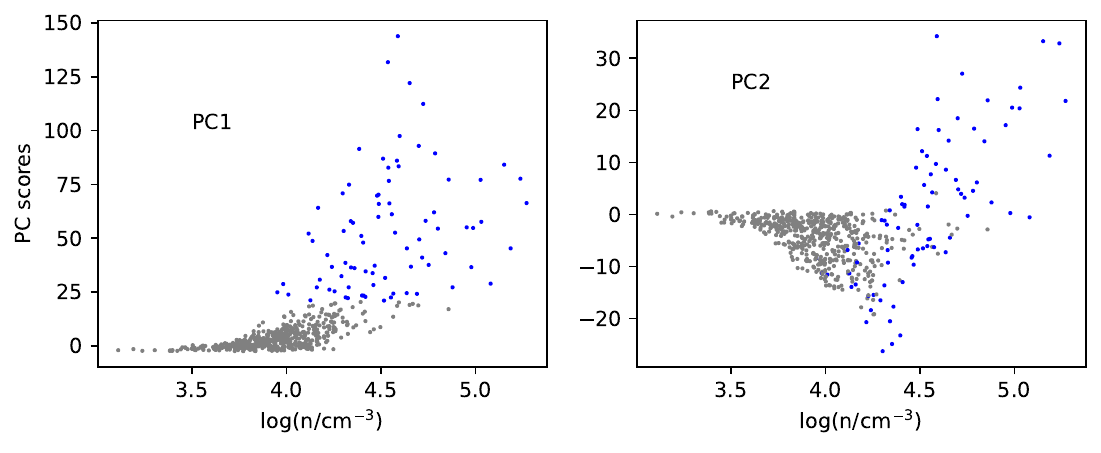}
\includegraphics[width=0.8\textwidth]{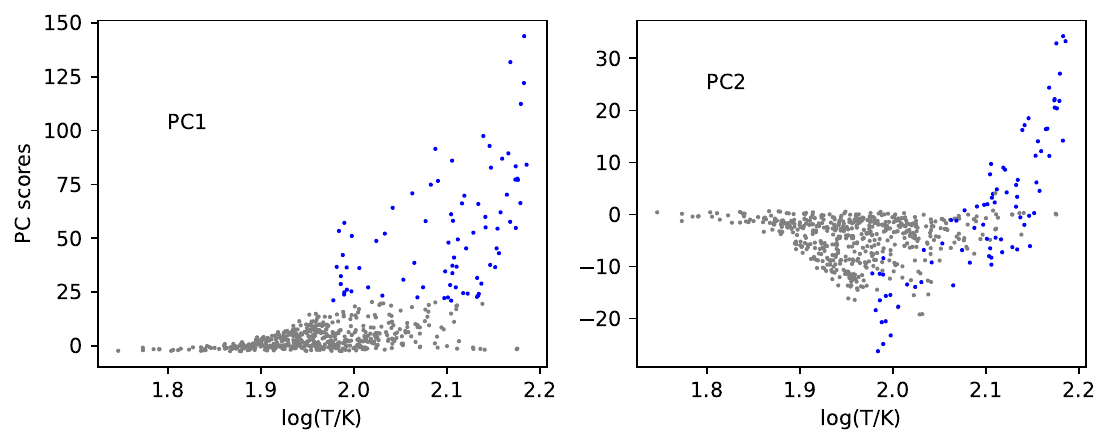}}
\caption{Physical parameters (top: column densities, middle: volume densities, bottom: temperatures) and PC scores in individual hexagonal pixels. Physical parameters are taken from ``high-HB" results from \citet{2023Tanaka_HB}, which are the ones derived from species with relatively high critical densities. Grey points represent data points that lie within one standard deviation of PC1 scores from zero, pixels with low signal-to-noise ratios. \label{fig:scatter_phys}}
\end{figure*}

Here we examine whether each PC has any clear relationship with physical parameters such as column densities, kinetic temperatures, and volume densities. Figure \ref{fig:scatter_phys} shows the relationship between the PC scores in our results and physical conditions derived by \citet{2023Tanaka_HB} from the ALCHEMI data. These physical parameters were obtained from the hierarchical Bayesian analysis method developed by \citet{2018ApJS..236...40T}. \citet{2023Tanaka_HB} present results from different sets of transitions. Among their results, we use their ``high-HB" model derived from species with higher critical densities such as HCN, HCO$^+$, and their isotopologues, \hctn, and SiO, instead of results derived from CO isotopologues. This is because our PCA used more species with high critical densities than CO isotopologues. Pixels from $N_{\rm H2}$-weighted averaged images along the velocity axis are binned to the same hexagonal pixels as our PCA. We omitted hexagonal pixels that do not have physical parameters derived for all the original pixels within the hexagon. We show pixels within one standard deviation of all the data points of PC1 scores from zero in grey because they represent pixels with low signal-to-noise ratios. The standard deviation of PC1 scores is 21. We note, however, that not all the grey points are noise. We only show relations with PCs 1 and 2 because there are no clear trends for PC3 and higher.

PC1 scores show a clear increasing trend with increasing column densities. They also have a similar, but weaker correlation with volume densities, and even to a lesser degree, with temperatures. These results suggest that PC1 primarily represents the overall molecular gas content. The volume densities tend to be higher in high column-density regions in \citet{2023Tanaka_HB}, and this is likely why PC1 and the volume densities have a positive correlation. The high-density regions are also regions with high star formation rates, which explains the trend of increasing PC1 scores with increasing temperatures. 

\begin{figure*}
\centering{
\includegraphics[width=0.49\textwidth]{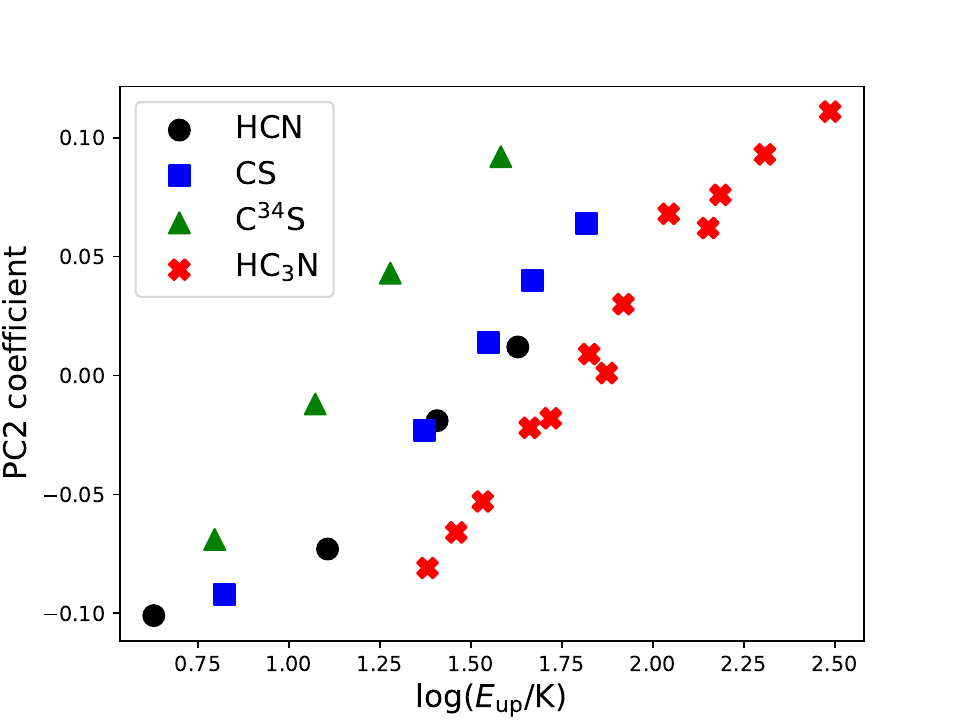}
\includegraphics[width=0.49\textwidth]{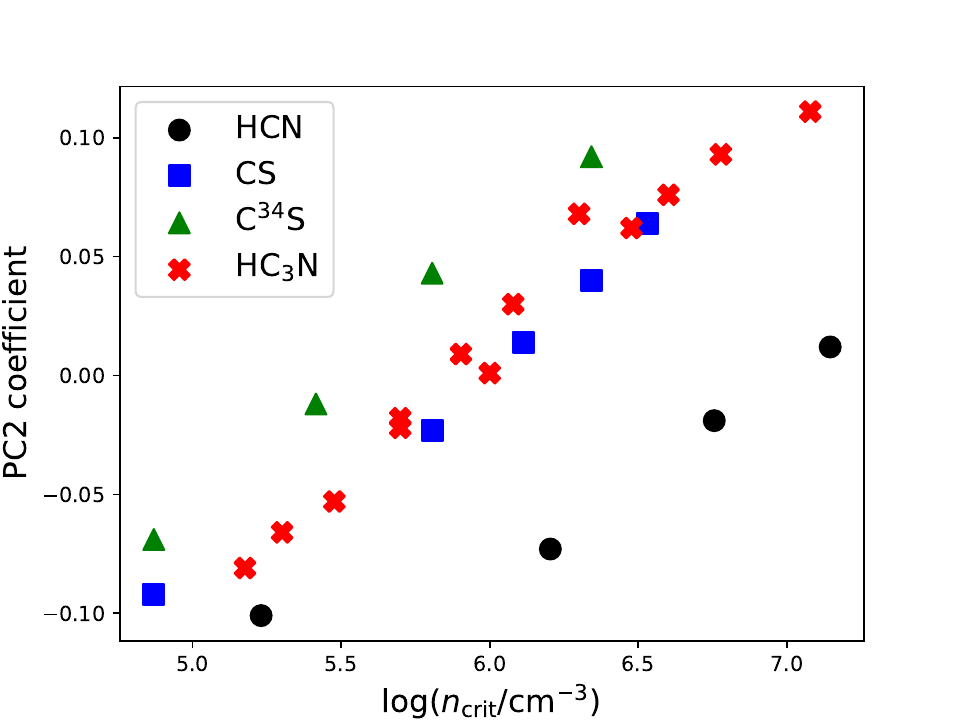}
}
\caption{Comparisons between PC2 coefficients and (left) upper state energies and (right) critical densities from the bright linear rotors, where the reasonable connection between $J$, $E_{\rm up}$, and $n_{\rm cr}$ is expected. For HCN and CS, values of critical densities are taken from \citet{2015PASP..127..299S} considering multiple levels. We assume the same critical densities as CS for C$^{34}$S. We take values of critical densities from \citet{2007A&A...464.1147W} for \hctn. \label{fig:pc2-par}}
\end{figure*}

PC2 scores increase with the increasing values of all of these physical parameters (column densities, volume densities, and temperature) if we focus on high S/N pixels (blue points in Figure \ref{fig:scatter_phys}). This correlation between PC2 scores and physical parameters is tightest in the temperature, while it gets slightly weaker in the density. The scatter becomes larger between PC2 scores and the column density. To consider the relationship between PC2 and physical parameters in a different perspective, we consider PC2 coefficients ($a_{i2}$ in the notation of Section \ref{sec:pca_expl}), instead of PC scores discussed above, of multiple transitions of HCN, CS, C$^{34}$S, and \hctn. Figure \ref{fig:pc2-par} shows the dependence of PC2 coefficients on upper-state energies (left figure) and critical densities (right figure). For all the species, PC2 coefficients tend to be higher with higher upper-state energies and critical densities. Transitions with high PC2 coefficients clearly emit preferentially from increasingly higher-excitation regions.

%%%%%%%%%%%%%%%%%%%%%
%\begin{figure*}[ht]
%\centering{
%\includegraphics[width=0.8\textwidth]{scatter_NH_low-H.pdf}
%\includegraphics[width=0.8\textwidth]{scatter_n_low-H.pdf}
%\includegraphics[width=0.8\textwidth]{scatter_T_low-H.pdf}}
%\caption{The same as Figure \ref{fig:scatter_phys}, but for ``low-H" results. These physical parameters are derived from CO isotopologue transitions. \label{fig:scatter_phys2}}
%\end{figure*}

%%%%%%%%%%%%%%%%%%%
%%%%%%%%%%%%%%%%%%%%%%%%
\subsection{PC coefficients of each species}\label{sec:comp}

\begin{figure*}[ht]
\includegraphics[width=0.9\textwidth]{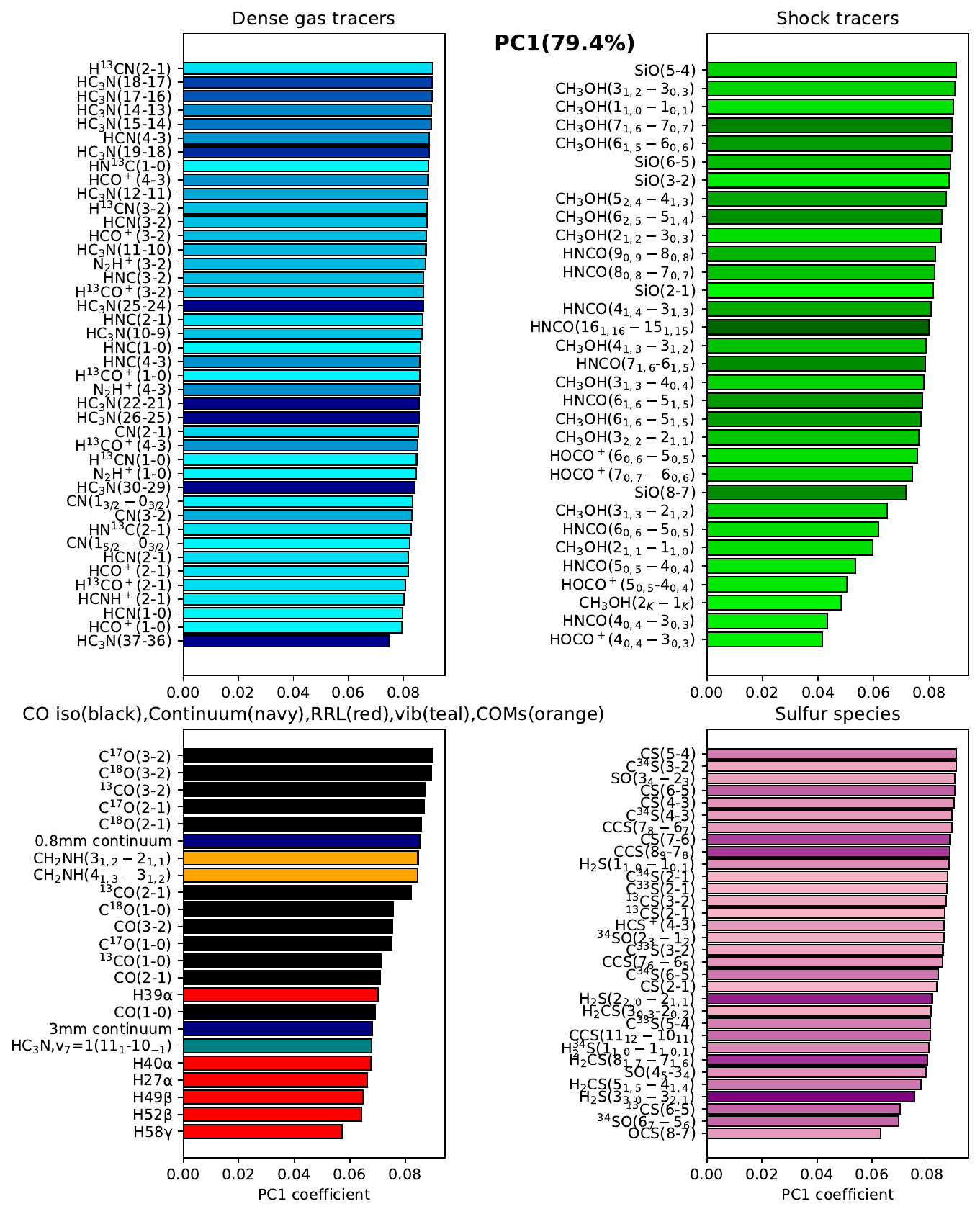}
\includegraphics[width=0.9\textwidth]{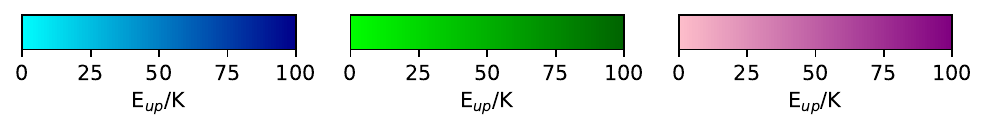}
\caption{PC1 coefficients for selected transitions are shown in a descending order.  For dense gas tracers, shock tracers, and sulfur species, upper-state energies are color-coded. Values from 0\,K to 100\,K are shown from cyan to dark blue (dense gas tracers), from lime to dark green (shock tracers), and pink to purple (sulfur species). Colors of CO, continuum, RRL, vibrationally excited lines, and COMs indicate categories, not energies. \label{fig:pc1}}
\end{figure*}

\begin{figure*}[ht]
\includegraphics[width=0.9\textwidth]{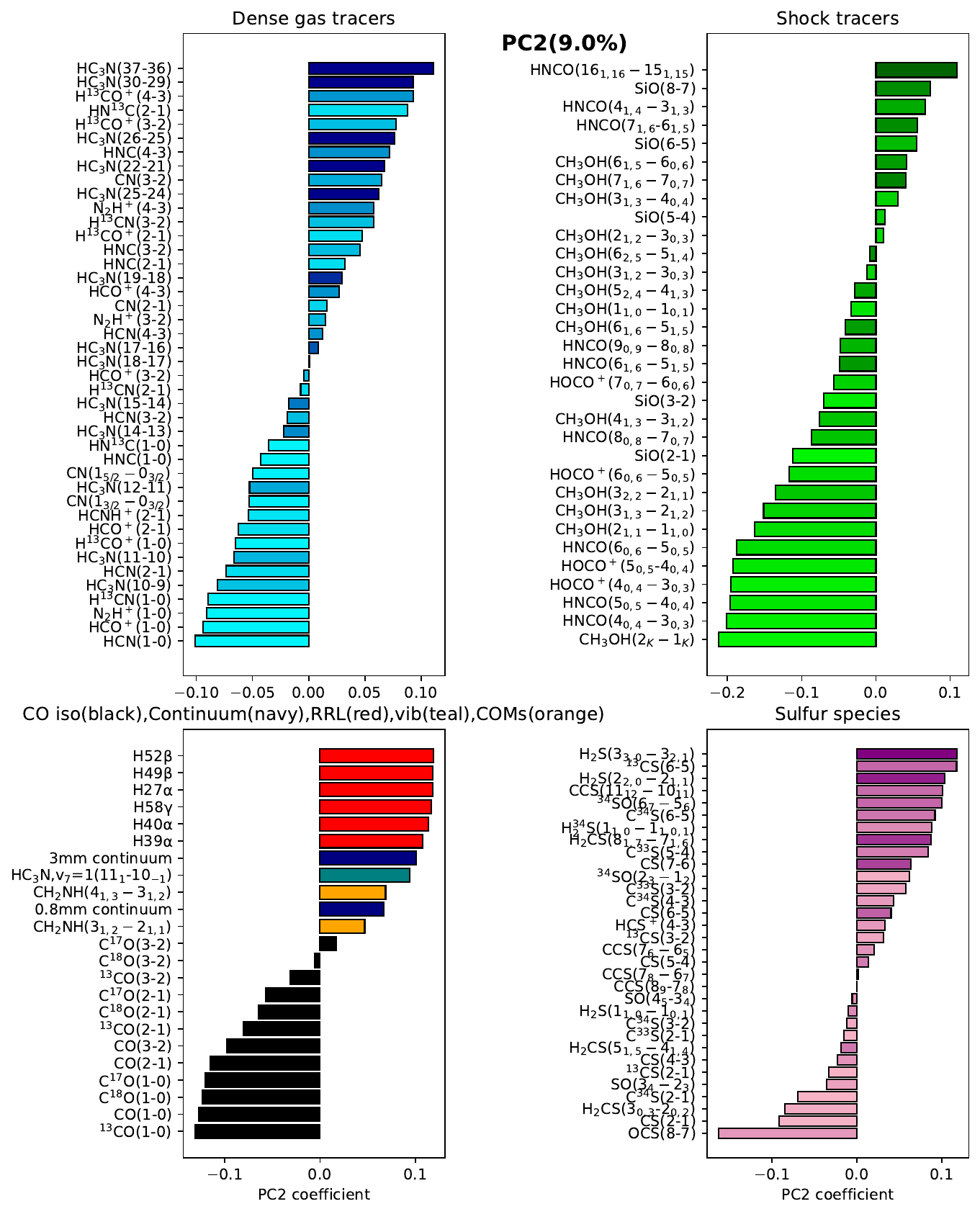}
\includegraphics[width=0.9\textwidth]{colorbar.pdf}
\caption{The same as Figure \ref{fig:pc1}, but for PC2. \label{fig:pc2}}
\end{figure*}

\begin{figure*}[ht]
\includegraphics[width=0.9\textwidth]{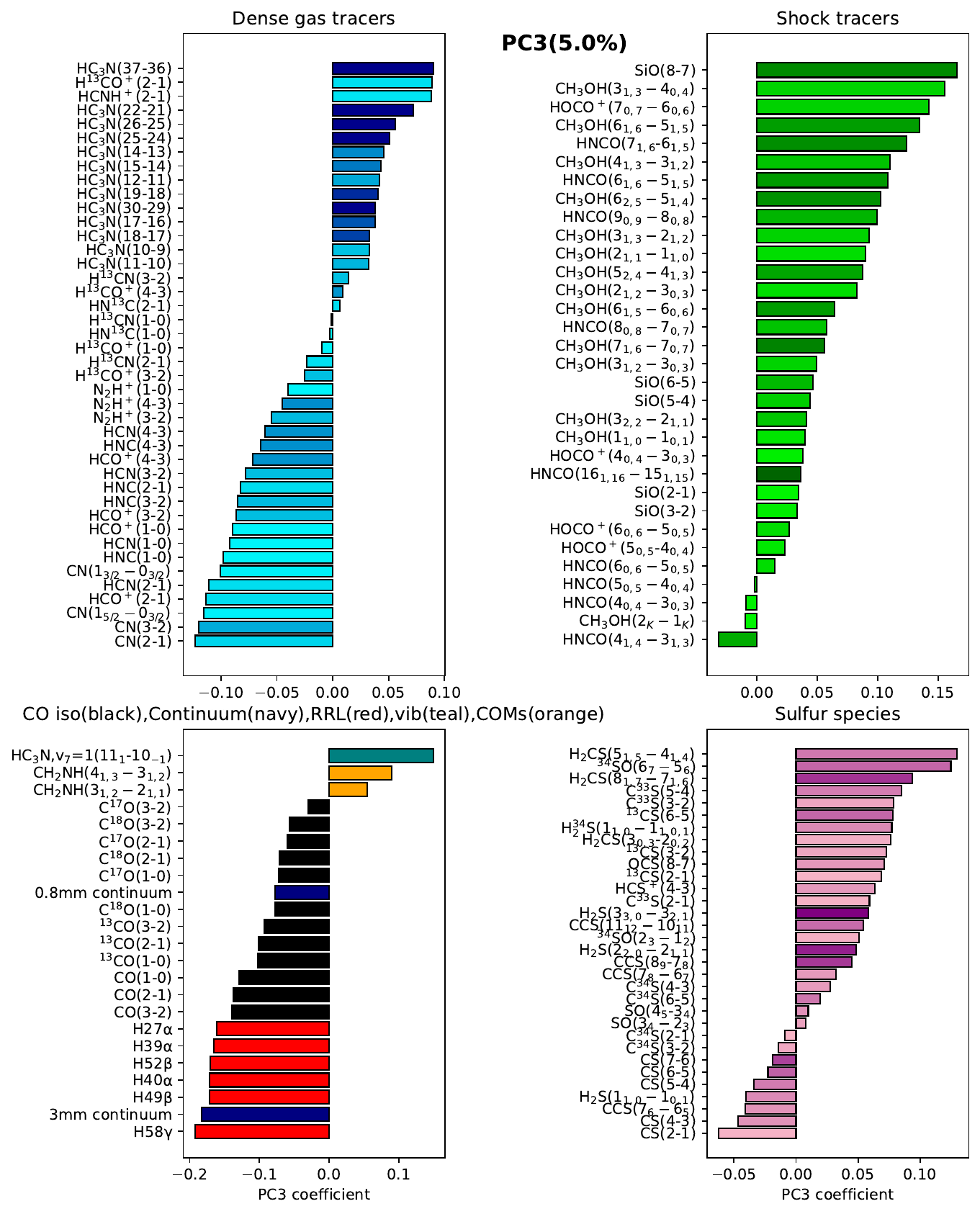}
\includegraphics[width=0.9\textwidth]{colorbar.pdf}
\caption{The same as Figure \ref{fig:pc1}, but for PC3. \label{fig:pc3}}
\end{figure*}

\begin{figure*}[ht]
\includegraphics[width=0.9\textwidth]{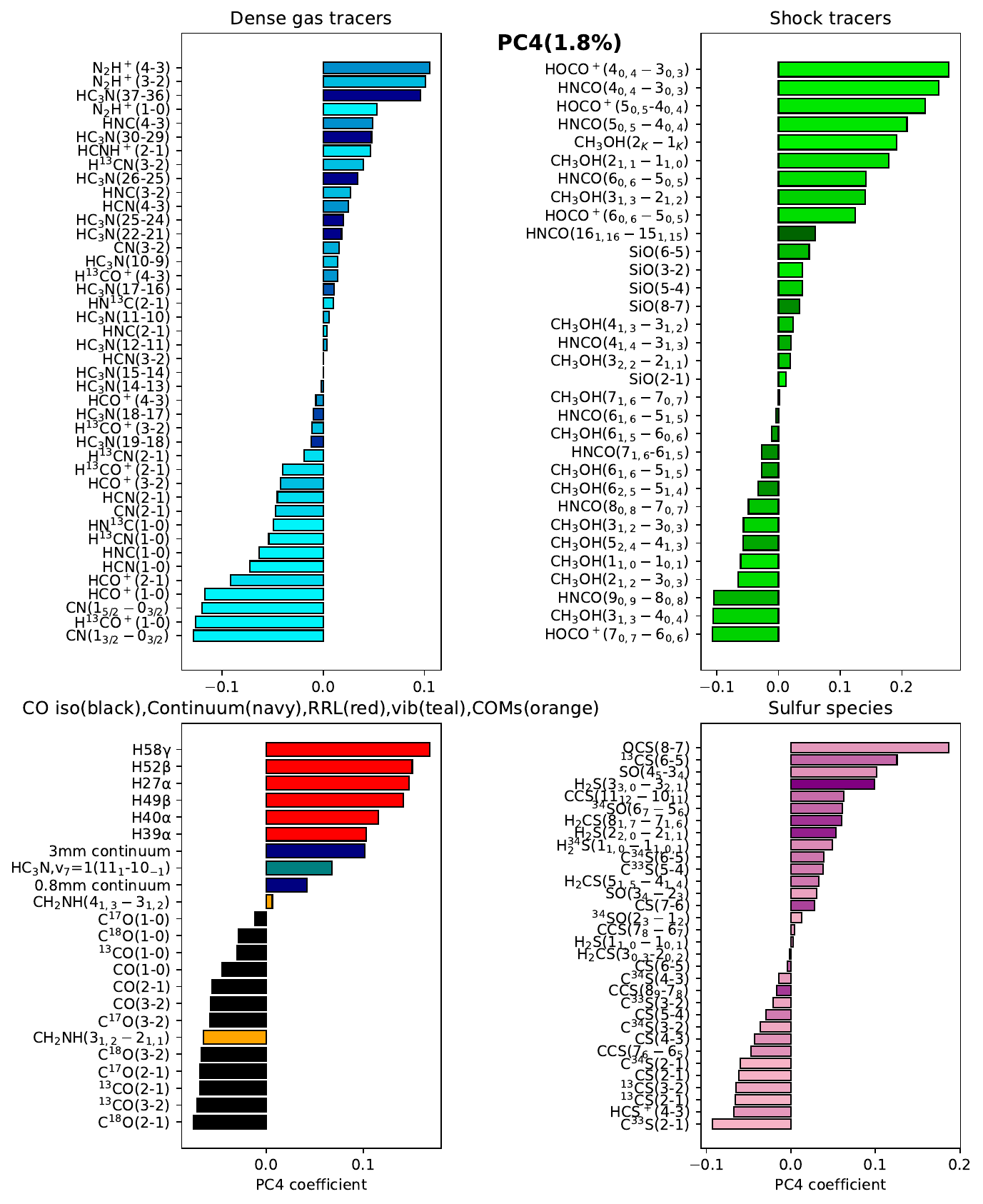}
\includegraphics[width=0.9\textwidth]{colorbar.pdf}
\caption{The same as Figure \ref{fig:pc1}, but for PC4. \label{fig:pc4}}
\end{figure*}

\begin{figure*}[ht]
\includegraphics[width=0.9\textwidth]{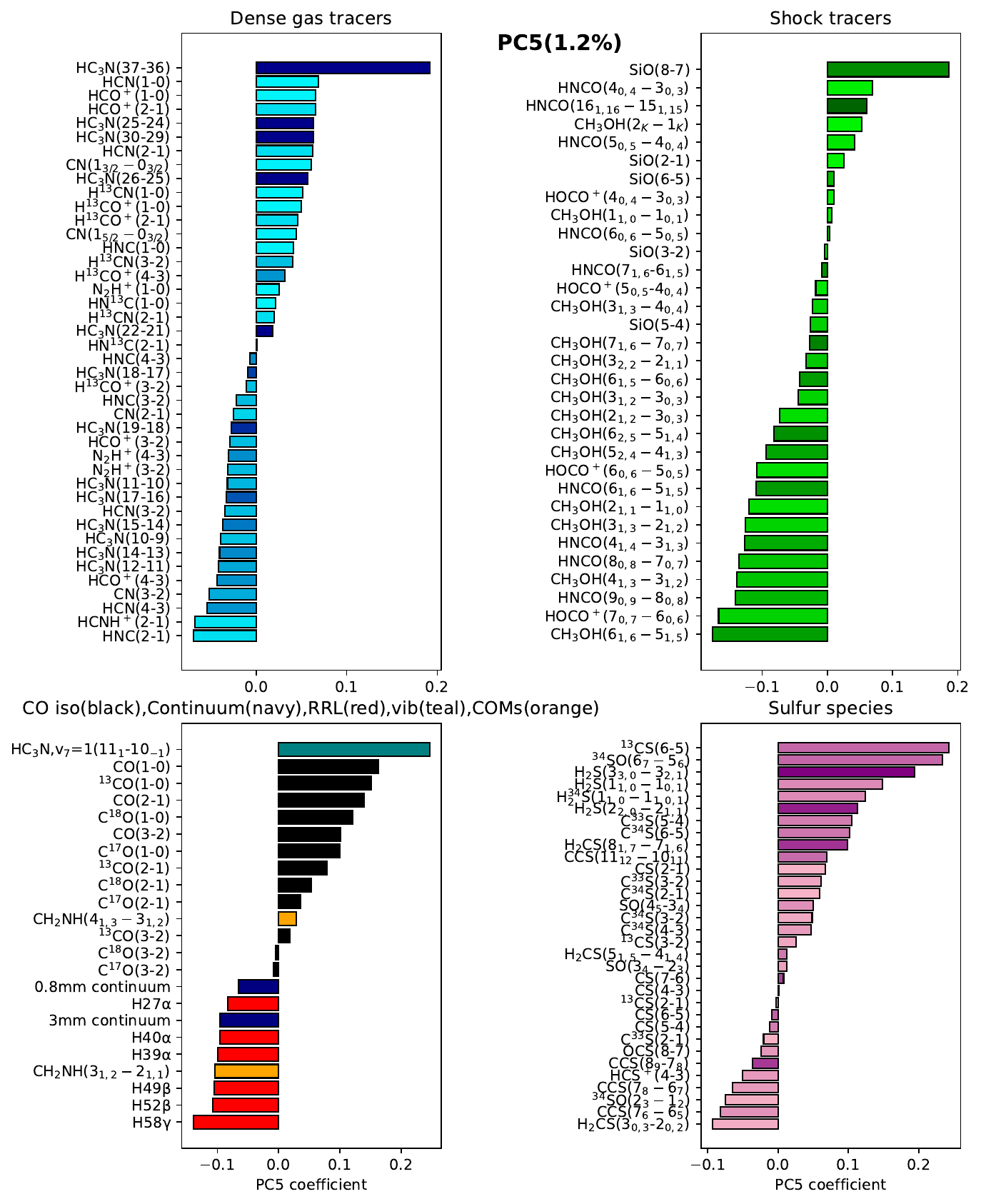}
\includegraphics[width=0.9\textwidth]{colorbar.pdf}
\caption{The same as Figure \ref{fig:pc1}, but for PC5. \label{fig:pc5}}
\end{figure*}

Figures \ref{fig:pc1} -- \ref{fig:pc5} show 
the coefficients of various transitions for PCs 1--5, i.e., $a_{ij}$ for a transition $i$ and PC$j$ for $j=$ 1--5. A table of these coefficients can be found in Appendix \ref{sec:app:PCcomp}. Figure \ref{fig:pcO} also shows the same information, but for transitions that are not shown in Figures \ref{fig:pc1} - \ref{fig:pc5}. Errors of the coefficients are evaluated in Appendix \ref{sec:app:boot} and are negligible. Figures \ref{fig:pc1} - \ref{fig:pc5} are separated into four panels. The first two panels are for dense gas tracers (HCN, HNC, \hcop, CN, and their isotopologues, as well as \nthp, \hctn) and shock tracers (\methanol, SiO, HNCO and \hocop)\footnote{While \hocop\ is not a conventional shock tracer, we consider it a shock tracer here because of the results obtained by \citet{2022ApJ...938...80H} who found an association between \hocop\ and shocks in NGC~253.}. Colors indicate the upper state energies of transitions from 0 to 100\,K in cyan to blue for dense gas tracers, and lime to green for shock gas tracers. The third panel shows CO isotopologues, continuuum, RRLs, vibrationally-excited \hctn, and CH$_2$NH, and the fourth panel is for sulfur-bearing molecules. The fourth panel also shows upper-state energies with colors (0 to 100\,K with pink to purple).

Figures \ref{fig:pc1} and \ref{fig:pcO} show that all the transitions have positive coefficients in PC1. The variation of PC1 coefficients among transitions is not large. This result can easily be explained by the fact that PC1 correlates with the emission brightness of these transitions. Consequently, PC1 in general traces overall gas column density as described in Section~\ref{sec:pca_phys} because regions with higher column densities tend to have higher brightness. Star formation tracers such as RRLs and the 3-mm continuum also have positive PC1 values because the star formation rate is generally higher towards the high column-density regions.

\begin{figure*}[ht]
\includegraphics[width=0.9\textwidth]{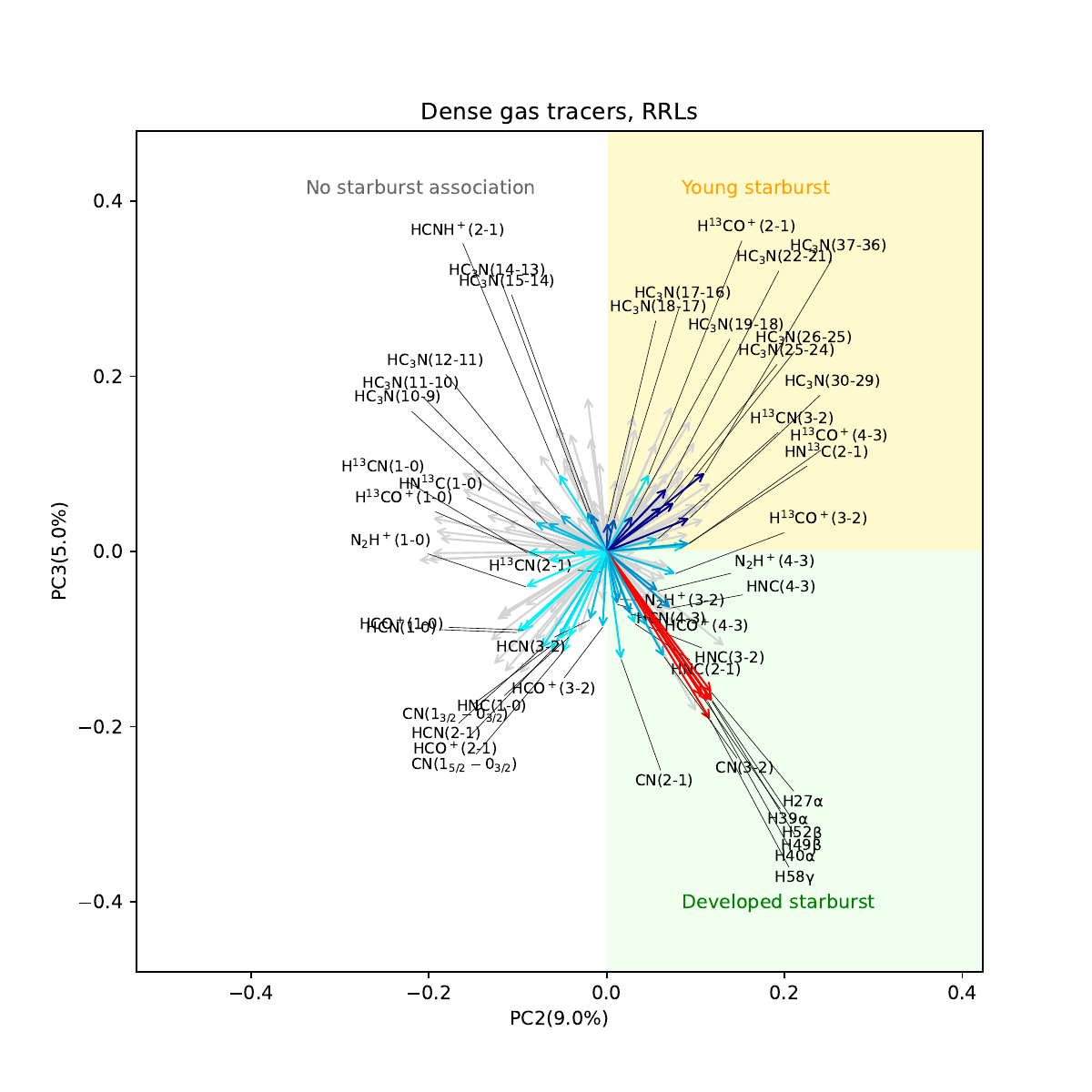}
\caption{PC2 and PC3 coefficients are shown in abscissa and ordinate, respectively, for dense gas tracers and RRLs. Other transitions are also shown as grey arrows. Arrow coloring is the same as in Figure \ref{fig:pc1}. Transitions in the first quadrant (filled with light yellow) are distributed near young starbursts ($<10^5\,$yr) while transitions in the fourth quadrant (filled with light green) are near well-developed starbursts ($>10^5\,$yr) as discussed in Section \ref{sec:offset}. \label{fig:pc2-3}}
\end{figure*}

High absolute values of PC3 scores (around GMCs 5 and 6) appear in regions with positive values of PC2 scores (Figure \ref{fig:pcmap1}).  Negative PC3 coefficients are seen for RRLs, while high-excitation dense gas tracers have positive PC3 coefficients (Figure \ref{fig:pc3}). It is interesting to note that most of the shock tracers and sulfur-bearing molecules have positive PC3 coefficients, unlike dense gas tracers. Meanwhile, CO isotopologues and continua all have negative PC3 coefficients. PC3 separates high-excitation transitions with high PC2 coefficients into different groups as illustrated in Figure \ref{fig:pc2-3}. High excitation transitions of CN, HNC, HCN, and RRLs have negative PC3 coefficients (the fourth quadrant in Figure \ref{fig:pc2-3})\footnote{We number quadrants counter-clockwise from the top right.} while \hctn\ transitions have positive ones (the first quadrant in Figure \ref{fig:pc2-3}). 

\begin{figure*}[ht]
\includegraphics[width=0.9\textwidth]{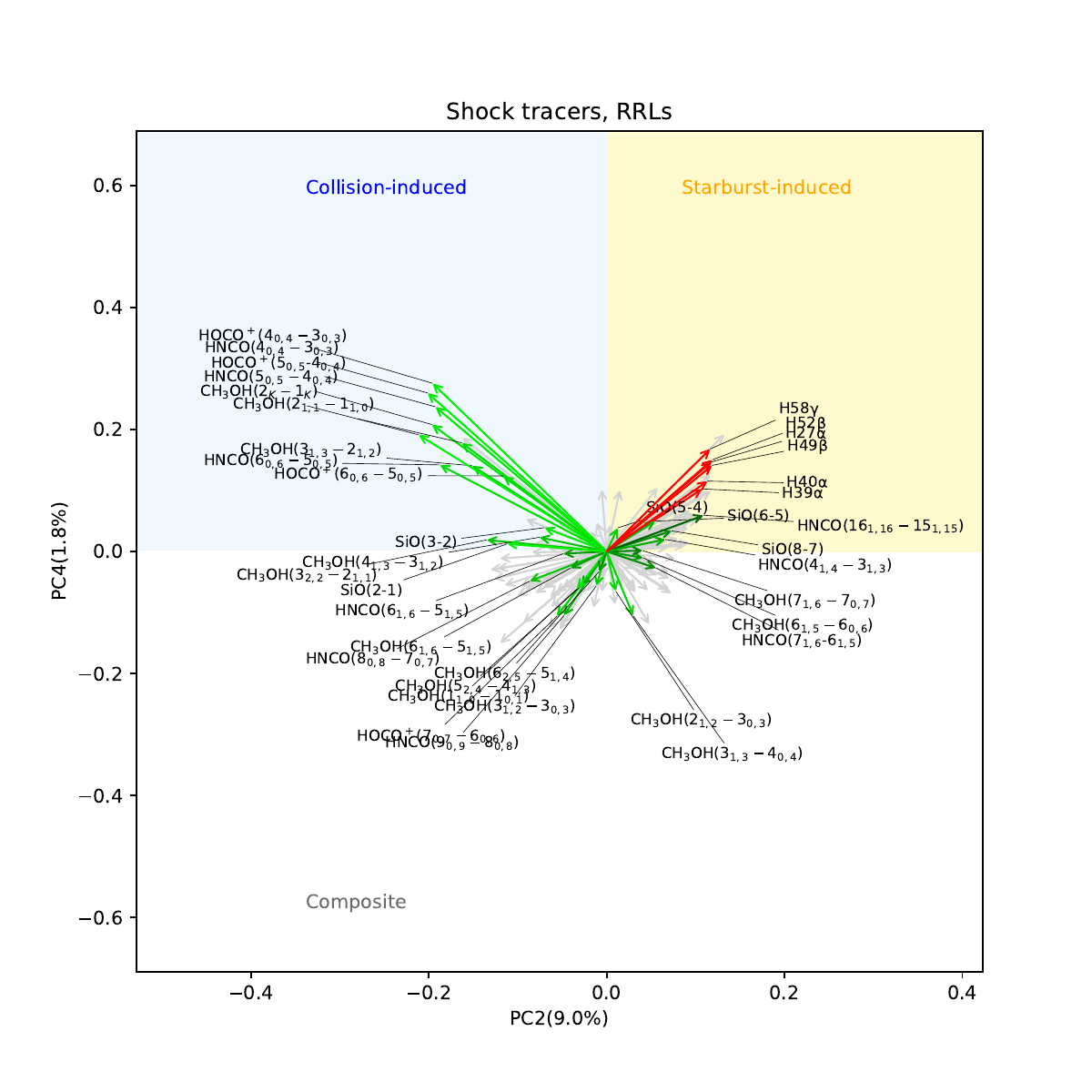}
\caption{Same as Figure \ref{fig:pc2-3}, but showing PC2 and PC4 for shock tracers. \label{fig:pc2-4}}
\end{figure*}

As mentioned previously, there are some similarities between the PC2 and PC4 score maps near the central starbursts. However, there are also interesting exceptions in the outskirts (Figure \ref{fig:pcmap1}). Therefore, if a transition has positive coefficients for PC2 and PC4 (the first quadrant in Figure \ref{fig:pc2-4}), it has stronger emission in GMCs 4, 5, and 6 than that can be accounted for by PC2 alone. On the other hand, if a transition has negative PC2 and positive PC4 coefficients (the second quadrant in Figure \ref{fig:pc2-4}), contributions from the CMZ outskirts are significant (GMCs 1, 2, 7, 8, 9).

\begin{figure*}[ht]
\includegraphics[width=0.9\textwidth]{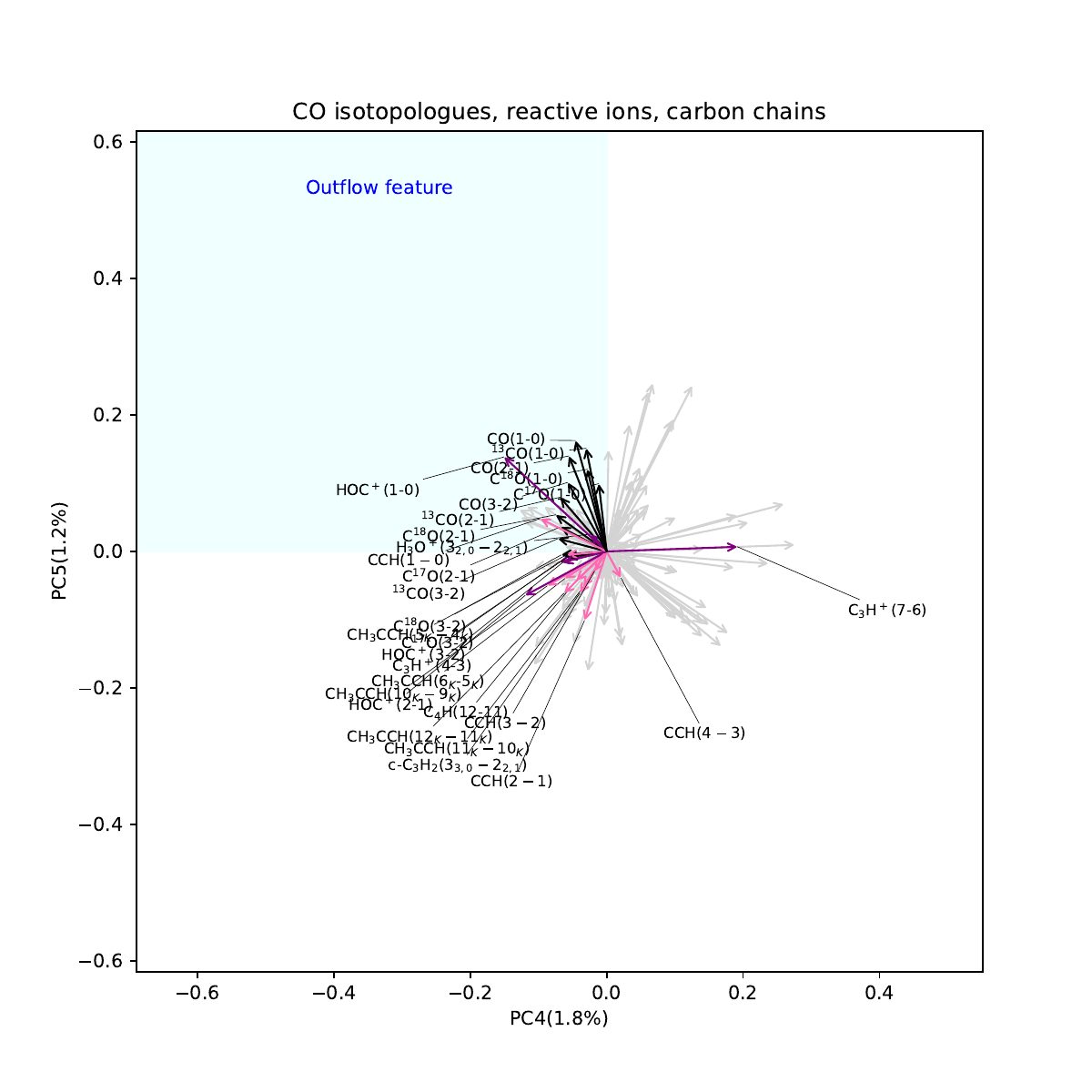}
\caption{Same as Figure \ref{fig:pc2-3}, but showing PC4 and PC5 for CO isotopologues, reactive ions, and carbon chains. Arrows show CO isotopologues in black, ions in violet, and carbon chains in pink. Transitions with outflow-like features should lie in the quadrant marked with light blue as discussed in Section \ref{sec:comp}. \label{fig:pc4-5}}
\end{figure*}

\begin{figure*}[ht]
\includegraphics[width=0.9\textwidth]{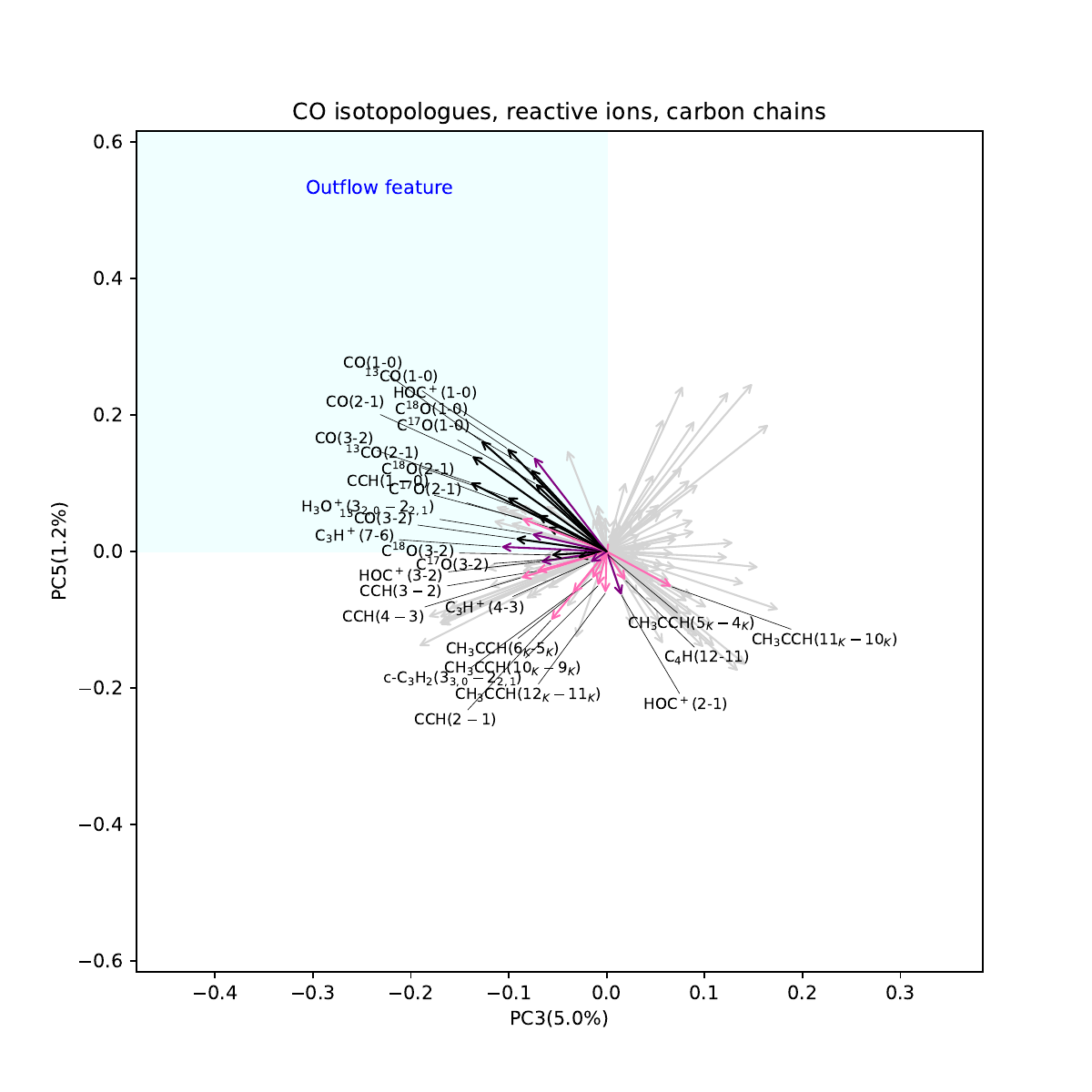}
\caption{Same as Figure \ref{fig:pc4-5}, but showing PC3 and PC5. Transitions with outflow-like features should lie in the quadrant marked with light blue as discussed in Section \ref{sec:comp}. These transitions are the same as ones in the second quadrant in Figure \ref{fig:pc4-5} except for C$_3$H$^+$(7-6). \label{fig:pc3-5}}
\end{figure*}

Positive PC5 scores are either at GMC6 or the component extended towards higher galactic latitudes likely associated with outflows, while the area of negative PC4 scores contain these outflow-related extended structures and GMC3 (Figure \ref{fig:pcmap1}). Because the vertically extended structure is likely caused by outflows and is of interest, we picked out transitions tracing this structure by comparing PC4 and PC5 coefficients in Figure \ref{fig:pc4-5}. The transitions of our interest should have negative PC4 and positive PC5 coefficients and in the second quadrant of the Figure.
%we extract transitions having this structure by comparing PC4 and PC5. Transitions with vertically extended structures should have negative PC4 and positive PC5 (the second quadrant in Figure \ref{fig:pc4-5}). 
Because these outflow features also appear weakly as negatives in the PC3 score map, we also compare PC3 and PC5 coefficients in Figure \ref{fig:pc3-5}, where the outflow tracers should be in the second quadrant. It turns out that most transitions that appear in the second quadrant in Figure \ref{fig:pc4-5} also appear in the second quadrant of Figure \ref{fig:pc3-5}. Transitions having this outflow feature are CO isotopologues, CCH(1--0), \hocp(1--0), and H$_3$O$^+$. Although they are not shown in the figure, dense gas tracers seen in an outflow (Southwest streamer) by \citet{2017ApJ...835..265W} such as HCN, \hcop, CS in J=1--0, and CN(N=1--0) also have negative PC4 and positive PC5 coefficients. We note that CCH(1--0) has been detected also in the outflow in Maffei 2 \citep{2012ApJ...755..104M} and from an AGN in NGC~1068 \citep{2017A&A...608A..56G,2022ApJ...935..155S}.

It is important to note that line intensities depend on column densities (or abundances), temperatures, and densities. Consequently, our results show very different PC coefficients, which means different distributions, even for the same species depending on transitions. This illustrates that caution is needed when using a particular species as a tracer of certain phenomenon simply by observing a single transition.

% Plots of coefficients
%%%%%%%%%%%%%%%%%%%%%
%%%%%%%%%%%%%%%%%%%%%
%%%%%%%%%%%%%%%%%%%%%
%\clearpage
%%%%%%%%%%%%%%%%%%%%%
%%%%%%%%%%%%%%%%%%%%%
%\clearpage
%%%%%%%%%%%%%%%%%%%%%
%%%%%%%%%%%%%%%%%%%%%
%\clearpage
%%%%%%%%%%%%%%%%%%%%%
%%%%%%%%%%%%%%%%%%%%%
%\clearpage
%%%%%%%%%%%%%%%%%%%%%
%%%%%%%%%%%%%%%%%%%%%
%\clearpage
\begin{figure*}[ht]
\includegraphics[width=0.9\textwidth]{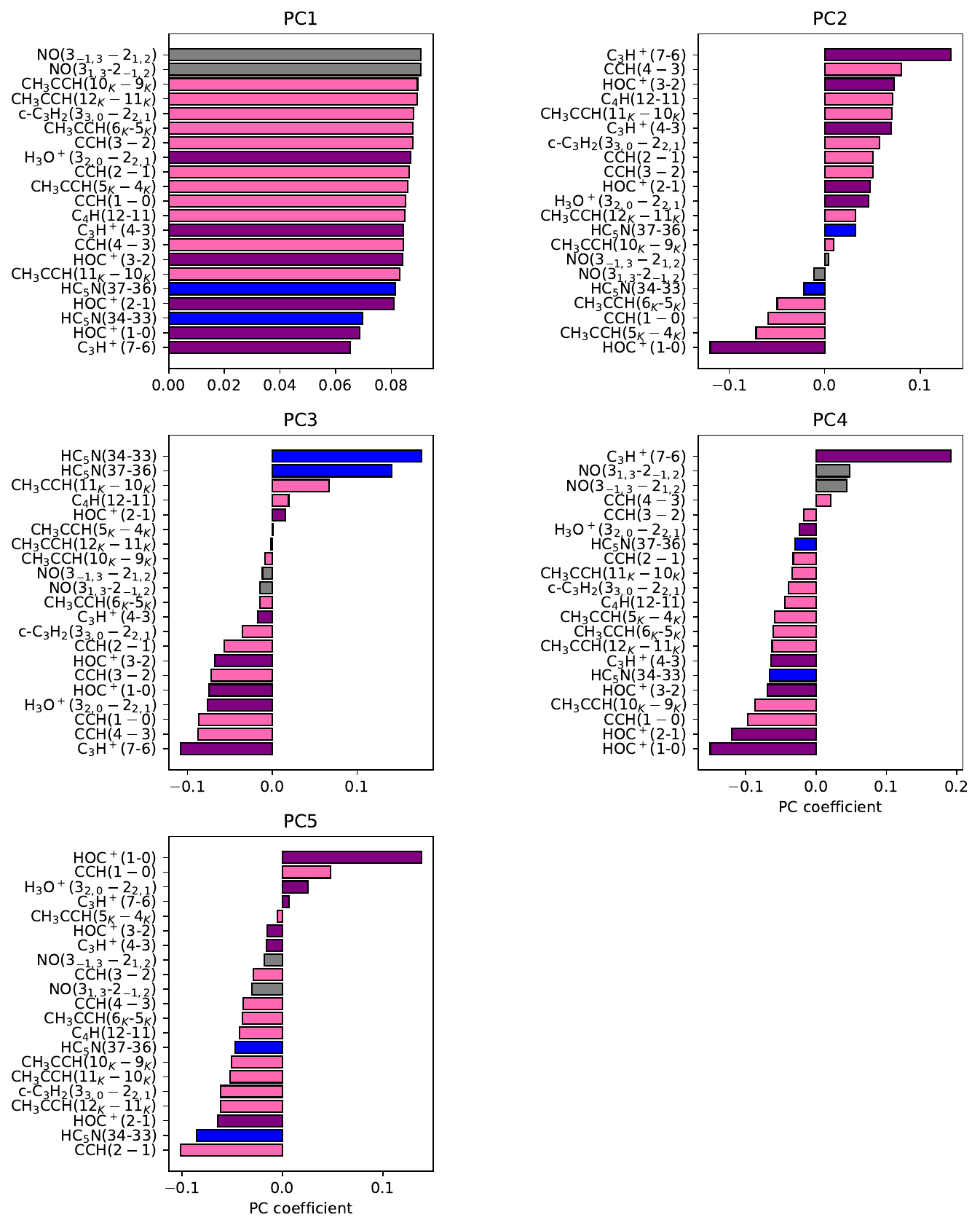}
\caption{The same as Figure \ref{fig:pc1}, but for PC1-5 of the rest of transitions. Dense gas tracers are shown with blue, reactive ions with violet, carbon chains with pink, and others with grey.  \label{fig:pcO}}
\end{figure*}
%%%%%%%%%%%%%%%%%%%%%

%\clearpage
%%%%%%%%%%%%%%%%%%%%%
%\clearpage
%%%%%%%%%%%%%%%%%%%%%
%\clearpage
%%%%%%%%%%%%%%%%%%%%%
%\clearpage
%%%%%%%%%%%%%%%%%%%%%
%\clearpage

\subsection{Similarities between species}
\begin{figure*}[ht]
\centering{
\includegraphics[width=0.49\textwidth]{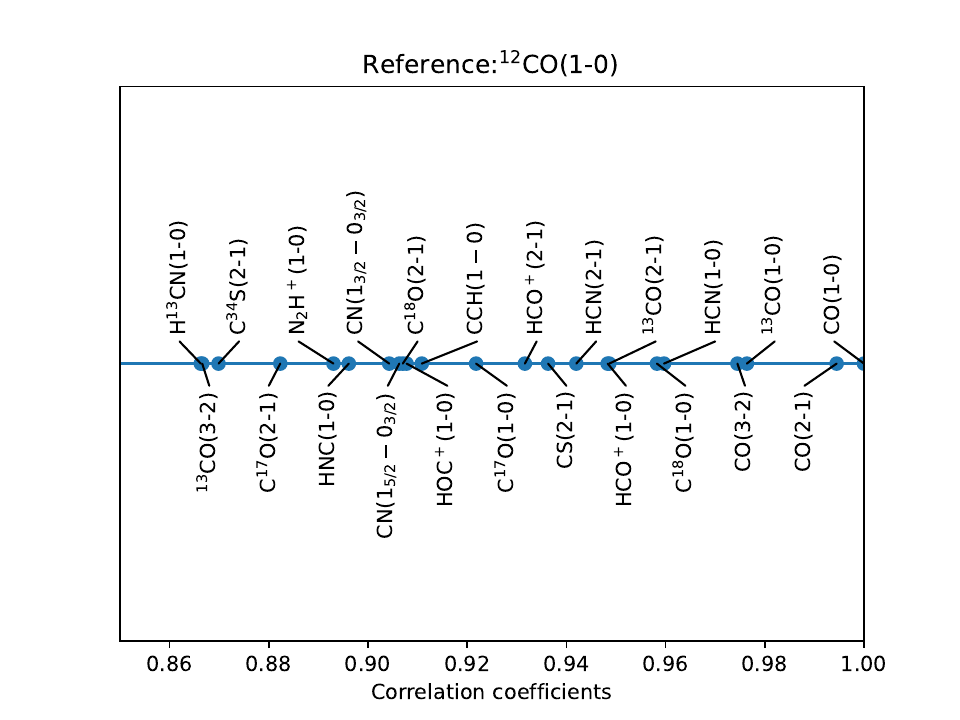}
\includegraphics[width=0.49\textwidth]{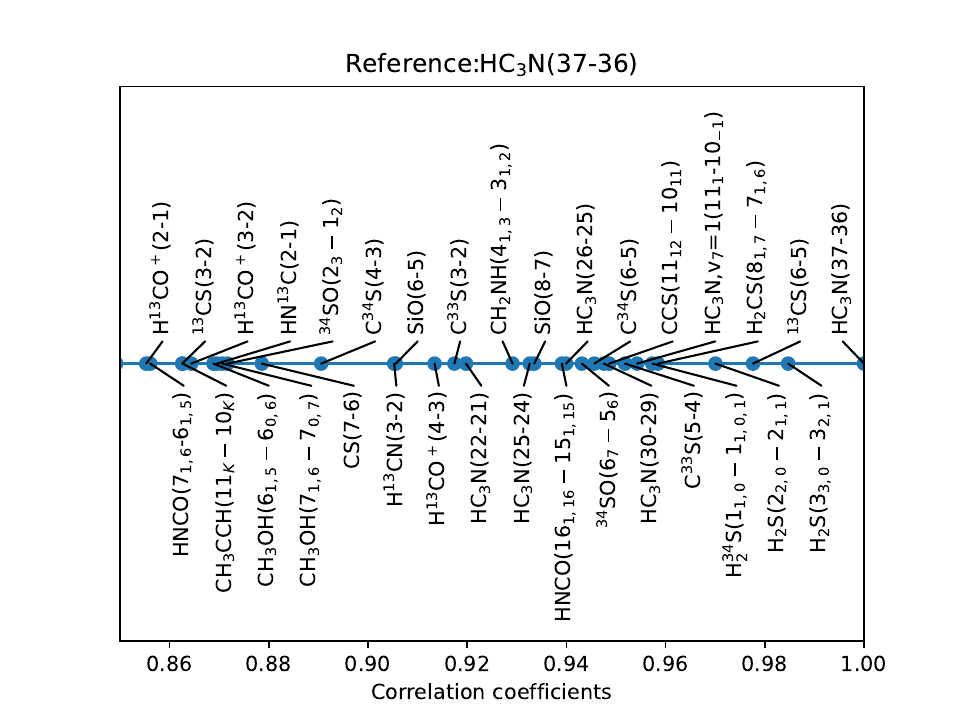}}
\centering{
\includegraphics[width=0.49\textwidth]{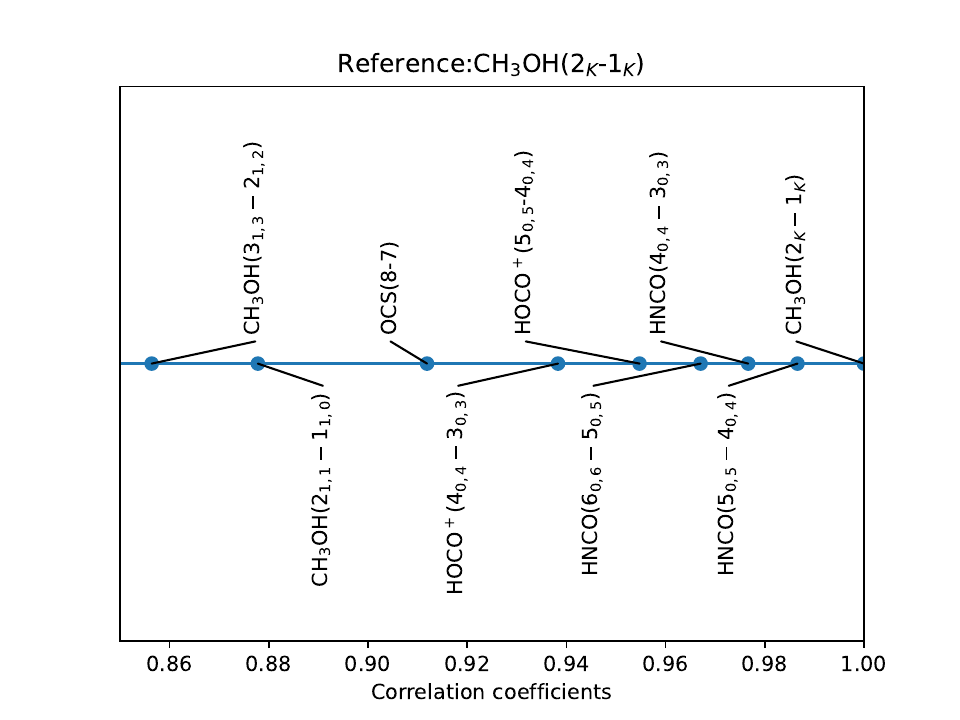}
\includegraphics[width=0.49\textwidth]{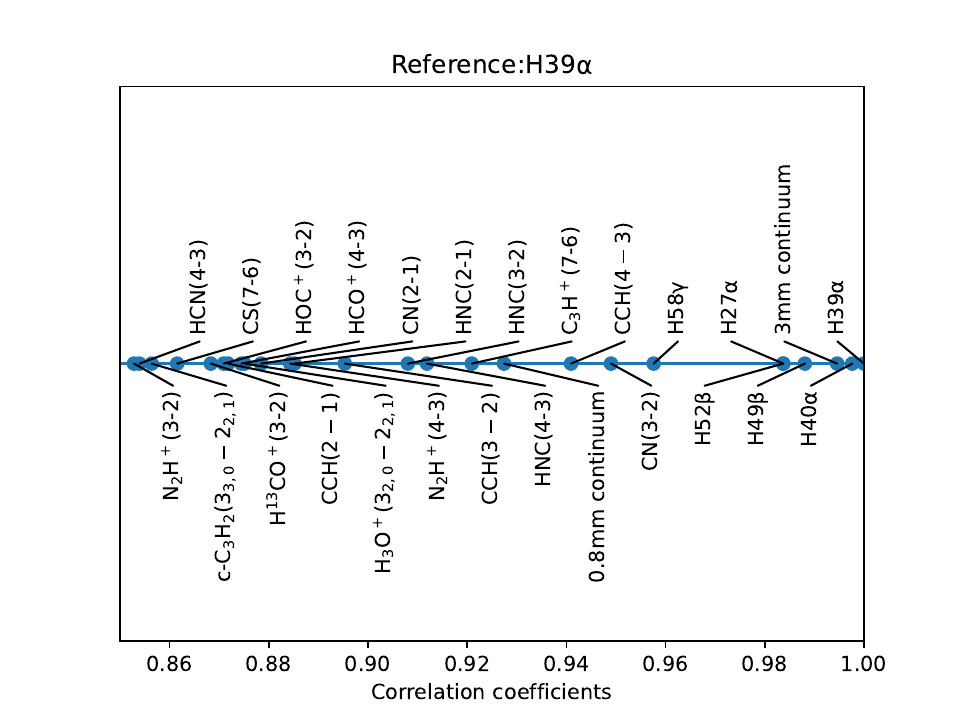}}
\caption{Covariance between reference transitions and other transitions. The reference transitions are \twco(1--0) (top left panel), \hctn(37--36) (top right panel), \methanol($2_K-1_K$) (bottom left panels), and H39$\alpha$ (right panels). Transitions that appear close to 1 have similar distributions as that of the reference transition. \label{fig:cov}}
\end{figure*}

The similarity between a pair of transitions can be evaluated by the correlation coefficient of their integrated intensity maps, or the relevant element of the matrix $\bm{\Sigma}$ used for the PCA\footnote{We also make the correlation matrix among all the transitions and continua available at \url{http://github.com/nanaseharada/alchemi\_pca}.}. Figure \ref{fig:cov} shows the correlation coefficients (or similarities) between a reference transition and other transitions, for reference transitions of \twco(1--0), \hctn(37--36), \methanol($2_K$--$1_K$) at 96.741 GHz, and H39$\alpha$. 
Transitions highly correlated with \twco(1--0) are other transitions of CO isotopologues, $J=1-0$ or $N=1-0$ transitions of HCN, \hcop, CN, and CCH, and CS(2--1) (Figure \ref{fig:cov}). These transitions are all ubiquitous in translucent ($n\sim 10^3$\,\cc) molecular gas. 
On the other hand, transitions that have similar distributions as that of \hctn(37--36) are highly excited. The examples are high-excitation transitions of other \hctn, sulfur-bearing species (rarer isotopologues of CS and SO, H$_2$S), SiO, and CH$_2$NH. Some of these transitions are known to trace shocks (SiO), but others are considered as hot core tracers. 
Transitions that are strongly correlated with \methanol(2--1) are low-excitation transitions of HNCO, \hocop, and OCS(8--7). Methanol and HNCO are both known slow shock tracers, and \hocop\ is shown to be enhanced in shocks in NGC~253 due to the ice sputtering of CO$_2$ \citep{2022ApJ...938...80H}. OCS may also be enhanced in shocks if OCS is a major constituent of ice. H39$\alpha$, other RRLs, and the 3-mm continuum are all very strongly correlated, all of which are known to be related to star formation. High-excitation radicals such as HOC$^+$(3--2), CN(3--2), CCH(4--3) are also highly correlated with RRLs. These species are radicals, and are abundant in PDRs or cosmic-ray dominated regions.

%In addition to the reference transitions above, there are some pairs of transitions that have good correlations. It turns out that the transitions of SiO and \hctn\ have tight correlations. This correlation may imply that the fractional abundance of \hctn\ is enhanced with shocks. Such enhancement may result from high-temperature gas-phase reactions or sublimation of ice to enhance acetylene. Acetylene can subsequently react with CN to form \hctn. Tanaka et al. (in preparation) also show that fractional abundances of SiO and \hctn\ are both elevated in shocked locations.

%%%%%%%%%%%%%%%%%%%%%
%%%%%%%%%%%%%%%%%%%%%

\subsection{GMCs in principal components}\label{sec:gmc_pcscore}

\begin{figure*}[ht]
\centering{
\includegraphics[width=0.49\textwidth]{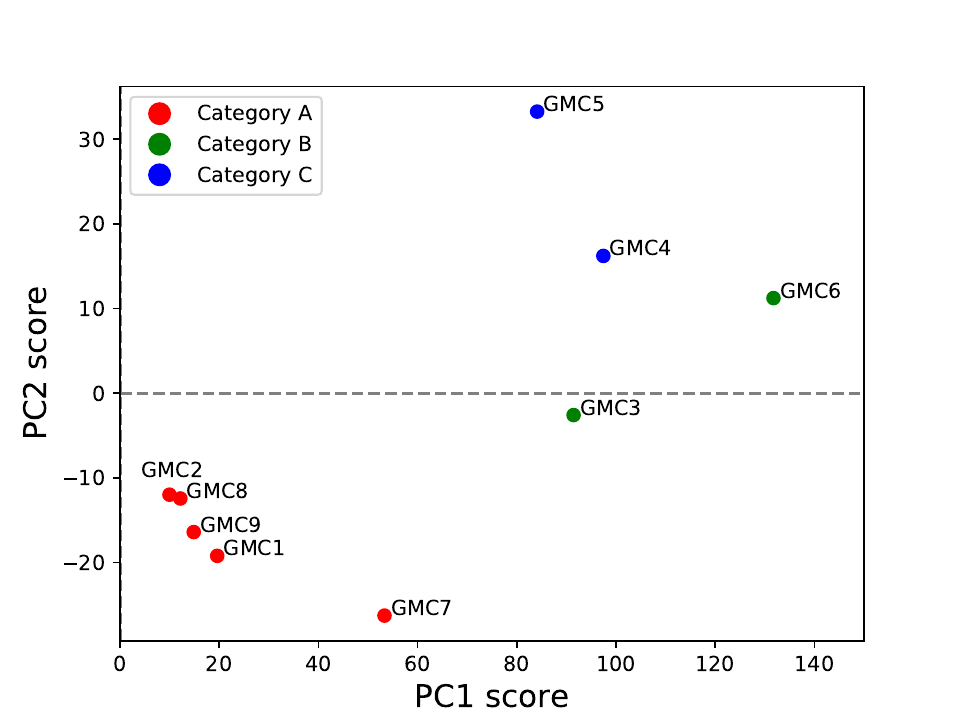}
\includegraphics[width=0.49\textwidth]{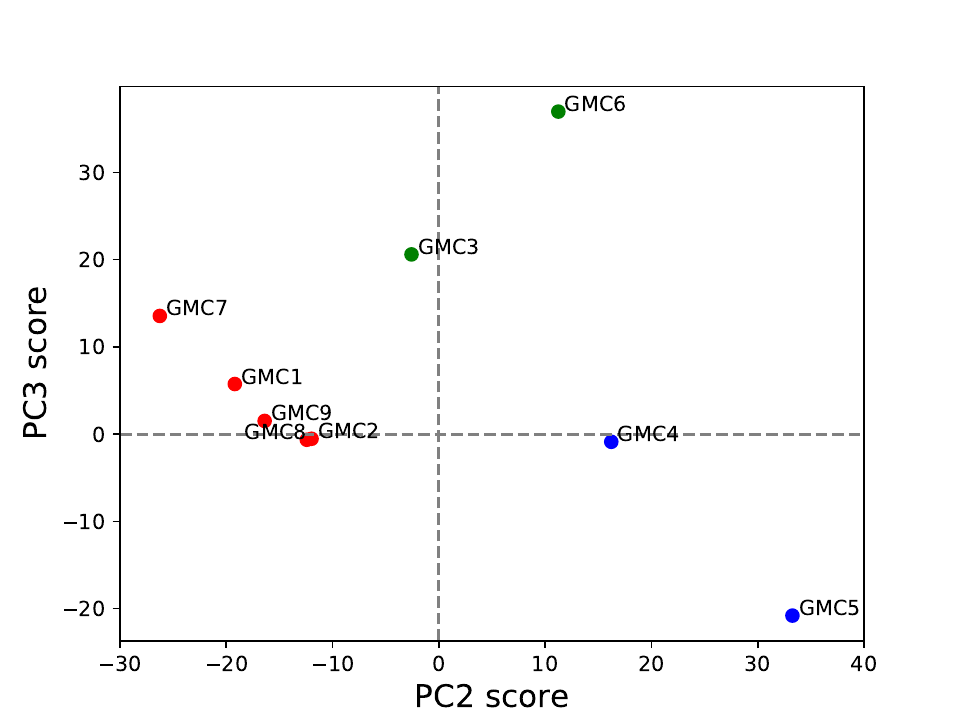}}
\centering{
\includegraphics[width=0.49\textwidth]{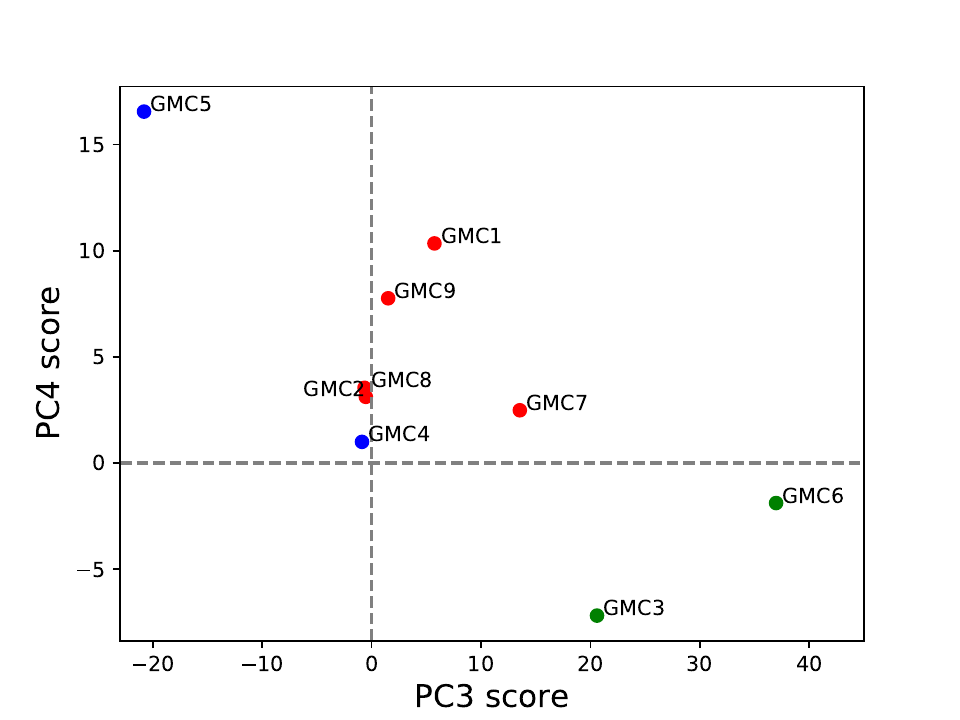}
\includegraphics[width=0.49\textwidth]{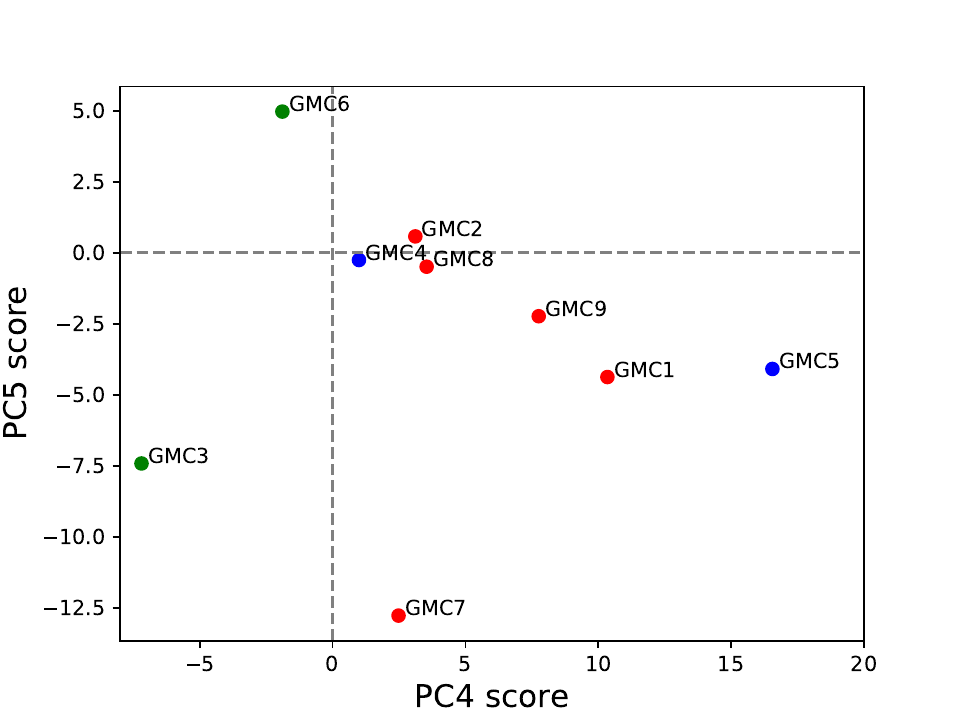}}
\caption{PC scores in GMCs. Different colors show three different groups categorized with the k-means clustering method. Each category is referred to as A (red), B (green), and C (blue) in the text.\label{fig:gmc_score}}
\end{figure*}

Figure~\ref{fig:gmc_score} shows the nine GMCs in their scores of PC1 through PC5. We measured the PC scores from the hexagonal pixels at the GMC positions. 
Each panel has dashed lines at the PC score of zero to indicate the mean score of the entire CMZ. All GMCs have positive PC1 scores since the GMCs are emission peaks by definition, and the PC1 score reflects the overall strength of the emissions, approximately the ISM column density, at each position. PCs 2--5 are defined to be orthogonal to PC1. Therefore, they disregard the effect of overall emission strengths or ISM quantity and should only reflect the qualities of the ISM. The nine GMCs are clearly different in PCs 2--5. For example, they spread in the PC2--PC3 plane with the range of PC scores matching that in the entire CMZ. Thus, the nine GMCs are distinct in their characteristics unrelated to their bulk emission strengths. 

%From PC score maps shown in Figures \ref{fig:pcmap1} and \ref{fig:pcmap2}, it is possible to categorize features in GMCs. Figure \ref{fig:gmc_score} shows PC scores for PC1 to PC5 taken at the pixels nearest to the GMCs 1-9 coordinates. 
We categorized the GMCs into three groups on the basis of their PC scores using the k-means clustering method (KMeans class in the {\tt scikit-learn} package). The k-means method is one of the simplest methods of categorization of data points into a desired number of clusters. This algorithm minimizes the following quantity:
\(\sum^k_{i=1} \sum_{z \in S_i} |z-\mu_i|^2\) for a set of clusters S = \{$S_1$, $S_2$, ..., $S_k$\} where $\mu_i$ is a mean of a set $S_i$ \citep{macqueen1967some}, that is, the sum of a square of a distance between a data point and a cluster mean for each data point. Computationally, this minimum is found by 1) randomly assigning data points into clusters\footnote{We use k-means$^{++}$, the default method in the KMeans class, where the initial assignment is not totally at random, but chosen so that the centers of mass are not close to each other.},  2) calculating the center of mass within each cluster, 3) re-assigning each data point into a cluster having the closest center of mass, and 4) iterating Step 2 and 3 until it converges. In our case, the distances are Euclidean in the five-dimensional PC space. 

Using the cluster number of three based on the inspection of Figure~\ref{fig:gmc_score},  we categorized the nine GMCs into the groups of GMCs [1,2,7,8,9], [3,6], and [4,5] and refer to them as categories A, B, and C, respectively. This categorization indeed appears to correspond to physical differences between the GMCs. The first GMC category (A) contains the locations where Class I methanol masers have been detected, which indicates shocked locations. The second category (B) contains GMCs with relatively high intensities of complex molecules. The third category (C) has locations where RRLs are strong. We discuss in Section \ref{sec:offset} that the difference between the second and third categories is likely attributed to the evolutionary stages of starbursts.

%%%%%%%%%%%%%%%%%%%%%
%%%%%%%%%%%%%%%%%%%%%

\section{Discussion}\label{sec:discuss}
\subsection{Overall gas content and star formation}
We suggest in Section \ref{sec:pca_phys} that PC1 is mostly associated with the molecular hydrogen column density. Figure \ref{fig:pcmap1} and Figure \ref{fig:pc1} show that all transitions have positive PC1 coefficients (in the range of 0.042--0.091, see Table \ref{tab:PCcomp}), which means that all transitions are correlated with the column density. Understandably, molecular transitions are correlated with the total column densities because the emission is expected to be stronger where there is more molecular gas. RRLs should also somewhat correlate with the total column density because when there is molecular gas, star formation is also expected in general \citep[e.g., Kenicutt-Schmidt Law,][]{1998ApJ...498..541K}. Although star formation and molecular gas content are not always co-spatial depending on the GMC evolutionary stage and spatial resolution \citep[e.g.,][]{2019ApJ...887...49S}, their correlation is relatively strong in the CMZ of NGC~253 where its star formation is relatively young. We note that this strong correlation may also be attributed to the edge-on inclination of this target.  

Although PC1 generally represents the overall molecular gas column density, transitions with the highest coefficients for PC1 are not what we expect given this relation. It is not \twco(1--0) that has the highest coefficients for PC1 among CO isotopologues; instead these are the $J=3-2$ transitions of C$^{18}$O and C$^{17}$O. This may be partly because \twco(1--0) is optically thick, and does not trace the overall gas column density properly. If the optical depth is the major effect, PC1 appears to trace the overall molecular gas better than \twco(1--0). On the other hand, there is a factor that could cause deviation of PC1 from overall molecular gas. Many high-excitation or optically-thin transitions or transitions from complex molecules show concentration on the central starburst regions with fewer transitions of optically thick, low-excitation, and extended transitions. That results in PC1 being biased towards central starburst regions, not purely indicating the total column density. Given the above factors, PC1 in general is a good tracer of overall molecular gas because many transitions have similar PC1 coefficients as C$^{18}$O and C$^{17}$O (3--2).

\subsection{Dense gas tracers and star formation}\label{sec:disc_dense}
Our PCA shows that PC2 is related to excitation (Figure \ref{fig:pc2-par}). PC2 coefficients are positive and large for RRLs (Figure \ref{fig:pc2}), which is also evident from the resemblance between the positive part of the PC2 score map (Figure \ref{fig:pcmap1}) and moment 0 maps of RRLs. PC2 scores are the highest in the GMCs of category C (GMCs 4 \& 5), and relatively high in category B (GMCs 3 \& 6), as shown in Figure \ref{fig:gmc_score}. They are GMCs with massive-star formation judging from the RRLs.  In addition to RRLs, dense gas tracers with high upper-state energies ($\gtrsim 50\,$K) also have large positive coefficients for PC2 (Figure \ref{fig:pc2}). Meanwhile, their lower excitation transitions (e.g., HCN $J=1-0$) do not have such a strong correlation to RRLs (e.g., see their PC2 coefficients in Figure \ref{fig:pc2}), and have negative PC2 coefficients comparable to the PC2 coefficients of CO isotopologues.

The correlation between star formation and low-excitation transitions with moderately high critical densities ($n_{\rm crit} \sim 10^5\,$cm$^{-3}$ in an optically-thin case, e.g. HCN $J=1-0$) is not very tight in our study, although such a correlation has been reported from extragalactic observations in single-dish resolution \citep[e.g.,][]{2004ApJ...606..271G}.
In fact, this correlation appears unlikely to be tight on $<100\,$pc scale in general. Galactic observations on the 1--10 parsec scale (in Orion A, Orion B, W51, W33(OH), and Perseus) have already shown that HCN(1--0) does not trace only high-density regions, but can also originate from translucent clouds ($n\gtrsim 10^3\,$\cc) \citep{2017A&A...599A..98P,2017ApJ...845..116W,2017A&A...605L...5K,2017ApJ...848...17N,2021A&A...646A..97T}. This may be in conflict with the results by \citet{2010ApJS..188..313W} who claim that the correlation between the HCN luminosity and star formation rate does not break down within $<100\,$pc. However, this difference may come from the fact that they focused on dense clumps where stars are already forming in the field of view of 4\,pc on average. Although the mean mass-weighted density of the molecular gas in NGC~253 CMZ is $\sim 10^{3.5}-10^{4}\,$\cc\ \citep{2023Tanaka_HB}, higher than in the GMCs in the Galactic disk \citep[$10^2-10^3\,$\cc;][]{2017A&A...599A..98P}, gas detected with low-excitation ``dense-gas tracers" should be low enough in density not to immediately form stars within the timescale for the gas to orbit to different locations in our observations. For example, it takes $\sim 10^6\,$yrs for a cloud to move 100\,pc if it is moving with a velocity of 100\,\kms, a few times the free-fall time of a cloud with $n=10^4\,$\cc. Note, though, that the star formation threshold is likely higher in the CMZ of NGC~253 than that in the Galactic center and the Galactic disk \citep{2023Tanaka_HB}.

The similarity between the high-$J$ dense gas tracers and star formation has at least two possible interpretations. One is that dense gas is intimately related to star formation. Because star formation takes place in dense gas, this high-density gas traced with high-$J$ transitions may co-exist with star formation. Alternatively, the high-excitation dense gas may have resulted from star formation. Star formation heats the gas, and molecules around young stars may be more excited owing to the high temperature. The distinction between these two scenarios is discussed further in Section \ref{sec:offset}.

\subsection{Evolutionary stages of proto super star clusters}\label{sec:offset}
%PC3 characterizes the difference between the peak of high-excitation gas and relatively large molecules such as \hctn\ (vibrationally ground state or $v_7=1$) or \hcfn (GMC6) with respect to star formation (GMC5). 
There are some works reported on the evolutionary different stages among GMCs in the center of NGC~253, and it is of interest to see whether it has an impact on the physical and chemical properties that appear in our data. \citet{2020MNRAS.491.4573R} showed different ages among these GMCs using the ratios between luminosities of protostars and zero-age main-sequence stars (ZAMS)\footnote{This age estimation by \citet{2020MNRAS.491.4573R} applies to young clusters up to the ZAMS stage.}, concluding that GMCs 3 (their clusters 1, 2, and 3) and 6 (their cluster 14) have proto-super star clusters that are younger (proto-super-star clusters $<10^5$\,yr) than most of the stars in GMC5 (their clusters 8-13; mostly ZAMS with $t_{\rm age} = 10^5 - 10^6\,$yr) although there are a couple of clusters that are young in GMC5 (their clusters 8 and 13). \citet{2021ApJ...919..105M} also concluded that their clusters 1, 2, 3, 8, 13, and 14 (our GMCs 3 and 6, a part of GMC 5) are younger than other clusters with the assumption that the stellar mass to dust mass ratios increases with age. 

The literature above indicates that GMCs in category B defined in Section \ref{sec:gmc_pcscore} host younger proto-super-star clusters, while GMCs in category C are more developed clusters (GMCs 4 and 5). It is therefore natural to suspect that the evolutionary stages of SSCs alter chemical characteristics in GMCs. The difference that appears in PC3 likely contributes to these characteristics in the PC score categorization. Positive PC3 scores seem to trace GMCs that have relatively young star formation (because its tracers have positive coefficients for PC3 as seen in the 1st quadrant of Figure \ref{fig:pc2-3}). There, relatively large molecules like cyanopolyynes (HC$_{2n+1}$N) especially vibrationally-excited lines are abundant. PC3 is also positive in GMC3, where \citet{2020MNRAS.491.4573R} show the youngest clusters. Meanwhile, negative PC3 values likely trace more well-developed star formation (the 4th quadrant of Figure \ref{fig:pc2-3}). GMC5 and a part of GMC4 have negative PC3 scores, and they contain more developed star formation ($>10^5$\,yr).   

We note that the age sequence proposed by \citet{2020MNRAS.491.4573R} and \citet{2021ApJ...919..105M} is not consistent with that reported by \citet{2020ApJ...897..176K}, who assumed that the HCN/\hctn\ ratios decrease with evolutionary stage owing to the increasing density. Instead, their HCN/\hctn\ intensity ratios in the 1-mm band tends to be higher in GMCs 5 and 6. We argue that the HCN/\hctn\ ratios may not follow the same stages as those measured by \citet{2020MNRAS.491.4573R}. \citet{2020ApJ...897..176K} used the HCN/\hctn\ ratios to estimate the ratios between the very dense gas and the dense gas. This ratio likely increases as a molecular cloud evolves to form stars, but it is unclear if the density continues to increase after stars form. In addition, the age differences among all the super-star clusters in NGC~253 CMZ may not be enough to cause chemical differentiation as already noted by \citet{2020ApJ...897..176K}.

These young and more developed starbursts embedded in the gas are characterized by different chemistry. In GMC6, high-excitation transitions of cyanopolyynes (\hctn\ and \hcfn), CH$_2$NH, and shock tracers have high intensities. On the other hand, in GMC5, there are high intensities of high excitation transitions of CN and CCH. These features can be attributed to differences in shock strengths/frequencies and/or the presence of hot molecular cores or PDRs. Embedded young starbursts may encounter a stronger influence from the interaction between the starburst and its surrounding medium, which causes shocks. Cyanopolyynes and CH$_2$NH may also be enhanced from the ice sublimation from shocks. Dense and hot molecular cores heated from star formation can also increase abundances of cyanopolyynes, CH$_2$NH, and methanol. These GMCs are hot and dense, having a similar environment as that of protostellar cores (``hot cores"), although individual protostellar cores are too small to resolve in our beams. If shocks or hot molecular cores are the only cause of this chemical differentiation, it does not fully explain the strong correlation between high-excitation radicals (CCH 4--3, CN 3--2) and RRLs. Dense PDRs around GMC5 may explain this trend as they can both increase the abundances of radicals and cause high excitation. Galactic PDRs around high-mass star-forming regions \citep[e.g., ][]{2022ApJ...931...99T} show decreased abundances of cyanopolyynes likely due to photodissociation. While low-excitation CN and CCH can be emitted from extended low-density gas or cosmic-ray dominated regions, we suggest that higher-excitation transitions such as CCH(4-3), CN(3-2), \hocp(3-2) may be relatively uncontaminated tracers of PDRs in the vicinity of starburst. Higher stellar-to-dust mass ratios in some clumps in GMC5 seen by \citet{2021ApJ...919..105M} may support this scenario because GMCs with more ionizing sources (massive stars) per unit mass are capable of creating more PDRs within them.

There are possible arguments against the strong influence of PDRs. For example, \citet{2021A&A...654A..55H,2022ApJ...931...89H} concluded that the total column densities of GMCs are extremely high ($N_{\rm H}\sim10^{24}\,$cm$^{-2}$), and PDRs should not have significant contributions assuming that the media are largely uniform, and not clumpy; instead, radicals can be enhanced with high cosmic-ray ionization rates ($\zeta \gtrsim 10^{-13}\,$s$^{-1}$). 
We note that the presence of PDRs, as previously claimed based on the detection of  significant abundances of \hocp\ and CO$^+$ \citep{2009ApJ...706.1323M}, does not necessarily exclude high cosmic-ray ionization rates. In fact, they are expected to be high in starburst galaxies, and we do not dispute that cosmic-ray ionization rates are in general much higher than that in the Galactic disk. Among the cosmic rays, lower-energy ones are attenuated even with low column densities while higher-energy cosmic rays are likely to influence a larger extent than individual GMC scale. If there are cosmic-ray sources within GMCs, cosmic rays can cause differences between individual GMCs. However, known supernovae lie outside of GMCs \citep{ulvestad_vla_1997}. Therefore, cosmic rays are unlikely to account for chemical differences among GMCs shown in this work. Moreover, cosmic rays have stronger influences in lower-density regions because the chemistry roughly scales as $\zeta/n$, and molecules can be dissociated with extremely high cosmic-ray ionization rates ($\zeta \gtrsim 10^{-13}\,$s$^{-1}$). One needs a reasonable explanation for the wide-spread presence of CO in low-density regions ($n\sim10^3\,$\cc) if the cosmic-ray ionization rates are that high over a large extent. Therefore, we suggest that PDRs can be a main driver of chemistry in some regions even if cosmic-ray ionization rates are also high.

How feedback is acting in these GMCs is an interesting question. Simulations have modeled quenching mechanisms of star formation in GMCs. Stellar feedback may include UV, optical, and infrared radiation pressure, stellar winds, supernovae, photoionization, and photoelectric heating \citep{2018MNRAS.475.3511G}. Photoionization feedback may be important in low-column-density clouds, but not in high-column-density clouds \citep{2021MNRAS.506.5512F}. Radiation pressure may instead be more important in many cases \citep{2018MNRAS.475.3511G}. Therefore, the incident radiation field producing the PDR-like emission characteristics that we see in our results may not directly measure the total feedback strength. In other words, a larger degree of feedback may be acting in young embedded proto-super-star clusters \citep[e.g., ][]{2021ApJ...912....4L} than in clusters showing PDR-like features. However, we expect such PDRs when molecular clouds are already disrupted by feedback and UV photons can travel spatially further away than when they are embedded (e.g., the ISM is made more porous).

We note that a strong continuum in GMC5 \citep[so-called ``TH2" from][]{turner_1_1985} causes absorption features for some transitions such as H$^{13}$CO$^+$(1-0), HC$^{13}$N(1-0), and SiO(2-1). The absorption features are not obvious for other transitions, and our PCA results are unlikely to be significantly affected. But this effect should be checked against higher angular resolution data.

\subsection{Origins of shocks and relationship with star formation}
The clustering analysis in the PC space found GMCs 1, 2, 7, 8, and 9 in the same group (category A, Figure \ref{fig:gmc_score}). These are locations with detected Class I methanol masers \citep{2022AA...663A..33H}, indicating the presence of shocks. Shocks are likely to be at intersections between orbits of bar ($x_1$ orbits) and circumnuclear ring ($x_2$ orbits) \citep{2015ApJ...801...63M,2022ApJ...938...80H,2022AA...663A..33H}. 
Transitions that are enhanced in these regions are low-$J$ transitions of \methanol, HNCO, and \hocop\ , tracers of weak shocks. Although star formation could also cause shocks, it is unlikely to be causing shocks here because there is no obvious evidence of star formation. Meanwhile, these shocks or cloud collisions could promote future star formation through the compression of molecular clouds \citep[e.g., ][]{2008AJ....135..479B,2019ApJ...884..100H}.
   
A well-known tracer of shocks, SiO, does not show strong enhancement near orbital intersections according to our result (e.g., PC2; Figure \ref{fig:pc2}). High excitation transitions of SiO (e.g., J=8-7, 6-5) correlate relatively well with RRLs when PC2 and PC4 are considered (Figure \ref{fig:pc2-4}). In addition to these SiO lines, some HNCO transitions are relatively well-correlated with RRLs. They are the $K_a\geq1$ transitions of HNCO$(J_{K_a,K_b}-J'_{K_a',K_b'})$, which \citet{1986ApJ...305..405C} suggested to be radiation-pumped, not collisionally pumped, in Galactic hot cores. For them to be radiation-pumped, a strong far-infrared radiation field needs to be present, which may come from embedded protostars. If these HNCO $K_a=1$ transitions originate from protostars, they do not necessarily need shocks. It has been suggested that HNCO can be abundant near young protostars \citep{2010ApJ...725.2101Q}. We note here that these SiO and HNCO transitions are similar to RRLs in contributions to PC2 and PC4, but not to PC3. This trend implies that these transitions are associated with younger starbursts, not developed ones (Section \ref{sec:offset}).

High-$J$ SiO transitions appear to be the only transitions of shock tracers that is centrally concentrated, if HNCO $K_a=1$ transitions are not related to shocks. This trend suggests the prevalence of strong shocks in the central regions. One has to be careful making this conclusion because we did not image high-$J$ HNCO transitions ($J>10$) due to the contamination criteria discussed in Section \ref{sec:mom0maps}, but the contamination to these transitions from neighboring transitions does not seem very severe except for $J=12$ \citep{2023arXiv230312685H}. However, it seems also true that SiO is more concentrated around the starburst regions, GMCs 4-6, compared with \methanol, HNCO, and \hocop\ of similar upper-state energy levels. The only exception to it is when the low-$J$ SiO line is affected by absorption against a strong continuum source. An LVG analysis by \citet{2023arXiv230312685H} found that SiO/HNCO column density ratios are higher in GMCs 4 and 6 compared with those in GMCs 3 and 7 although this difference is not significant considering errors and there are some in the outskirts (GMCs 2 and 9) that have high SiO/HNCO ratios as well. Note that they did not model GMC5 because of the self-absorption. It is not clear whether fast shocks are more dominant than slow shocks in this central starburst region as \citet{2023arXiv230312685H}
 concluded, but there are indeed strong shocks near the starburst regions.
 
What are the origins of the shocks that caused the high-$J$ SiO line emission near the central starbursts? One scenario is again cloud collisions. \citet{2022ApJ...935...19L} suggested another $x_2$ orbits connecting GMCs 3-6 in addition to the larger one shown in Figure \ref{fig:rgb}. If this orbit crosses with some $x_1$ orbits, cloud collisions could occur. However, features from such cloud collisions should appear in GMCs 3 and 6 where $x_1$ orbits likely crosses with the inner $x_2$ orbits instead of GMCs 4-6. Another explanation is that the medium in NGC~253 CMZ is in general turbulent, and high-excitation transitions are seen simply because the gas is excited near the starbursts. This scenario is hard to exclude, but it does not solely explain variations among SiO and other shock tracers. The other cause of shocks is the starburst itself. Cluster-scale outflows have been found in high-angular-resolution observations through P-Cygni profiles in GMCs 4 and 6 \citep{2021ApJ...912....4L} and through broad line wings of H40$\alpha$ in GMCs 4 and 5 \citep{2021ApJ...919..105M}. Therefore, outflows from young stellar objects may cause strong shocks, which yield high-excitation SiO emission. In addition to such cluster-scale outflows, fast shocks are common around young massive stars from stellar winds. We consider this starburst-induced shock scenario is most plausible for now because of the signatures of outflows, but this claim is still speculative.

\subsection{Vertically extended structures likely associated with outflows}
Outflows on the galactic scale have been detected in the center of NGC~253 in previous studies, and some of them contribute to the vertically extended structures or emission at high galactic altitudes seen in PC4 and PC5 \citep{1985ApJ...299..312T,bolatto_suppression_2013,2017ApJ...835..265W,krieger_molecular_2019}. In Section \ref{sec:comp}, we show that CO isotopologues, $J$ or $N$=1--0 transitions of HCN, \hcop, CN, CS, CCH, \hocp, and H$_3$O$^+$($3_{2,0}-2_{2,1}$) have emission in this structure. Among them, \hocp\ (Figure \ref{fig:mom0_8}) and \waterp\ (Figure \ref{fig:mom0_18}) have relatively weak overall intensities but still show this extended structure. This is likely because these species are reactive ions that are enhanced by strong UV radiation or cosmic rays \citep{2021ApJ...923...24H,2022ApJ...931...89H}. This result suggests a high ionization degree in these outflows.

In general, there are multiple possible mechanisms creating starburst-driven and AGN-driven outflows such as thermal energy, radiation, cosmic rays, and radio jets \citep{2020A&ARv..28....2V}. While our observations do not suggest which mechanisms are most effective in producing energy, they do suggest an influence of cosmic rays because \waterp\ is preferentially enhanced with cosmic rays compared to UV photons \citep{2022ApJ...931...89H}. 

\subsection{Comparisons with other PCA studies}
%IC 342 Meier, saito N1068, ungerechts - OMC-1, Gratier Orion B, Chidiac M82,
We compare our study with previous image-based PCA studies conducted for nearby galaxies \citep{2005ApJ...618..259M,2012ApJ...755..104M,2020PhDT........55C,2022ApJ...935..155S} and Galactic molecular clouds \citep{1997ApJ...482..245U,2017A&A...599A.100G}. Comparisons are not necessarily straightforward because the number of molecules and transitions in this work is unprecedented. The results of PCAs are affected by the transitions used, their intensities, and the normalization method. In addition, the field of view and angular resolutions are different in all studies. Here, our discussion takes these factors into consideration. Similar to our results, most studies found positive PC1 coefficients for all input variables used in PCA, except for \citet{2022ApJ...935..155S}. This is because they used not only molecular transitions or continuum but also other quantities such as the [S{\sc iii}]/[S{\sc ii}] ratio and [C{\sc i}] intensity that do not necessarily trace overall molecular content. Note also that \citet{2017A&A...599A.100G} showed that the interstellar radiation field anticorrelates with all the molecular emission intensities. 

\citet{2005ApJ...618..259M} observed the center of the modest starburst galaxy IC~342, with a similar field of view and twice larger beam size (50\,pc) in comparison with our study. They used \twco(1-0), C$^{18}$O(1-0), CCH, C$^{34}$S(2-1), \nthp(1-0), \methanol, HNCO, HNC, \hctn(10-9), SO, and continuum in the 3-mm band. Their study shows that \methanol\ and HNCO emission also peaks at the outskirts of the CMZ, and has a high absolute value in their PC2. This is in agreement with our study; high-excitation transitions have high absolute values of PC2 while \methanol\ and HNCO have the opposite sign. In the 3-mm band observations of \citet{2005ApJ...618..259M}, there were no such high-excitation transitions. Instead, C$^{34}$S and CCH showed different degrees of contribution in PC2 from \methanol\ and HNCO coming from high column density regions (C$^{34}$S) or PDRs (CCH). In our study, these transitions have somewhat different coefficients for PC2 compared to CH$_3$OH and HNCO, but the difference is not as large as for the high-excitation transitions. Similar results are found by \citet{2012ApJ...755..104M} in Maffei 2.

In the study of M82 by \citet{2020PhDT........55C}, they used CCH, CN, CS, HCN, \hcop, HNC, and \ttco\ in the 3-mm band. Because of the limited number of transitions, \ttco\ and other transitions had different contributions to PC2 likely because of the difference in the critical densities.

The nature of PCA in \citet{2022ApJ...935..155S} is quite different from ours in the sense that they focused on the region around the AGN in NGC 1068. They used sixteen transitions in the 3-mm band line survey together with [C{\sc i}] and [S{\sc iii}]/[S{\sc ii}] ratio from archival data. They separated out the circumnuclear disk in PC1 and the AGN-driven outflow in PC2. In their result, CN, HNC, and CCH showed high contribution from the outflow while \nthp, \hctn, and CO isotopologues show the opposite trend according to the categorization from PC2. In our result, CN, HNC, and CCH also have outflow features, but so do the CO isotopologues, HCN, and \hcop.

\citet{1997ApJ...482..245U} mapped the Orion BN-KL regions with \twco, CS, HCN, HNC, \hcop\ and their $^{13}$C isotopologues, \ceto, C$^{34}$S, \nthp, CN, CCH, \hctn, \propyne, C$_3$H$_2$, SO, and \methanol. Because their line widths are much narrower than ours, their observations spectrally resolved hyperfine structure lines of \nthp\ and various $K$-ladders of \propyne. Their field of view is at the pc scale with the beam size of 0.1\,pc and so is qualitatively different from the extragalactic studies. Their results separate among transitions enhanced near the high-mass star-forming regions of BN-KL such as \hctn\ and \methanol, ones enhanced around the Ridge such as CS and CN, and ones enhanced near a relatively cold cloud such as \nthp. This distinction is not obvious in our result. One reason may be the difference in spatial resolution. Our 27-pc beam is not small enough to separately resolve dense clouds and star-forming regions within a GMC. Another reason may be that the difference between dense clouds and star-forming regions is better highlighted in our study by the difference between higher-$J$ \hctn\ and \nthp\ rather than \hctn(10-9) and \nthp\ in the 3-mm band. It is also important to note that NGC~253 contains more extreme star formation compared with Orion-KL.

\citet{2017A&A...599A.100G} conducted a PCA study using maps of \twco, \ttco, \ceto, CS, HCN, \hcop, SO, CN, HNC, CCH, \nthp, and \methanol\ in the Orion B molecular cloud with the linear resolution of $0.05-0.07\,$pc covering the area of $\sim 6$\,pc$\times 8\,$pc. In addition to these line intensities, they also examined the correlation between these molecular line emission intensities and logarithms of the H$_2$ column density $N_{\rm H2}$ obtained from dust continuum flux, volume density $n_{\rm H}$ from the dust mass and core sizes, and UV radiation field normalized by the mean value $U/\bar{U}$ from PAH emissivity. Their coefficients for PC2 show a large difference between \nthp\ or \methanol\ vs. CO isotopologues. This is likely due to the difference in the density and temperature as \nthp\ and \methanol\ should be more abundant in cold and dense regions. Our result also shows a similarity between \nthp\ and \methanol. They also showed a contrast between CCH, CN and other species. CCH and CN are known as PDR tracers, and their PC3 should have contributions from the UV radiation field. In fact, they suggested a tight positive correlation between PC1 and $N_{\rm H2}$, a positive but weak correlation between PC2 and $n_H$, and a moderate positive correlation between PC3 and the UV radiation field.

The above studies show that PCA is a useful way to identify coupled chemistries, reveal the underlying dynamics driving them, and summarize rich molecular line datasets. Extragalactic PCA studies seem to be consistent in showing distinct contributions in low-excitation transitions of methanol and HNCO, especially near the intersection of the bar orbit and the central region. Galactic and extragalactic PCAs may show different categorizations of transitions, likely due to the difference in spatial scale. 

\section{Summary}\label{sec:summary}
We have presented integrated-intensity images of isolated line emission from the ALCHEMI survey, a wide-frequency imaging spectral scan of the CMZ of the starburst galaxy NGC~253. With these images, we have performed a principal component analysis (PCA) as a statistical approach to reduce the dimensions of the dataset to investigate underlying physical conditions that cause differences in line intensities, and to relate these physical conditions to starburst activities. 

We succeeded in reducing the dimension of the data space from 150 variables to 5 principal components through PCA while retaining more than 96\% of the original information. It reflects the fact that spatial distributions of various sub/millimeter emissions are highly correlated in the CMZ. One needs only a handful of uncorrelated parameters to explain the most of the standardized emission maps. Our PCA found that the nine GMCs in the CMZ are distinct from each other in their qualities after subtracting the effect of their overall emission strengths. We interpreted the PCs in physical terms and assessed various emissions with their contributions to the PCs. We list our findings below.

\begin{itemize}
    \item As expected, the largest factor that changes line intensities is the amount of molecular gas. There is an overall correlation among all the molecular emission lines, radio recombination lines (RRLs), and continuum images. This is indicated by the first principal component (PC1). Locations with larger molecular gas column densities in general emit stronger molecular transitions. These locations are also likely to have higher star formation rates (Kennicut-Schmidt Law), which causes the higher intensity of RRLs. We also note that there are some biases that cause PC1 to deviate from an accurate measure of the total column density as discussed in the text. The deviation from this correlation appears in PCs 2 or higher. 
    
    \item High-excitation dense gas tracers and RRLs have similar spatial distributions and large positive contributions to PC2. Meanwhile, low-excitation ($J=1-0$) dense gas tracers have negative coefficients for PC2; indeed, we found strong correlations between contributions to PC2 and critical densities as well as upper-state energies of the transitions of dense gas tracers. The difference between low-excitation dense gas and star formation tracers suggests that these low-excitation dense gas tracers also emit from less dense regions than indicated by their critical densities. This result is consistent with Galactic large-scale cloud observations in the literature.
    
    \item Evolutionary stages of giant molecular clouds cause differences in the chemistry. Young starbursts (GMCs 3 and 6) tend to have peaks of high-excitation transitions of \hctn\ and COMs, while more developed starbursts (GMCs 4, 5) have peaks of RRLs and high-excitation transitions of radicals (CN, CCH). This difference appears as positive and negative contributions to PC3, respectively. The lack of \hctn\ and COMs in developed starbursts may be caused by photodissociation, which shows the effect of stellar feedback. These PDRs near the starbursts can be traced by high-excitation transitions of CN and CCH. These transitions should be observable for some sources at $z=2-3$ in the 3-mm band, and can be useful for studying the ISM variation during $z=0-2$.
    
    \item Comparisons between PC2 and PC4 show that low-excitation shock tracers (\methanol, HNCO, \hocop, and OCS) do not correlate with star formation, and likely trace shocks caused by cloud collisions of $x_1$ and $x_2$ orbits. On the other hand, high-excitation transitions of SiO may originate from star formation as their emissions are from locations of cluster-scale outflows in the literature.
    
    \item Vertically-extended structures, some of which must originate from outflows, have emissions of CO isotopologues, CCH, \hocp, \waterp, HCN, \hcop, CS, and CN. The enhanced emission of \hocp, \waterp\ in outflows indicates that these outflows are highly ionized, possibly due to cosmic rays.
    
    \item We have also made the first extragalactic detection of \hcnhp, and the first extragalactic images for \cthp\, NO, and HCS$^+$.  From the characteristics of these species and their distributions, we suggest that \cthp\ emission preferentially comes from PDRs near the starburst while \hcnhp\ emits in cold clouds. 
\end{itemize}

We have identified some categories of molecular transitions that are characteristics of physical conditions through the extensive spectral scan ALCHEMI. While not many observations can afford a wide-band scan like ALCHEMI, overall chemical structures can be found by observing sets of transitions such as CO isotopologues, shock tracers (SiO, \methanol, HNCO), high-excitation dense gas tracers (\hctn) or COMs, high-excitation radicals (CN, CCH) as starburst tracers. Reactive ions such as \hocp\ and \waterp\ are also useful in tracing highly ionized regions as our previous studies found.

\section{acknowledgements}
%\begin{acknowledgements}
We appreciate the constructive comments by the anonymous referee. We thank Francesco Costagliola for setting the path to the ALCHEMI survey. We are also grateful to the ALMA staff at JAO and ARCs for their dedicated work to keep ALMA running, which made this survey possible. This paper makes use of the following ALMA data: ADS/JAO.ALMA\#2017.1.00161.L, ADS/JAO.ALMA\#2018.1.00162.S. ALMA is a partnership of ESO (representing its member states), NSF (USA) and NINS (Japan), together with NRC (Canada), MOST and ASIAA (Taiwan), and KASI (Republic of Korea), in cooperation with the Republic of Chile. The National Radio Astronomy Observatory is a facility of the National Science Foundation operated under cooperative agreement by Associated Universities, Inc. The Joint ALMA Observatory is operated by ESO, AUI/NRAO and NAOJ. Data analysis was in part carried out on the Multi-wavelength Data Analysis System operated by the Astronomy Data Center (ADC), National Astronomical Observatory of Japan. N.H. acknowledges support from JSPS KAKENHI grant No. JP21K03634, and the ALMA Japan Research Grant of NAOJ ALMA Project, NAOJ-ALMA-304.
K.S. acknowledges the support from the Ministry of Science and Technology (MOST) of Taiwan through the grant MOST 111-2112-M-001-039.
This work was supported by NAOJ ALMA Scientific
Research grant No. 2021-18A. 
L.C. acknowledges financial support through the Spanish
grant PID2019-105552RB-C41 funded by MCIN/AEI/10.13039/501100011033. S.V., M.B., and K.-Y.H. acknowledge support from  the European Research Council (ERC) Advanced Grant MOPPEX 833460. K.K. acknowledges the support by JSPS KAKENHI Grant Number JP17H06130 and the NAOJ ALMA Scientific Research Grant Number 2017-06B.
%\end{acknowledgements}


\vspace{5mm}
\facilities{ALMA}


\software{astropy \citep{2013A&A...558A..33A,2018AJ....156..123A}, scikit-learn \citep{scikit-learn}, CASA \citep{2007ASPC..376..127M} 
          }



\appendix
\startlongtable

\section{Transitions properties of blended with the same species}\label{sec:app:blend}
Transitions of species such as CN, CCH are blended in our observations due to large line widths. We combined these blended neighboring transitions to make velocity-integrated images of these species. Table \ref{tab:spec_blend} lists these transitions.
%%%%%%%%%%%%%%%%%%%%%
\startlongtable
\begin{deluxetable*}{ccccc} 
\tablecolumns{5} 
\tablewidth{0pc} 
%\tabletypesize{\scriptsize} 
\tablecaption{Spectroscopic Properties for transitions blended with the same species \label{tab:spec_blend}}
\tablehead{\colhead{Species} & \colhead{Transition} &\colhead{$\nu_{rest}^{(a)}$} &\colhead{$E_{\rm up}$$^{(b)}$}  &\colhead{$log(A_{\rm ul})$$^{(c)}$} \\
\colhead{} &\colhead{} & \colhead{(GHz)} &\colhead{(K)} &\colhead{(s$^{-1}$)} }
\startdata
\hline
CN & N=1-0, J=1/2-1/2, F=1/2-3/2 & 113.144 & 5.43 & -4.98 \\
& N=1-0, J=1/2-1/2, F=3/2-1/2 & 113.170 & 5.43 & -5.29 \\
& N=1-0, J=1/2-1/2, F=3/2-3/2 & 113.191 & 5.43 & -5.18 \\
\hline
CN & N=1-0, J=3/2-1/2, F=3/2-1/2 & 113.488 & 5.45 & -5.17 \\
& N=1-0, J=3/2-1/2, F=5/2-3/2 & 113.491 & 5.45 & -4.92 \\
& N=1-0, J=3/2-1/2, F=1/2-1/2 & 113.500 & 5.45 & -4.97 \\
& N=1-0, J=3/2-1/2, F=3/2-3/2 & 113.509 & 5.45 & -5.28 \\
\hline
CN & N= 2-1, J=3/2-3/2, F=1/2-1/2 & 226.287 & 16.3 & -4.99 \\
& N= 2-1, J=3/2-3/2, F=1/2-3/2 & 226.299 & 16.3 & -5.08 \\
& N= 2-1, J=3/2-3/2, F=3/2-1/2 & 226.303 & 16.3 & -5.38 \\
& N= 2-1, J=3/2-3/2, F=3/2-3/2 & 226.315 & 16.3 & -5.00 \\
& N= 2-1, J=3/2-3/2, F=3/2-5/2 & 226.332 & 16.3 & -5.34 \\
 & N= 2-1, J=3/2-3/2, F=5/2-3/2 & 226.342 & 16.3 & -5.50 \\
& N= 2-1, J=3/2-3/2, F=5/2-5/2 & 226.360 & 16.3 & -4.79 \\
& N= 2-1, J=3/2-1/2, F=1/2-3/2 & 226.617 & 16.3 & -4.97 \\
& N= 2-1, J=3/2-1/2, F=3/2-3/2 & 226.632 & 16.3 & -4.37 \\
& N= 2-1, J=3/2-1/2, F=5/2-3/2 & 226.660 & 16.3 & -4.02 \\
& N= 2-1, J=3/2-1/2, F=1/2-1/2 & 226.664 & 16.3 & -4.07 \\
& N= 2-1, J=3/2-1/2, F=3/2-1/2 & 226.679 & 16.3 & -4.28 \\
& N= 2-1, J=5/2-3/2, F=5/2-3/2 & 226.874 & 16.3 & -4.02 \\
& N= 2-1, J=5/2-3/2, F=7/2-5/2 & 226.875 & 16.3 & -3.94 \\
& N= 2-1, J=5/2-3/2, F=3/2-1/2 & 226.876 & 16.3 & -4.07 \\
& N= 2-1, J=5/2-3/2, F=3/2-3/2 & 226.887 & 16.3 & -4.56 \\
& N= 2-1, J=5/2-3/2, F=5/2-5/2 & 226.892 & 16.3 & -4.74 \\
& N= 2-1, J=5/2-3/2, F=3/2-5/2 & 226.905 & 16.3 & -5.95 \\
\hline
CN & N= 3-2, J=5/2-5/2, F=3/2-3/2 & 339.447 & 32.6 & -4.64 \\
& N= 3-2, J=5/2-5/2, F=3/2-5/2 & 339.460 & 32.6 & -5.36 \\
& N= 3-2, J=5/2-5/2, F=5/2-3/2 & 339.463 & 32.6 & -5.53 \\
& N= 3-2, J=5/2-5/2, F=5/2-5/2 & 339.476 & 32.6 & -4.67 \\
& N= 3-2, J=5/2-5/2, F=5/2-7/2 & 339.493 & 32.6 & -5.52 \\
& N= 3-2, J=5/2-5/2, F=7/2-5/2 & 339.499 & 32.6 & -5.63 \\
& N= 3-2, J=5/2-5/2, F=7/2-7/2 & 339.517 & 32.6 & -4.60 \\
& N= 3-2, J=5/2-3/2, F=3/2-5/2 & 339.992 & 32.6 & -5.41 \\
& N= 3-2, J=5/2-3/2, F=5/2-5/2 & 340.008 & 32.6 & -4.21 \\
& N= 3-2, J=5/2-3/2, F=3/2-3/2 & 340.020 & 32.6 & -4.03 \\
& N= 3-2, J=5/2-3/2, F=7/2-5/2 & 340.032 & 32.6 & -3.41 \\
& N= 3-2, J=5/2-3/2, F=3/2-1/2 & 340.035 & 32.6 & -3.54 \\
& N= 3-2, J=5/2-3/2, F=3/2-1/2 & 340.035 & 32.6 & -3.49 \\
& N= 3-2, J=7/2-5/2, F=7/2-5/2 & 340.248 & 32.7 & -3.42 \\
& N= 3-2, J=7/2-5/2, F=7/2-5/2 & 340.248 & 32.7 & -3.38 \\
& N= 3-2, J=7/2-5/2, F=5/2-3/2 & 340.249 & 32.7 & -3.43 \\
& N= 3-2, J=7/2-5/2, F=5/2-5/2 & 340.262 & 32.7 & -4.35 \\
& N= 3-2, J=7/2-5/2, F=7/2-7/2 & 340.265 & 32.7 & -4.47 \\
\hline
CH$_3$CCH & J=5-4, K=4 & 85.431 & 128. & -6.65\\
 & J=5-4, K=3 & 85.443 & 77.3 & -6.40\\
 & J=5-4, K=2 & 85.451 & 41.2 & -6.28\\
 & J=5-4, K=1 & 85.456 & 19.5 & -6.23\\
 & J=5-4, K=0 & 85.457 & 12.3 & -6.21\\
\hline
CH$_3$CCH & J=6-5, K=5 & 102.499 & 198. & -6.48\\
 & J=6-5, K=4 & 102.517 & 133. & -6.22\\
 & J=6-5, K=3 & 102.530 & 82.3 & -6.09\\
 & J=6-5, K=2 & 102.540 & 46.1 & -6.02\\
 & J=6-5, K=1 & 102.546 & 24.5 & -5.98\\
 & J=6-5, K=0 & 102.548 & 17.2 & -5.96\\
\hline
CH$_3$CCH & J=10-9, K=5 & 170.824 & 226. & -5.41\\
 & J=10-9, K=4 & 170.854 & 161. & -5.36\\
 & J=10-9, K=3 & 170.876 & 110. & -5.33\\
 & J=10-9, K=2 & 170.893 & 74.0 & -5.30\\
 & J=10-9, K=1 & 170.903 & 52.3 & -5.29\\
 & J=10-9, K=0 & 170.906 & 45.1 & -5.29\\
\hline
CH$_3$CCH & J=11-10, K=5 & 187.904 & 235. & -5.26\\
 & J=11-10, K=4 & 187.936 & 170. & -5.22\\
 & J=11-10, K=3 & 187.961 & 119. & -5.19\\
 & J=11-10, K=2 & 187.979 & 83.0 & -5.17\\
 & J=11-10, K=1 & 187.990 & 61.4 & -5.16\\
 & J=11-10, K=0 & 187.994 & 54.1 & -5.16\\
\hline
\methanol\ & 2( 1, 2)- 1( 1, 1) E  & 96.739 & 12.5 & -5.59\\
 & 2( 0, 2)- 1( 0, 1) A  & 96.741 & 6.96 & -5.47\\
 & 2(- 0, 2)- 1(- 0, 1) E  & 96.745 & 20.1 & -5.47\\
 & 2(- 1, 1)- 1(- 1, 0) E  & 96.756 & 28.0 & -5.58\\
\hline
CCH & N= 1-0, J=3/2-1/2, F= 1- 1 & 87.284 & 4.19 & -6.59 \\
 & N= 1-0, J=3/2-1/2, F= 2- 1 & 87.317 & 4.19 & -5.82 \\
 & N= 1-0, J=3/2-1/2, F= 1- 0 & 87.329 & 4.19 & -5.90 \\
 & N= 1-0, J=1/2-1/2, F= 1- 1 & 87.402 & 4.20 & -5.90 \\
 & N= 1-0, J=1/2-1/2, F= 0- 1 & 87.407 & 4.20 & -5.81 \\
 & N= 1-0, J=1/2-1/2, F= 1- 0 & 87.446 & 4.20 & -6.58 \\ 
\hline 
CCH & N= 2-1, J=5/2-3/2, F= 3- 2 & 174.663 & 12.6 & -4.83 \\
 & N= 2-1, J=5/2-3/2, F= 2- 1 & 174.668 & 12.6 & -4.87 \\
 & N= 2-1, J=3/2-1/2, F= 2- 1 & 174.722 & 12.6 & -4.95 \\
 & N= 2-1, J=3/2-1/2, F= 1- 0 & 174.728 & 12.6 & -5.09 \\
 & N= 2-1, J=3/2-1/2, F= 1- 1 & 174.733 & 12.6 & -5.29 \\
 & N= 2-1, J=3/2-3/2, F= 2- 2 & 174.807 & 12.6 & -5.57 \\
\hline
CCH & N= 3-2, J=7/2-5/2, F= 4- 3 & 262.004 & 25.1 & -4.28\\
 & N= 3-2, J=7/2-5/2, F= 3- 2 & 262.006 & 25.1 & -4.29\\
 & N= 3-2, J=5/2-3/2, F= 3- 2 & 262.065 & 25.2 & -4.31\\
 & N= 3-2, J=5/2-3/2, F= 2- 1 & 262.067 & 25.2 & -4.35\\
 & N= 3-2, J=5/2-3/2, F= 2- 2 & 262.079 & 25.2 & -5.22\\
 & N= 3-2, J=5/2-5/2, F= 3- 3 & 262.209 & 25.2 & -5.40\\
\hline
CCH & N= 4-3, J=9/2-7/2, F= 5- 4 & 349.338 & 41.9 & -3.88 \\
 & N= 4-3, J=9/2-7/2, F= 4- 3 & 349.339 & 41.9 & -3.89 \\
 & N= 4-3, J=7/2-5/2, F= 4- 3 & 349.399 & 41.9 & -3.90 \\
 & N= 4-3, J=7/2-5/2, F= 3- 2 & 349.401 & 41.9 & -3.92 \\
\hline
C$_4$H & N = 12-11, J = 25/2 - 23/2, F = 12 - 11 & 114.183 & 35.6 & -4.44\\
 & N = 12-11, J = 25/2 - 23/2, F = 13 - 12 & 114.183 & 35.6 & -4.44\\
 & N = 12-11, J = 23/2 - 21/2, F = 11 - 10 & 114.221 & 35.6 & -4.44\\
 & N = 12-11, J = 23/2 - 21/2, F = 12 - 11 & 114.221 & 35.6 & -4.44\\
\enddata 
\tablecomments{$(a)$ Rest frequency; $(b)$ Upper level energy of the transition; $(c)$ Einstein $A$ coefficient; 
All values were taken from the Cologne Database for Molecular Spectroscopy 
\citep[CDMS; https://cdms.astro.uni-koeln.de;][]{2001AA370L49M,2005JMoSt.742..215M}} 
\end{deluxetable*} 
%%%%%%%%%%%%%%%%%%%%%%%%%%%%%%

\section{Transition properties}\label{sec:tran}

Table \ref{tab:spec} shows the spectroscopic properties of transitions used in our PCA in an ascending order in frequency. 
%%%%%%%%%%%%%%%%%%%%
\startlongtable
\begin{deluxetable}{ccc|ccc} 
\tablecolumns{6} 
\tablewidth{0pc} 
%\tabletypesize{\scriptsize} 
\tablecaption{A list of molecular transitions and RRLs used in PCA\label{tab:spec}}
\tablehead{\colhead{Transition} &\colhead{$\nu_{\rm rest}^{(a)}$} &\colhead{$E_{\rm up}$$^{(b)}$} &\colhead{Transition} &\colhead{$\nu_{\rm rest}^{(a)}$} &\colhead{$E_{\rm up}$$^{(b)}$} \\
\colhead{} & \colhead{(GHz)} &\colhead{(K)} & \colhead{} & \colhead{(GHz)} &\colhead{(K)} }
\startdata 
CH$_3$CCH($5_K-4_K$) &  85.457 & 12.3 & HOCO$^+$($4_{0,4}-3_{0,3}$) &  85.531 & 10.26\\ 
CCS($7_6-6_5$)$\dagger$ &  86.181 & 23.35 & H$^{13}$CN(1-0)$\dagger$ &  86.34 & 4.14\\ 
H$^{13}$CO$^+$(1-0) &  86.754 & 4.16 & SiO(2-1) &  86.847 & 6.25\\ 
HN$^{13}$C(1-0) &  87.091 & 4.18 & CCH($1-0$) &  87.317 & 4.19\\ 
HNCO($4_{1,4}-3_{1,3}$) &  87.597 & 53.78 & HNCO($4_{0,4}-3_{0,3}$) &  87.925 & 10.55\\ 
H52$\beta$ &  88.406 &  & HCN(1-0) &  88.632 & 4.25\\ 
HCO$^+$(1-0) &  89.189 & 4.28 & HOC$^+$(1-0) &  89.487 & 4.29\\ 
C$_3$H$^+$(4-3) &  89.958 & 10.79 & HC$_5$N(34-33) &  90.526 & 76.03\\ 
HNC(1-0) &  90.664 & 4.35 & HC$_3$N(10-9) &  90.979 & 24.01\\ 
$^{13}$CS(2-1) &  92.494 & 6.65 & N$_2$H$^+$(1-0) &  93.173 & 4.47\\ 
H58$\gamma$ &  93.776 &  & CCS($7_8-6_7$)$\dagger$ &  93.87 & 19.89\\ 
C$^{34}$S(2-1) &  96.413 & 6.25 & CH$_3$OH($2_K-1_K$) &  96.741 & 6.96\\ 
C$^{33}$S(2-1) &  97.172 & 7.0 & OCS(8-7)$\dagger$ &  97.301 & 21.01\\ 
CH$_3$OH($2_{1,1}-1_{1,0}$) &  97.583 & 21.56 & $^{34}$SO($2_3-1_2$)$\dagger$ &  97.715 & 9.09\\ 
CS(2-1)$\dagger$ &  97.981 & 7.1 & HC$_5$N(37-36) &  98.513 & 89.83\\ 
H40$\alpha$ &  99.023 &  & HC$_3$N(11-10) &  100.076 & 28.82\\ 
HC$_3$N,v$_7$=1(11$_1$-10$_{-1}$) &  100.322 & 349.73 & CH$_3$CCH(6$_K$-5$_K$) &  102.548 & 17.23\\ 
H$_2$CS(3$_{0,3}$-2$_{0,2}$)$\dagger$ &  103.04 & 9.89 & H49$\beta$ &  105.302 & \\ 
CCS(8$_9$-7$_8$)$\dagger$ &  106.348 & 70.39 & H39$\alpha$ &  106.737 & \\ 
HOCO$^+$(5$_{0,5}$-4$_{0,4}$) &  106.914 & 15.39 & CH$_3$OH($3_{1,3}-4_{0,4}$) &  107.014 & 28.35\\ 
HC$_3$N(12-11) &  109.174 & 34.06 & C$^{18}$O(1-0)$\dagger$ &  109.782 & 5.27\\ 
HNCO($5_{0,5}-4_{0,4}$) &  109.906 & 15.82 & $^{13}$CO(1-0)$\dagger$ &  110.201 & 5.29\\ 
C$^{17}$O(1-0)$\dagger$ &  112.359 & 5.39 & CN($1_{3/2}-0_{3/2}$) &  113.191 & 5.43\\ 
CN($1_{5/2}-0_{3/2}$) &  113.491 & 5.45 & C$_4$H(12-11) &  114.221 & 35.64\\ 
CO(1-0) &  115.271 & 5.53 & HC$_3$N(14-13) &  127.368 & 45.85\\ 
HOCO$^+$($6_{0,6}-5_{0,5}$) &  128.295 & 21.55 & SiO(3-2) &  130.269 & 12.5\\ 
HNCO($6_{1,6}-5_{1,5}$) &  131.394 & 65.35 & HNCO($6_{0,6}-5_{0,5}$) &  131.886 & 22.15\\ 
HC$_3$N(15-14) &  136.464 & 52.4 & SO($3_4-2_3$) &  138.179 & 15.86\\ 
$^{13}$CS(3-2) &  138.739 & 13.32 & CH$_3$OH($3_{1,3}-2_{1,2}$) &  143.866 & 28.35\\ 
CCS($11_{12}-10_{11}$)$\dagger$ &  144.245 & 43.94 & C$^{34}$S(3-2) &  144.617 & 11.8\\ 
C$^{33}$S(3-2) &  145.756 & 13.99 & HCNH$^+$(2-1) &  148.221 & 10.67\\ 
HOCO$^+$($7_{0,7}-6_{0,6}$) &  149.676 & 28.73 & HNCO(7$_{1,6}$-6$_{1,5}$) &  154.415 & 72.92\\ 
HC$_3$N(17-16) &  154.657 & 66.8 & CH$_3$OH($2_{1,2}-3_{0,3}$) &  156.602 & 21.44\\ 
C$_3$H$^+$(7-6) &  157.419 & 30.22 & HC$_3$N(18-17) &  163.753 & 74.66\\ 
H$_2^{34}$S($1_{1,0}-1_{1,0,1}$)$\dagger$ &  167.911 & 27.83 & H$_2$S($1_{1,0}-1_{0,1}$)$\dagger$ &  168.763 & 27.88\\ 
H$_2$CS($5_{1,5}-4_{1,4}$)$\dagger$ &  169.114 & 37.52 & CH$_3$OH($3_{2,2}-2_{1,1}$) &  170.061 & 36.17\\ 
HCS$^+$(4-3) &  170.692 & 20.48 & CH$_3$CCH($10_K-9_K$) &  170.906 & 45.11\\ 
H$^{13}$CN(2-1)$\dagger$ &  172.678 & 12.43 & HC$_3$N(19-18) &  172.849 & 82.96\\ 
H$^{13}$CO$^+$(2-1)$\dagger$ &  173.507 & 12.49 & HN$^{13}$C(2-1) &  174.179 & 12.54\\ 
CCH($2-1$) &  174.663 & 12.57 & HNCO($8_{0,8}-7_{0,7}$) &  175.844 & 37.98\\ 
HCN(2-1) &  177.261 & 12.76 & HCO$^+$(2-1) &  178.375 & 12.84\\ 
SO(4$_5$-3$_4$)$\dagger$ &  178.605 & 24.43 & HOC$^+$(2-1) &  178.972 & 12.88\\ 
HNC(2-1) &  181.325 & 13.05 & CH$_3$CCH($11_K-10_K$) &  187.994 & 54.13\\ 
C$^{34}$S(4-3) &  192.818 & 18.98 & CH$_3$OH($4_{1,3}-3_{1,2}$) &  195.147 & 37.95\\ 
CS(4-3)$\dagger$ &  195.954 & 23.5 & HNCO($9_{0,9}-8_{0,8}$) &  197.821 & 47.47\\ 
CH$_2$NH($3_{1,2}-2_{1,1}$) &  199.823 & 27.07 & HC$_3$N(22-21) &  200.135 & 110.46\\ 
CH$_3$CCH($12_K-11_K$) &  205.081 & 63.98 & c-C$_3$H$_2$($3_{3,0}-2_{2,1}$) &  216.279 & 13.53\\ 
H$_2$S($2_{2,0}-2_{1,1}$)$\dagger$ &  216.71 & 83.98 & SiO(5-4) &  217.105 & 31.26\\ 
C$^{18}$O(2-1) &  219.56 & 15.81 & $^{13}$CO(2-1) &  220.399 & 15.87\\ 
C$^{17}$O(2-1)$\dagger$ &  224.714 & 16.18 & CN(2-1) &  226.66 & 16.31\\ 
HC$_3$N(25-24) &  227.419 & 141.89 & CO(2-1) &  230.538 & 16.6\\ 
HC$_3$N(26-25) &  236.513 & 153.25 & C$^{33}$S(5-4) &  242.914 & 34.97\\ 
CS(5-4) &  244.936 & 35.3 & NO($3_{1,3}$-2$_{-1,2}$) &  250.437 & 19.23\\ 
NO($3_{-1,3}-2_{1,2}$) &  250.796 & 19.28 & $^{34}$SO($6_7-5_6$)$\dagger$ &  256.878 & 46.71\\ 
H$^{13}$CN(3-2)$\dagger$ &  259.012 & 24.86 & H$^{13}$CO$^+$(3-2) &  260.255 & 24.98\\ 
SiO(6-5) &  260.518 & 43.76 & CCH($3-2$) &  262.004 & 25.15\\ 
HCN(3-2) &  265.886 & 25.52 & CH$_2$NH($4_{1,3}-3_{1,2}$) &  266.27 & 39.84\\ 
CH$_3$OH($5_{2,4}-4_{1,3}$) &  266.838 & 57.07 & HCO$^+$(3-2) &  267.558 & 25.68\\ 
HOC$^+$(3-2) &  268.451 & 25.77 & HNC(3-2) &  271.981 & 26.11\\ 
HC$_3$N(30-29) &  272.885 & 203.01 & $^{13}$CS(6-5) &  277.455 & 46.61\\ 
H$_2$CS($8_{1,7}-7_{1,6}$)$\dagger$ &  278.888 & 73.41 & N$_2$H$^+$(3-2) &  279.512 & 26.83\\ 
CH$_3$OH($6_{1,6}-5_{1,5}$) &  287.671 & 62.87 & C$^{34}$S(6-5) &  289.209 & 38.19\\ 
CS(6-5) &  293.912 & 49.4 & H$_2$S(3$_{3,0}-3_{2,1}$)$\dagger$ &  300.506 & 168.9\\ 
CH$_3$OH($1_{1,0}-1_{0,1}$) &  303.367 & 16.88 & CH$_3$OH($3_{1,2}-3_{0,3}$) &  305.473 & 28.59\\ 
CH$_3$OH($6_{1,5}-6_{0,6}$) &  311.853 & 63.71 & CH$_3$OH($7_{1,6}-7_{0,7}$) &  314.86 & 80.09\\ 
CH$_3$OH($6_{2,5}-5_{1,4}$) &  315.267 & 71.0 & H27$\alpha$ &  316.415 & \\ 
C$^{18}$O(3-2) &  329.331 & 31.61 & $^{13}$CO(3-2) &  330.588 & 31.73\\ 
HC$_3$N(37-36) &  336.52 & 306.91 & C$^{17}$O(3-2)$\dagger$ &  337.061 & 32.35\\ 
CN(3-2) &  340.032 & 32.66 & CS(7-6) &  342.883 & 65.8\\ 
CO(3-2) &  345.796 & 33.19 & H$^{13}$CO$^+$(4-3) &  346.998 & 41.63\\ 
SiO(8-7) &  347.331 & 75.02 & CCH($4-3$) &  349.338 & 41.91\\ 
HNCO($16_{1,16}-15_{1,15}$) &  350.333 & 186.2 & HCN(4-3) &  354.505 & 42.53\\ 
HCO$^+$(4-3) &  356.734 & 42.8 & HNC(4-3) &  362.63 & 43.51\\ 
H$_3$O$^+$($3_{2,0}-2_{2,1}$) &  364.797 & 97.06 & N$_2$H$^+$(4-3) &  372.672 & 44.71\\ 
\enddata 
\tablecomments{$(a)$ Rest frequency; $(b)$ Upper level energy of the transition; All values were taken from the Cologne Database for Molecular Spectroscopy 
\citep[CDMS; https://cdms.astro.uni-koeln.de;][]{2001AA370L49M,2005JMoSt.742..215M}. $\dagger$ Maps that are used in PCA, but not shown in this paper because they are included in the future publications.} 
\end{deluxetable} 
%%%%%%%%%%%%%%%%%%%%%%%%%%%%%%

\section{PC coefficients}\label{sec:app:PCcomp}
The coefficients of all the 150 emissions for PC1 through PC 10 are 
%All the transitions projected to principal coefficients are shown 
in Table \ref{tab:PCcomp}.
%%%%%%%%%%%%%%%%%%%%%
\startlongtable
\begin{deluxetable*}{ccccccccccc} 
\tablecolumns{11} 
\tablewidth{0pc} 
%\tabletypesize{\scriptsize} 
\tablecaption{PC coefficients for each species \label{tab:PCcomp}}
\tablehead{\colhead{Transition} & \colhead{PC1} & \colhead{PC2} & \colhead{PC3} & \colhead{PC4} & \colhead{PC5} & \colhead{PC6} & \colhead{PC7} & \colhead{PC8} & \colhead{PC9} & \colhead{PC10}} 
\startdata
 CO(1-0) &0.069 &-0.127 &-0.129 &-0.046 &0.163 &0.029 &0.161 &-0.001 &-0.116 &-0.065\\ 
CO(2-1) &0.071 &-0.115 &-0.138 &-0.055 &0.140 &-0.017 &0.170 &-0.024 &-0.132 &-0.058\\ 
CO(3-2) &0.075 &-0.097 &-0.139 &-0.057 &0.101 &-0.060 &0.125 &-0.011 &-0.087 &-0.035\\ 
$^{13}$CO(1-0) &0.071 &-0.131 &-0.102 &-0.030 &0.151 &0.133 &0.081 &0.064 &-0.001 &-0.110\\ 
$^{13}$CO(2-1) &0.082 &-0.080 &-0.102 &-0.068 &0.079 &0.053 &0.034 &0.038 &0.015 &-0.059\\ 
$^{13}$CO(3-2) &0.087 &-0.031 &-0.093 &-0.071 &0.019 &0.015 &-0.024 &0.046 &0.034 &-0.022\\ 
C$^{18}$O(1-0) &0.076 &-0.123 &-0.078 &-0.028 &0.120 &0.145 &0.055 &0.065 &0.052 &-0.079\\ 
C$^{18}$O(2-1) &0.086 &-0.064 &-0.071 &-0.075 &0.054 &0.058 &0.014 &0.043 &0.066 &-0.015\\ 
C$^{18}$O(3-2) &0.090 &-0.005 &-0.057 &-0.067 &-0.005 &0.031 &-0.044 &0.047 &0.056 &-0.013\\ 
C$^{17}$O(1-0) &0.075 &-0.120 &-0.073 &-0.012 &0.099 &0.188 &-0.008 &0.090 &0.102 &-0.074\\ 
C$^{17}$O(2-1) &0.087 &-0.057 &-0.060 &-0.068 &0.036 &0.067 &-0.027 &0.056 &0.072 &-0.057\\ 
C$^{17}$O(3-2) &0.090 &0.017 &-0.030 &-0.058 &-0.009 &0.019 &-0.073 &0.051 &0.082 &-0.017\\ 
HCN(1-0) &0.080 &-0.101 &-0.093 &-0.073 &0.068 &-0.063 &0.089 &-0.109 &-0.093 &-0.001\\ 
HCN(2-1) &0.082 &-0.073 &-0.111 &-0.045 &0.063 &-0.085 &0.125 &-0.053 &-0.094 &0.037\\ 
HCN(3-2) &0.088 &-0.019 &-0.078 &0.000 &-0.035 &-0.112 &0.016 &-0.032 &-0.033 &0.046\\ 
HCN(4-3) &0.089 &0.012 &-0.060 &0.025 &-0.054 &-0.098 &-0.021 &-0.023 &-0.011 &0.037\\ 
HCO$^+$(1-0) &0.079 &-0.094 &-0.090 &-0.117 &0.066 &-0.033 &0.106 &-0.128 &-0.077 &-0.035\\ 
HCO$^+$(2-1) &0.082 &-0.063 &-0.113 &-0.091 &0.066 &-0.056 &0.138 &-0.063 &-0.062 &0.011\\ 
HCO$^+$(3-2) &0.088 &-0.005 &-0.087 &-0.042 &-0.029 &-0.063 &0.011 &-0.029 &-0.001 &0.025\\ 
HCO$^+$(4-3) &0.089 &0.027 &-0.072 &-0.007 &-0.043 &-0.015 &-0.041 &-0.015 &0.031 &0.029\\ 
CN($1_{3/2}-0_{3/2}$) &0.083 &-0.053 &-0.101 &-0.128 &0.061 &-0.047 &-0.008 &-0.073 &-0.025 &0.038\\ 
CN($1_{5/2}-0_{3/2}$) &0.082 &-0.050 &-0.116 &-0.120 &0.045 &-0.067 &0.058 &-0.099 &-0.057 &0.023\\ 
CN(2-1) &0.085 &0.016 &-0.123 &-0.047 &-0.025 &-0.070 &-0.019 &-0.015 &-0.009 &0.039\\ 
CN(3-2) &0.083 &0.065 &-0.120 &0.016 &-0.052 &-0.052 &-0.053 &0.023 &-0.003 &0.019\\ 
HNC(1-0) &0.086 &-0.043 &-0.098 &-0.063 &0.041 &-0.008 &-0.014 &-0.067 &-0.015 &0.040\\ 
HNC(2-1) &0.087 &0.032 &-0.083 &0.004 &-0.069 &-0.040 &-0.075 &-0.068 &0.073 &0.053\\ 
HNC(3-2) &0.087 &0.045 &-0.085 &0.027 &-0.022 &-0.034 &-0.062 &-0.069 &0.009 &0.027\\ 
HNC(4-3) &0.086 &0.072 &-0.065 &0.048 &-0.007 &-0.023 &-0.089 &-0.094 &0.007 &0.015\\ 
N$_2$H$^+$(1-0) &0.085 &-0.091 &-0.040 &0.053 &0.026 &-0.017 &-0.015 &-0.136 &-0.033 &-0.032\\ 
N$_2$H$^+$(3-2) &0.088 &0.015 &-0.055 &0.101 &-0.031 &-0.043 &-0.029 &-0.162 &0.017 &-0.051\\ 
N$_2$H$^+$(4-3) &0.086 &0.058 &-0.045 &0.105 &-0.030 &-0.083 &-0.053 &-0.150 &0.003 &-0.060\\ 
CS(2-1) &0.084 &-0.092 &-0.062 &-0.062 &0.067 &-0.045 &0.046 &-0.013 &0.016 &0.003\\ 
CS(4-3) &0.090 &-0.023 &-0.047 &-0.043 &0.002 &-0.082 &0.001 &0.019 &0.018 &0.017\\ 
CS(5-4) &0.091 &0.014 &-0.034 &-0.029 &-0.012 &-0.075 &-0.011 &0.032 &0.024 &0.016\\ 
CS(6-5) &0.090 &0.040 &-0.023 &-0.004 &-0.009 &-0.070 &-0.013 &0.065 &0.015 &0.005\\ 
CS(7-6) &0.088 &0.064 &-0.019 &0.028 &0.008 &-0.068 &-0.027 &0.092 &-0.008 &0.014\\ 
CCH($1-0$) &0.085 &-0.059 &-0.087 &-0.097 &0.048 &0.009 &-0.007 &-0.006 &0.005 &0.017\\ 
CCH($2-1$) &0.086 &0.051 &-0.057 &-0.032 &-0.101 &0.003 &-0.118 &0.002 &0.022 &-0.009\\ 
CCH($3-2$) &0.088 &0.051 &-0.072 &-0.018 &-0.029 &-0.004 &-0.067 &0.036 &-0.015 &0.005\\ 
CCH($4-3$) &0.084 &0.080 &-0.088 &0.021 &-0.039 &0.006 &-0.056 &0.015 &-0.006 &0.007\\ 
H$^{13}$CN(1-0) &0.085 &-0.089 &-0.001 &-0.054 &0.051 &-0.070 &-0.108 &-0.125 &-0.049 &0.006\\ 
H$^{13}$CN(2-1) &0.091 &-0.007 &-0.023 &-0.019 &0.020 &-0.079 &-0.069 &-0.047 &-0.045 &0.021\\ 
H$^{13}$CN(3-2) &0.089 &0.058 &0.014 &0.040 &0.040 &-0.085 &-0.071 &-0.025 &0.004 &0.008\\ 
H$^{13}$CO$^+$(1-0) &0.086 &-0.065 &-0.010 &-0.126 &0.050 &0.020 &-0.118 &-0.085 &-0.006 &-0.009\\ 
H$^{13}$CO$^+$(2-1) &0.081 &0.048 &0.089 &-0.040 &0.046 &0.144 &0.027 &0.078 &0.066 &-0.173\\ 
H$^{13}$CO$^+$(3-2) &0.087 &0.077 &-0.025 &-0.011 &-0.011 &0.045 &-0.070 &-0.028 &-0.022 &-0.011\\ 
H$^{13}$CO$^+$(4-3) &0.085 &0.093 &0.009 &0.014 &0.032 &0.067 &-0.071 &-0.063 &-0.021 &-0.013\\ 
HN$^{13}$C(1-0) &0.089 &-0.036 &-0.003 &-0.049 &0.022 &0.009 &-0.171 &-0.100 &0.042 &0.005\\ 
HN$^{13}$C(2-1) &0.083 &0.088 &0.006 &0.010 &0.001 &0.124 &-0.089 &-0.170 &-0.010 &-0.004\\ 
$^{13}$CS(2-1) &0.086 &-0.033 &0.069 &-0.066 &-0.003 &-0.111 &-0.048 &0.036 &0.152 &-0.016\\ 
$^{13}$CS(3-2) &0.087 &0.031 &0.073 &-0.065 &0.025 &-0.070 &-0.026 &0.020 &0.153 &-0.017\\ 
$^{13}$CS(6-5) &0.070 &0.117 &0.078 &0.126 &0.243 &-0.063 &0.059 &0.039 &0.131 &0.093\\ 
C$^{34}$S(2-1) &0.087 &-0.069 &-0.009 &-0.060 &0.059 &-0.041 &0.018 &0.009 &0.061 &0.013\\ 
C$^{34}$S(3-2) &0.090 &-0.012 &-0.014 &-0.036 &0.048 &-0.057 &0.011 &0.043 &0.096 &0.033\\ 
C$^{34}$S(4-3) &0.089 &0.043 &0.027 &-0.014 &0.047 &-0.084 &-0.025 &0.039 &0.087 &0.019\\ 
C$^{34}$S(6-5) &0.084 &0.092 &0.019 &0.039 &0.102 &-0.075 &-0.044 &0.085 &0.073 &0.027\\ 
C$^{33}$S(2-1) &0.087 &-0.015 &0.059 &-0.093 &-0.020 &-0.086 &-0.046 &0.024 &0.171 &-0.053\\ 
C$^{33}$S(3-2) &0.086 &0.057 &0.078 &-0.021 &0.061 &-0.045 &-0.015 &-0.011 &0.079 &0.003\\ 
C$^{33}$S(5-4) &0.081 &0.084 &0.085 &0.038 &0.105 &0.021 &-0.024 &-0.002 &-0.036 &0.013\\ 
HOC$^+$(1-0) &0.069 &-0.120 &-0.075 &-0.151 &0.138 &0.070 &0.174 &0.184 &0.206 &0.197\\ 
HOC$^+$(2-1) &0.081 &0.048 &0.015 &-0.120 &-0.065 &0.205 &-0.103 &0.293 &-0.075 &0.023\\ 
HOC$^+$(3-2) &0.084 &0.073 &-0.068 &-0.069 &-0.015 &0.049 &-0.182 &0.008 &-0.109 &0.065\\ 
SiO(2-1) &0.081 &-0.112 &0.034 &0.012 &0.025 &-0.058 &-0.117 &-0.118 &-0.102 &0.060\\ 
SiO(3-2) &0.087 &-0.070 &0.033 &0.039 &-0.004 &-0.057 &-0.063 &-0.071 &-0.087 &0.060\\ 
SiO(5-4) &0.090 &0.013 &0.044 &0.039 &-0.026 &-0.095 &-0.061 &0.046 &-0.072 &0.024\\ 
SiO(6-5) &0.088 &0.055 &0.046 &0.050 &0.010 &-0.080 &-0.036 &0.079 &-0.084 &0.020\\ 
SiO(8-7) &0.072 &0.073 &0.166 &0.034 &0.186 &-0.072 &0.034 &0.083 &0.007 &0.018\\ 
CH$_3$OH($2_K-1_K$) &0.048 &-0.211 &-0.010 &0.191 &0.053 &0.032 &0.040 &0.023 &-0.054 &0.013\\ 
CH$_3$OH($2_{1,1}-1_{1,0}$) &0.060 &-0.163 &0.090 &0.178 &-0.120 &-0.103 &0.049 &-0.066 &0.058 &0.057\\ 
CH$_3$OH($3_{1,3}-4_{0,4}$) &0.078 &0.030 &0.155 &-0.106 &-0.023 &0.166 &0.051 &-0.004 &-0.139 &-0.019\\ 
CH$_3$OH($3_{1,3}-2_{1,2}$) &0.065 &-0.151 &0.093 &0.140 &-0.125 &-0.137 &0.075 &-0.077 &0.026 &0.074\\ 
CH$_3$OH($2_{1,2}-3_{0,3}$) &0.084 &0.010 &0.082 &-0.065 &-0.073 &0.084 &0.115 &0.119 &0.114 &0.133\\ 
CH$_3$OH($3_{2,2}-2_{1,1}$) &0.076 &-0.135 &0.041 &0.019 &-0.033 &0.002 &0.050 &0.059 &0.135 &0.077\\ 
CH$_3$OH($4_{1,3}-3_{1,2}$) &0.079 &-0.076 &0.110 &0.023 &-0.139 &-0.139 &0.115 &-0.125 &-0.042 &0.014\\ 
CH$_3$OH($5_{2,4}-4_{1,3}$) &0.086 &-0.029 &0.087 &-0.057 &-0.094 &-0.048 &0.006 &0.049 &-0.048 &-0.013\\ 
CH$_3$OH($6_{1,6}-5_{1,5}$) &0.077 &-0.041 &0.135 &-0.028 &-0.176 &-0.106 &0.220 &-0.197 &0.030 &-0.001\\ 
CH$_3$OH($1_{1,0}-1_{0,1}$) &0.089 &-0.033 &0.040 &-0.061 &0.007 &0.076 &-0.033 &0.042 &-0.028 &0.106\\ 
CH$_3$OH($3_{1,2}-3_{0,3}$) &0.089 &-0.012 &0.049 &-0.057 &-0.045 &0.084 &-0.049 &-0.011 &-0.045 &0.067\\ 
CH$_3$OH($6_{1,5}-6_{0,6}$) &0.088 &0.041 &0.064 &-0.011 &-0.042 &0.035 &-0.017 &-0.115 &-0.079 &0.006\\ 
CH$_3$OH($7_{1,6}-7_{0,7}$) &0.088 &0.041 &0.056 &0.001 &-0.028 &0.019 &-0.027 &-0.071 &-0.145 &0.025\\ 
CH$_3$OH($6_{2,5}-5_{1,4}$) &0.085 &-0.009 &0.102 &-0.033 &-0.082 &0.031 &0.002 &-0.015 &-0.214 &0.078\\ 
HNCO($4_{1,4}-3_{1,3}$) &0.081 &0.066 &-0.032 &0.020 &-0.127 &0.106 &0.175 &-0.112 &0.031 &-0.093\\ 
HNCO($4_{0,4}-3_{0,3}$) &0.043 &-0.201 &-0.009 &0.260 &0.069 &0.130 &-0.061 &0.066 &-0.121 &0.062\\ 
HNCO($5_{0,5}-4_{0,4}$) &0.053 &-0.197 &-0.002 &0.208 &0.042 &0.097 &-0.023 &0.052 &-0.015 &0.045\\ 
HNCO($6_{1,6}-5_{1,5}$) &0.077 &-0.049 &0.108 &-0.004 &-0.110 &0.161 &0.101 &-0.151 &0.004 &-0.027\\ 
HNCO($6_{0,6}-5_{0,5}$) &0.062 &-0.188 &0.015 &0.142 &0.003 &0.065 &0.026 &0.041 &0.076 &0.048\\ 
HNCO(7$_{1,6}$-6$_{1,5}$) &0.078 &0.056 &0.124 &-0.027 &-0.009 &0.181 &0.213 &0.047 &-0.091 &-0.018\\ 
HNCO($8_{0,8}-7_{0,7}$) &0.082 &-0.087 &0.058 &-0.049 &-0.136 &-0.003 &0.010 &0.056 &0.217 &0.032\\ 
HNCO($9_{0,9}-8_{0,8}$) &0.082 &-0.048 &0.099 &-0.105 &-0.141 &0.014 &0.093 &0.050 &0.142 &-0.019\\ 
HNCO($16_{1,16}-15_{1,15}$) &0.080 &0.109 &0.036 &0.059 &0.061 &0.163 &0.058 &-0.166 &-0.067 &0.023\\ 
HOCO$^+$($4_{0,4}-3_{0,3}$) &0.042 &-0.196 &0.038 &0.276 &0.010 &0.123 &-0.157 &0.069 &-0.117 &0.050\\ 
HOCO$^+$(5$_{0,5}$-4$_{0,4}$) &0.050 &-0.193 &0.023 &0.237 &-0.018 &0.076 &-0.097 &0.045 &-0.023 &-0.032\\ 
HOCO$^+$($6_{0,6}-5_{0,5}$) &0.076 &-0.116 &0.027 &0.124 &-0.108 &0.002 &-0.026 &-0.000 &0.173 &-0.036\\ 
HOCO$^+$($7_{0,7}-6_{0,6}$) &0.074 &-0.056 &0.142 &-0.107 &-0.167 &-0.008 &0.111 &-0.035 &0.226 &-0.155\\ 
H40$\alpha$ &0.068 &0.113 &-0.171 &0.115 &-0.096 &0.015 &0.053 &0.073 &0.009 &0.040\\ 
H39$\alpha$ &0.070 &0.107 &-0.165 &0.103 &-0.099 &0.011 &0.060 &0.062 &0.020 &0.034\\ 
H27$\alpha$ &0.066 &0.118 &-0.161 &0.147 &-0.083 &0.042 &0.124 &0.018 &0.017 &0.053\\ 
H52$\beta$ &0.064 &0.119 &-0.170 &0.150 &-0.108 &0.002 &0.097 &0.065 &-0.014 &-0.017\\ 
H49$\beta$ &0.065 &0.119 &-0.171 &0.141 &-0.106 &0.026 &0.075 &0.027 &-0.011 &0.029\\ 
H58$\gamma$ &0.057 &0.116 &-0.192 &0.168 &-0.139 &0.039 &0.166 &0.031 &0.025 &0.002\\ 
CH$_3$CCH($5_K-4_K$) &0.086 &-0.072 &0.001 &-0.059 &-0.005 &0.172 &-0.052 &-0.018 &0.072 &0.021\\ 
CH$_3$CCH(6$_K$-5$_K$) &0.088 &-0.050 &-0.015 &-0.061 &-0.040 &0.146 &-0.035 &0.001 &0.096 &0.044\\ 
CH$_3$CCH($10_K-9_K$) &0.089 &0.010 &-0.009 &-0.087 &-0.050 &0.097 &-0.039 &-0.040 &-0.019 &0.039\\ 
CH$_3$CCH($11_K-10_K$) &0.083 &0.070 &0.067 &-0.034 &-0.052 &0.232 &0.078 &-0.035 &-0.126 &-0.035\\ 
CH$_3$CCH($12_K-11_K$) &0.089 &0.033 &-0.001 &-0.062 &-0.061 &0.097 &-0.020 &-0.027 &-0.022 &0.011\\ 
SO($3_4-2_3$) &0.090 &-0.036 &0.008 &0.030 &0.012 &-0.041 &-0.057 &0.039 &-0.035 &-0.012\\ 
SO(4$_5$-3$_4$) &0.080 &-0.006 &0.010 &0.102 &0.050 &-0.134 &0.002 &0.140 &-0.128 &-0.748\\ 
$^{34}$SO($2_3-1_2$) &0.086 &0.061 &0.050 &0.013 &-0.075 &0.049 &0.114 &-0.021 &0.040 &-0.106\\ 
$^{34}$SO($6_7-5_6$) &0.070 &0.100 &0.125 &0.061 &0.234 &0.042 &-0.031 &-0.240 &-0.008 &0.017\\ 
H$_2$S($1_{1,0}-1_{0,1}$) &0.088 &-0.010 &-0.040 &0.003 &0.149 &-0.090 &-0.057 &-0.030 &0.013 &-0.010\\ 
H$_2$S($2_{2,0}-2_{1,1}$) &0.082 &0.103 &0.048 &0.053 &0.113 &0.034 &-0.020 &-0.063 &0.029 &0.042\\ 
H$_2$S(3$_{3,0}-3_{2,1}$) &0.075 &0.118 &0.058 &0.099 &0.194 &-0.022 &-0.044 &-0.047 &0.066 &0.065\\ 
OCS(8-7) &0.063 &-0.162 &0.071 &0.187 &-0.024 &0.079 &-0.046 &0.003 &0.097 &0.012\\ 
H$_2^{34}$S($1_{1,0}-1_{1,0,1}$) &0.081 &0.088 &0.077 &0.049 &0.124 &0.020 &0.006 &-0.099 &0.146 &-0.006\\ 
HC$_3$N(10-9) &0.087 &-0.081 &0.033 &0.014 &-0.040 &-0.031 &0.008 &-0.009 &-0.027 &0.032\\ 
HC$_3$N(11-10) &0.088 &-0.066 &0.032 &0.006 &-0.032 &-0.035 &0.004 &-0.014 &-0.018 &0.017\\ 
HC$_3$N(12-11) &0.089 &-0.053 &0.042 &0.003 &-0.042 &-0.054 &0.016 &0.001 &-0.031 &0.015\\ 
HC$_3$N(14-13) &0.090 &-0.022 &0.046 &-0.002 &-0.040 &-0.046 &0.006 &0.005 &-0.049 &0.017\\ 
HC$_3$N(15-14) &0.090 &-0.018 &0.043 &-0.000 &-0.037 &-0.067 &0.013 &0.035 &-0.090 &0.002\\ 
HC$_3$N(17-16) &0.090 &0.009 &0.038 &0.011 &-0.033 &-0.070 &-0.003 &0.083 &-0.077 &0.033\\ 
HC$_3$N(18-17) &0.090 &0.001 &0.033 &-0.010 &-0.009 &-0.086 &0.006 &0.078 &-0.057 &0.025\\ 
HC$_3$N(19-18) &0.089 &0.030 &0.041 &-0.012 &-0.028 &-0.094 &-0.017 &0.094 &-0.089 &0.004\\ 
HC$_3$N(22-21) &0.086 &0.068 &0.072 &0.018 &0.018 &-0.056 &0.031 &0.127 &-0.072 &-0.004\\ 
HC$_3$N(25-24) &0.087 &0.062 &0.051 &0.020 &0.063 &-0.087 &0.021 &0.100 &0.005 &0.042\\ 
HC$_3$N(26-25) &0.086 &0.076 &0.056 &0.034 &0.057 &-0.078 &0.014 &0.104 &-0.050 &-0.003\\ 
HC$_3$N(30-29) &0.084 &0.093 &0.038 &0.048 &0.063 &-0.027 &0.015 &0.110 &0.006 &0.003\\ 
HC$_3$N(37-36) &0.075 &0.111 &0.090 &0.096 &0.192 &0.013 &0.060 &0.023 &0.067 &0.050\\ 
HC$_3$N,v$_7$=1(11$_1$-10$_{-1}$) &0.068 &0.094 &0.149 &0.068 &0.246 &0.019 &0.052 &-0.061 &0.073 &0.053\\ 
H$_3$O$^+$($3_{2,0}-2_{2,1}$) &0.087 &0.046 &-0.077 &-0.023 &0.025 &0.004 &-0.091 &0.065 &-0.004 &0.090\\ 
CCS($7_6-6_5$) &0.086 &0.020 &-0.041 &-0.048 &-0.081 &-0.113 &-0.090 &0.125 &0.080 &-0.025\\ 
CCS($7_8-6_7$) &0.089 &0.001 &0.032 &0.004 &-0.064 &0.016 &-0.084 &0.016 &0.031 &-0.094\\ 
CCS(8$_9$-7$_8$) &0.088 &-0.000 &0.045 &-0.017 &-0.036 &-0.051 &-0.041 &-0.024 &0.072 &-0.075\\ 
CCS($11_{12}-10_{11}$) &0.081 &0.101 &0.054 &0.062 &0.069 &0.046 &0.059 &0.048 &-0.035 &-0.076\\ 
H$_2$CS(3$_{0,3}$-2$_{0,2}$) &0.081 &-0.085 &0.076 &-0.001 &-0.093 &-0.076 &-0.110 &0.018 &0.046 &0.010\\ 
H$_2$CS($5_{1,5}-4_{1,4}$) &0.078 &-0.018 &0.130 &0.033 &0.013 &-0.169 &0.025 &0.142 &-0.028 &-0.118\\ 
H$_2$CS($8_{1,7}-7_{1,6}$) &0.080 &0.087 &0.094 &0.060 &0.098 &0.011 &0.142 &-0.011 &-0.014 &0.044\\ 
HC$_5$N(34-33) &0.070 &-0.022 &0.176 &-0.066 &-0.085 &-0.106 &0.151 &0.260 &-0.335 &0.255\\ 
HC$_5$N(37-36) &0.082 &0.032 &0.140 &-0.030 &-0.047 &0.019 &0.166 &-0.034 &-0.014 &-0.060\\ 
C$_3$H$^+$(4-3) &0.084 &0.070 &-0.017 &-0.064 &-0.016 &0.107 &-0.212 &0.034 &-0.053 &0.018\\ 
C$_3$H$^+$(7-6) &0.065 &0.132 &-0.108 &0.192 &0.007 &0.012 &0.168 &-0.082 &0.126 &0.034\\ 
CH$_2$NH($3_{1,2}-2_{1,1}$) &0.085 &0.047 &0.054 &-0.064 &-0.104 &0.150 &0.050 &-0.152 &-0.090 &-0.029\\ 
CH$_2$NH($4_{1,3}-3_{1,2}$) &0.085 &0.069 &0.090 &0.006 &0.029 &0.053 &0.103 &-0.007 &-0.001 &0.013\\ 
C$_4$H(12-11) &0.085 &0.071 &0.019 &-0.045 &-0.042 &0.132 &-0.040 &0.106 &-0.117 &-0.055\\ 
HCNH$^+$(2-1) &0.080 &-0.054 &0.088 &0.047 &-0.067 &0.130 &-0.104 &-0.038 &0.085 &-0.043\\ 
HCS$^+$(4-3) &0.086 &0.033 &0.063 &-0.068 &-0.050 &-0.042 &-0.059 &-0.025 &-0.046 &-0.083\\ 
c-C$_3$H$_2$($3_{3,0}-2_{2,1}$) &0.088 &0.057 &-0.035 &-0.039 &-0.061 &0.032 &-0.096 &-0.086 &0.005 &-0.016\\ 
NO($3_{1,3}$-2$_{-1,2}$) &0.091 &-0.011 &-0.015 &0.048 &-0.030 &0.008 &-0.091 &0.024 &-0.053 &-0.043\\ 
NO($3_{-1,3}-2_{1,2}$) &0.091 &0.004 &-0.012 &0.043 &-0.018 &0.026 &-0.102 &0.048 &-0.047 &-0.037\\ 
3mm continuum &0.068 &0.101 &-0.182 &0.101 &-0.096 &-0.008 &0.096 &0.112 &0.026 &0.036\\ 
0.8mm continuum &0.085 &0.067 &-0.078 &0.042 &-0.065 &0.101 &-0.039 &-0.036 &0.040 &-0.024\\ 
\enddata 
%\tablecomments{} 
\end{deluxetable*} 
%%%%%%%%%%%%%%%%%%%%%%%%%%%%%%

\section{Estimation of errors in PCA}\label{sec:app:boot}
To estimate errors, we tried the so-called bootstrapping method. We first created the rms error map, then multiplied it with a random number distribution that has the same size as the number of pixels, mean of 0 and standard distribution of 1. We added this noise map to the original image, and ran PCA. We examined how results changed for 1000 different sets of random number distributions for each transition tested. We only used selected transitions for bootstrapping: HCN(1-0) and \ttco(1-0) for high-intensity transitions, H39$\alpha$ for intermediate, HC$_3$N v$_7$=1 (11$_1$-10$_{-1}$), CH$_2$NH(4$_{1,3}$--3$_{1,2}$), and OCS(8-7) for low-intensity transitions. Results are shown in Table \ref{tab:bootstrap}.  For high-intensity transitions, addition of noise did not change results in any significant way, and results are not shown. Variation among 1000 different noise map added is overall small. However, we also note that, for a compact transition such as HC$_3$N v$_7$=1 (11$_1$-10$_{-1}$), an act of adding noise itself changes the PCs more than the variation among different random noise sets.  Nonetheless, qualitative results do not change even for these transitions.

%%%%%%%%%%%%%%%%%%%%%
\begin{deluxetable*}{cccc} 
\tablecolumns{4} 
\tablewidth{0pc} 
%\tabletypesize{\scriptsize} 
\tablecaption{Bootstrapping results \label{tab:bootstrap}}
\tablehead{\colhead{PC} &\colhead{Original} & \colhead{Mean(Bootstrap)} & \colhead{Standard deviation(Bootstrap)}}
\startdata 
\hline
\multicolumn{4}{c}{H39$\alpha$}\\
\hline 
PC1 & 0.070 & 0.070 & 0.000 \\ 
PC2 & 0.107 & 0.107 & 0.001 \\ 
PC3 & -0.165 & -0.165 & 0.001 \\ 
PC4 & 0.103 & 0.102 & 0.001 \\ 
PC5 & -0.099 & -0.099 & 0.002 \\ 
PC6 & 0.011 & 0.011 & 0.003 \\ 
PC7 & 0.060 & 0.060 & 0.003 \\ 
PC8 & 0.062 & 0.063 & 0.004 \\ 
PC9 & 0.020 & 0.021 & 0.004 \\ 
PC10 & 0.034 & 0.034 & 0.005 \\ 
\hline
\multicolumn{4}{c}{HC$_3$N v7=1 (11$_1$-10$_{-1}$)}\\ 
\hline
PC1 & 0.068 & 0.066 & 0.000 \\ 
PC2 & 0.094 & 0.092 & 0.002 \\ 
PC3 & 0.149 & 0.146 & 0.003 \\ 
PC4 & 0.068 & 0.067 & 0.005 \\ 
PC5 & 0.246 & 0.247 & 0.005 \\ 
PC6 & 0.019 & 0.019 & 0.008 \\ 
PC7 & 0.052 & 0.057 & 0.012 \\ 
PC8 & -0.061 & -0.070 & 0.016 \\ 
PC9 & 0.073 & 0.082 & 0.014 \\ 
PC10 & 0.053 & 0.072 & 0.029 \\ 
\hline
\multicolumn{4}{c}{CH$_2$NH(4$_{1,3}$-3$_{1,2}$)}\\ 
\hline
PC1 & 0.085 & 0.084 & 0.000 \\ 
PC2 & 0.069 & 0.068 & 0.001 \\ 
PC3 & 0.090 & 0.089 & 0.002 \\ 
PC4 & 0.006 & 0.006 & 0.003 \\ 
PC5 & 0.029 & 0.030 & 0.003 \\ 
PC6 & 0.053 & 0.054 & 0.005 \\ 
PC7 & 0.103 & 0.105 & 0.006 \\ 
PC8 & -0.007 & -0.007 & 0.008 \\ 
PC9 & -0.001 & -0.002 & 0.007 \\ 
PC10 & 0.013 & 0.013 & 0.010 \\ 
\hline
\multicolumn{4}{c}{OCS(8-7)}\\
\hline
PC1 & 0.063 & 0.063 & 0.000 \\ 
PC2 & -0.162 & -0.162 & 0.000 \\ 
PC3 & 0.071 & 0.071 & 0.001 \\ 
PC4 & 0.187 & 0.187 & 0.001 \\ 
PC5 & -0.024 & -0.023 & 0.001 \\ 
PC6 & 0.079 & 0.079 & 0.002 \\ 
PC7 & -0.046 & -0.045 & 0.003 \\ 
PC8 & 0.003 & 0.002 & 0.004 \\ 
PC9 & 0.097 & 0.099 & 0.003 \\ 
PC10 & 0.012 & 0.011 & 0.005 \\ 
\enddata 
\tablecomments{Transitions with strong emission such as $^{13}$CO(1-0) and HCN(1-0) did not show any significant difference between the original and bootstrapped cases, with very little standard deviation from the errors. Therefore, these transitions are not shown.} 
\end{deluxetable*} 
%%%%%%%%%%%%%%%%%%%%%%%%%%%%%%


\bibliography{pca}{}
\bibliographystyle{aasjournal}



\end{document}

%% file: ALCHEMICollabAuthList.tex
\author[0000-0002-6824-6627]{Nanase Harada}
\affiliation{\naoj}
\affiliation{\asiaa}
\affiliation{\sokendai}

\author[0000-0001-9436-9471]{David S.~Meier}
\affiliation{\nmimt}
\affiliation{\nraos}
%\email{david.meier@nmt.edu}

\author[0000-0001-9281-2919]{Sergio Mart\'in}
\affiliation{\esoc}
\affiliation{\jao}
%\email{Sergio.Martin@eso.org}

\author[0000-0002-9931-1313]{Sebastien Muller}
\affiliation{\oso}
%\email{mullers@chalmers.se}

\author[0000-0001-5187-2288]{Kazushi Sakamoto}
\affiliation{\asiaa}
%\email{ksakamoto@asiaa.sinica.edu.tw}

\author[0000-0002-2501-9328]{Toshiki Saito}
\affiliation{\naoj}

\author[0000-0001-9300-354X]{Mark D.~Gorski}
\affiliation{\oso}
\affiliation{\ciera}

\author[0000-0002-7495-4005]{Christian Henkel}
\affiliation{\mpifr}
\affiliation{\kau}
%\email{chenkel@mpifr-bonn.mpg.de}

\author[0000-0001-8153-1986]{Kunihiko Tanaka}
\affiliation{\keio}
%\email{ktanaka@phys.keio.ac.jp}

\author[0000-0003-1183-9293]{Jeffrey G.~Mangum}
\affiliation{\nraoc}
%\email{jmangum@nrao.edu}

%alphabetical order from here -----------------------------------
\author[0000-0002-5828-7660]{Susanne Aalto}
\affiliation{\chalmers}

\author[0000-0002-1316-1343]{Rebeca Aladro}
\affiliation{\mpifr}
%\email{aladro@mpifr-bonn.mpg.de}

\author[0000-0003-0167-0746]{Mathilde Bouvier}
\affiliation{\leiden}

\author[0000-0001-8064-6394]{Laura Colzi}
\affiliation{\cisc}

\author[0000-0001-6527-6954]{Kimberly L. Emig}
\altaffiliation{Jansky Fellow of the National Radio Astronomy Observatory}
\affiliation{\nraoc}

\author[0000-0002-7758-8717]{Rub\'en~Herrero-Illana}
\affiliation{\esoc}
\affiliation{Institute of Space Sciences (ICE, CSIC), Campus UAB, Carrer de Magrans, E-08193 Barcelona, Spain}
%\email{rherrero@eso.org}

\author[0000-0002-1227-8435]{Ko-Yun Huang}
\affiliation{\leiden}

\author[0000-0002-4052-2394]{Kotaro Kohno}
\affiliation{\ioa}

\author[0000-0001-6174-8467]{Sabine K\"onig}
\affiliation{\oso}

\author[0000-0002-6939-0372]{Kouichiro Nakanishi}
\affiliation{\naoj}
\affiliation{\sokendai}
%\email{nakanisi.k@nao.ac.jp}

\author[0000-0003-0563-067X]{Yuri Nishimura}
\affiliation{\utastro}
\affiliation{\ioa}
\affiliation{ALMA Project, National Astronomical Observatory of Japan, 2-21-1, Osawa, Mitaka, Tokyo 181-8588, Japan}
%\email{yuri@ioa.s.u-tokyo.ac.jp}

\author[0000-0001-6788-7230]{Shuro Takano}
\affiliation{\nihon}
%\email{takano.shuro@nihon-u.ac.jp}

\author[0000-0002-2887-5859]{V\'ictor M. Rivilla}
\affiliation{\cisc}

\author[0000-0001-8504-8844]{Serena Viti}
\affiliation{\leiden}
%\affiliation{\ucl}
%\email{pvdwerf@strw.leidenuniv.nl}

\author[0000-0002-9668-3592]{Yoshimasa Watanabe}
\affiliation{\shibaura}

\author[0000-0001-5434-5942]{Paul P.~van der Werf}
\affiliation{\leiden}

\author[0000-0002-1413-1963]{Yuki Yoshimura}
\affiliation{\ioa}